\documentclass{JHEP3}
\preprint{LU-TP 12-27}

\usepackage{cite}
\usepackage{epsfig}
\usepackage{psfrag}
\usepackage{bbm}
\usepackage{color}
\let\normalcolor\relax
\usepackage{graphics}
\usepackage{axodraw}
\usepackage{inputenc}
\usepackage{xspace}
\usepackage[vcentermath]{youngtab}
\usepackage{amsmath}
\usepackage{amssymb}
\usepackage{url}
\usepackage{rotating}

\inputencoding{latin1}

\usepackage{def} 

\title{\boldmath 
Orthogonal multiplet bases in $\SU(\Nc)$ color space}

\author{Stefan Keppeler$^a$ and Malin Sjödahl$^b$\\
  $^a$Universität Tübingen,  Mathematisches Institut,
  Auf der Morgenstelle~10, 72076~Tübingen, Germany\\
  $^b$Dept. of Astronomy and Theoretical Physics, Lund University, 
  Sölvegatan 14A, 223\,62~Lund, Sweden\\
  E-mail: \email{stefan.keppeler@uni-tuebingen.de}
  and \email{Malin.Sjodahl@thep.lu.se}}
  
\abstract{
We develop a general recipe for constructing orthogonal bases for
the calculation of color structures appearing in QCD for any
number of partons and arbitrary $\Nc$. The bases are constructed using
hermitian gluon projectors onto irreducible subspaces invariant under
$\SU(\Nc)$.  Thus, each basis vector is associated with an irreducible
representation of $\SU(\Nc)$. The resulting multiplet bases are not
only orthogonal, but also minimal for finite $\Nc$. As a
consequence, for calculations involving many colored particles, the
number of basis vectors is reduced significantly compared to standard
approaches employing overcomplete bases. We exemplify the
method by constructing multiplet bases for all processes involving a
total of $6$ external colored partons.

}

\allowdisplaybreaks 
\begin{document}

\vspace*{-1 cm}

\section{Introduction}
\label{sec:motivation}

With the start of the Large Hadron Collider follows an increased demand
for accurately calculated processes in perturbative quantum
chromodynamics (QCD), as the higher energies open up for events with
more colored partons. A major challenge for these calculations is the
complication brought about by the non-abelian gauge structure in QCD.

Several methods have been developed for treating the color structure
in special cases
\cite{Zeppenfeld:1988bz,DelDuca:1999rs,Dokshitzer:2005ig,
  Kyrieleis:2005dt, Sjodahl:2008fz}. The most general, and probably
most widely used approach for exact calculations employs a
decomposition of the color space into open and closed quark-lines
\cite{Paton:1969je,Dittner:1972hm,Cvi76,Cvitanovic:1980bu,Mangano:1987xk,Mangano:1988kk,Nagy:2007ty,Platzer:2012np,Sjodahl:2009wx}, i.e.\ linear
combinations of terms like
\begin{equation}
\label{eq:trace_basis_example}
  \tr[t^{g_1}t^{g_2}t^{g_3}]\;(t^{g_4}t^{g_5}t^{g_6})^{q_1}_{q_2} 
  = \ \parbox{7cm}{\epsfig{file=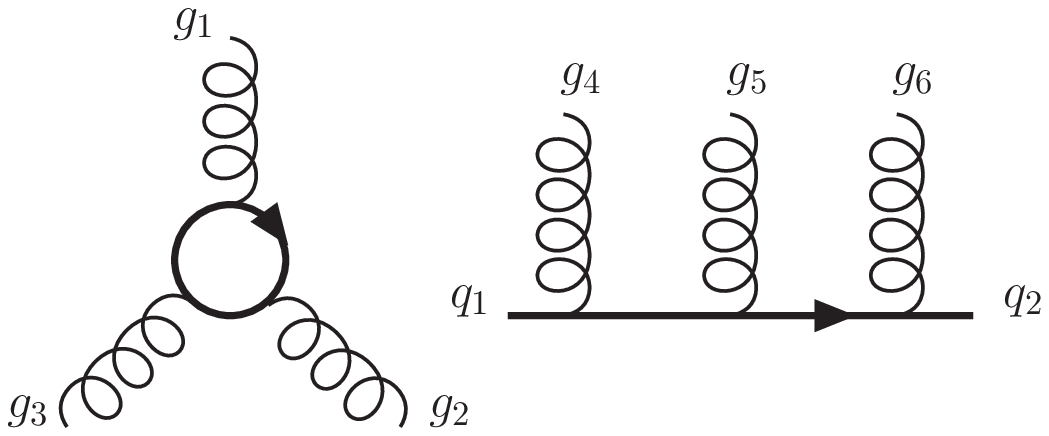,width=7cm}} \, , 
\end{equation}
where the involved partons may be combined to form any structures
allowed in QCD.  Here $t^{g_j}$, $g_j=1,\hdots,\Nc^2-1$ denotes a
generator in the fundamental representation of $\SU(\Nc)$,
$\Nc=3$ for QCD,  
and $q_{1,2}=1,...,\Nc$.  We refer to this type of basis as a
{\it trace basis}.  Any amplitude, at tree level and beyond, can be
decomposed in this way, and for fixed order calculations only a small
fraction of all possible products of open and closed quark-lines have
non-vanishing amplitudes.  For calculations involving many partons,
approximative Monte Carlo techniques
\cite{Caravaglios:1998yr,Maltoni:2002mq,Papadopoulos:2005ky,
  Duhr:2006iq,Giele:2009ui,vanHameren:2009dr} and, for higher
efficiency, the color-flow basis \cite{Maltoni:2002mq}, may be
employed.

Two drawbacks of the trace basis are that it is non-orthogonal 
and, in general, overcomplete, i.e.\ it is not a proper basis but just a
spanning set. In an alternative approach the state spaces of incoming
and outgoing partons are decomposed into multiplets, i.e.\ into
irreducible subspaces invariant under $\SU(3)$, or, more generally,
$\SU(\Nc)$. It is then possible to construct (minimal) orthogonal
bases for color spaces. We refer to this kind of basis as a {\it
  multiplet basis}. These bases have the potential to significantly
speed up QCD color calculations. However, to the best of our
knowledge, multiplet bases have so far only been employed for
processes with up to five colored partons
\cite{Sotiropoulos:1993rd,Kidonakis:1998nf,Kyrieleis:2005dt,
  Dokshitzer:2005ig,Sjodahl:2008fz,Beneke:2009rj}, typically in the
context of resummation. One reason is that, in general, the
construction of these bases is far from obvious. This is the problem
we want to shed light on in this article.

Our main result is a general recipe for constructing orthogonal
multiplet bases for QCD processes with an arbitrary number of quarks
and gluons, to arbitrary order in perturbation theory and for
arbitrary $\Nc$. We explicitly demonstrate the method 
by constructing bases for all processes with six colored partons.

This article is organized as follows: In the remainder of the
introduction we discuss the notion of color space
(\secref{sec:ColorSpace}), review the trace basis approach
(\secref{sec:TraceBasis}) and set the stage for our method with an
example in \secref{sec:illustration_gg_to_gg}.  Thereafter we address
the construction of projection operators
(\secref{sec:QuarkProjectors}) and basis vectors
(\secref{sec:QuarkBasis}) in the quarks-only case. We discuss the
importance of hermitian projection operators, and define quark
projectors which we use in the subsequent construction of gluon
projectors. In \secref{sec:Gluons} we address the considerably more
involved task of constructing projection operators for an arbitrary
number of gluons. Starting from these projectors, we outline in
\secref{sec:GluonBasis} how to build orthogonal bases for processes
involving only gluons. After having addressed the construction in the
gluon-only case, we find in \secref{sec:QuarksAndGluons} that the
extension to processes involving both quarks and gluons is
straightforward. We conclude with some remarks in
\secref{sec:conclusions}.
 
\subsection{Color space}
\label{sec:ColorSpace}

Consider a process with a certain number of incoming and outgoing
quarks, anti-quarks and gluons. We denote by $\Nq$ the number of
outgoing quarks plus the number of incoming anti-quarks, and by $\Ng$
the number gluons (incoming plus outgoing).  Due to the QCD Feynman
rules the number of incoming quarks plus the number of outgoing
anti-quarks also has to equal $\Nq$. Focusing on the color degrees of
freedom, i.e.\ ignoring spin and momentum, quark states are elements of
$V = \mathbb{C}^{\Nc}$ and transform under the fundamental or defining
representation of $\SU(\Nc)$, anti-quarks states are elements of the
dual space $\overline{V} \cong \mathbb{C}^{\Nc}$ and transform in the
complex conjugate of the fundamental representation, whereas gluons
transform in the adjoint representation, i.e.\ gluon states are
elements of a real $\Nc^2-1$-dimensional vector space which we
complexify to $A \cong \mathbb{C}^{\Nc^2-1}$. Thus, with a QCD
amplitude is associated a tensor $\Col \in (V \otimes
\overline{V})^{\otimes \Nq} \otimes A^{\otimes \Ng}$, its {\it color
  structure}.

Let us briefly remark on some conventions. We refer to
$\SU(\Nc)$-invariant irreducible subspaces as {\it multiplets}. A
multiplet carries an irreducible representation of $\SU(\Nc)$.
As there is a unique irreducible representation associated with each
multiplet, we often use the two terms interchangeably, e.g.\ we refer
to the trivial representation as singlet or to the adjoint
representation of $\SU(3)$ as octet.

\FIGURE[t]{\epsfig{file=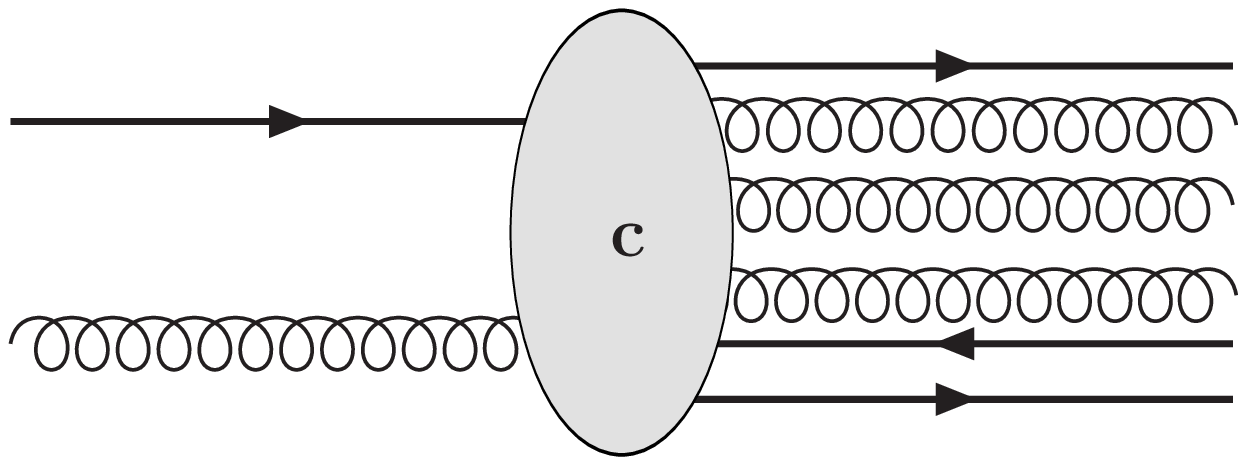,width=.7\textwidth}
  \caption{\label{fig:ColorStructure} A color structure for a process
    with $\Nq=2$ and $\Ng=4$.}}

Eventually we will use Cvitanovi\'c's birdtrack notation
\cite{Cvi76,Cvi08} representing a tensor $\Col \in (V \otimes
\overline{V})^{\otimes \Nq} \otimes A^{\otimes \Ng}$ as a blob with
$2\Nq+\Ng$ legs, where straight lines with outward pointing
arrows correspond to outgoing quarks (or incoming anti-quarks),
straight lines with inward pointing arrows correspond to incoming
quarks (or outgoing anti-quarks) and curly lines correspond to
gluons\footnote{Cvitanovi\'c \cite{Cvi76,Cvi08} represents gluons by thin instead of
  curly lines.}, see \figref{fig:ColorStructure}. Inside the blob the
lines can be connected directly or via any number of
\begin{equation}
\label{eq:t_and_f_def}
\begin{split} 
  \text{quark-gluon vertices,} \qquad 
  \epsfig{file=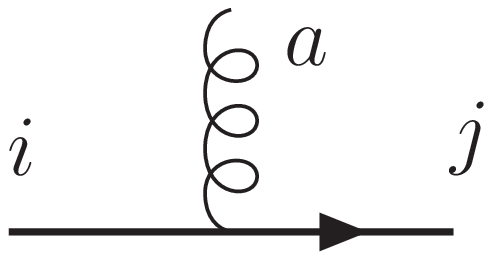,width=2.5cm} &= (t^a)^i_{\phantom{i}j} 
  \qquad \text{and} \\
  \text{triple-gluon vertices,} \qquad 
  \parbox{2.5cm}{\epsfig{file=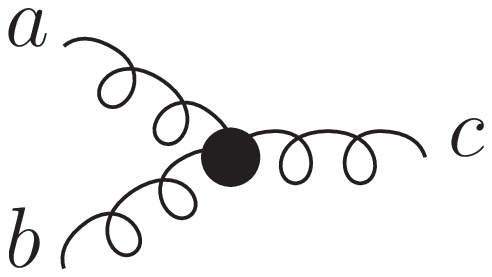,width=2.5cm}} 
  &= \ui f_{abc} 
  \, , 
\end{split}
\end{equation} 
where $i,j = 1,\hdots,\Nc$, $a,b,c = 1,\hdots,\Nc^2-1$ and the indices
in the triple gluon vertex are to be read {\it anti-clockwise}.  Here
$t^a$ denotes a generator of $\SU(\Nc)$ in the fundamental
representation and $f_{abc}$ are the $\SU(\Nc)$ structure
constants. We do not include the four gluon vertex in this list since
its color part can be built from linear combinations of (one-gluon)
contracted products of two triple-gluon vertices.

Since QCD-processes conserve color we are only interested in color
structures that are overall singlets, 
i.e.\ invariant
tensors, see \appref{sec:invariant_tensors}. Thus, we define 
the {\it color space} as the color singlet subspace of
$(V\otimes\overline{V})^{\otimes \Nq} \otimes A^{\otimes\Ng}$,
i.e.\ the span of all tensors that transform under the trivial
representation of $\SU(\Nc)$.  For instance, consider a process with
two incoming and two outgoing quarks, $qq \to qq$ for short. The color
space for this process is spanned by the singlets in $\overline{V}
\otimes \overline{V} \otimes V \otimes V$ and as such has dimension 2,
with a possible basis being given by the linear operators
$\Proj_s,\ \Proj_a : V \otimes V \to V \otimes V$ projecting onto the
symmetric (sextet) and anti-symmetric (anti-triplet) tensors in $V
\otimes V$, respectively.
 
We are only interested in color summed (averaged) cross sections, which depend on 
the norm squared of the color structure, 
\begin{equation}
\label{eq:norm}
  \|\Col\|^2 = \langle \Col | \Col \rangle \, , 
\end{equation}
where the scalar product is given by summing over all external color
indices, i.e.
\begin{equation}
  \left\langle \Col_1 | \Col_2 \right\rangle
  =\sum_{a_1,\,a_2,\,...}\Col_1^{* a_1\,a_2...} \, \Col_2^{a_1\,a_2...} 
\label{eq:scalar_product}
\end{equation}
with $a_i=1,\hdots,\Nc$ if parton $i$ is a quark or anti-quark and
$a_i=1,\hdots,\Nc^2-1$ if parton $i$ is a gluon. For $\Ng$ even,
i.e.\ $\Ng=2\ng$, color structures $\Col \in
(V\otimes\overline{V})^{\otimes \Nq} \otimes A^{\otimes2\ng}$ can be
viewed as linear operators $\Col: V^{\otimes\Nq} \otimes
A^{\otimes\ng} \to V^{\otimes\Nq} \otimes A^{\otimes\ng}$
and the
scalar product (\ref{eq:scalar_product}) reads
\begin{equation}
\label{eq:sp_trace}
  \left\langle \Col_1 | \Col_2 \right\rangle
  = \tr ( \Col_1^\dag \Col_2 ) \, .
\end{equation}
Out of these operators the hermitian projectors, 
\begin{equation}
  \Proj_{i_1\,...i_{n_q},o_1...o_{n_q}}=
  (\Proj^\dagger)_{i_1\,...i_{n_q},o_1...o_{n_q}}=
  (\Proj_{o_1...o_{n_q},i_1\,...i_{n_q}})^*
\end{equation}
onto $\SU(\Nc)$-invariant subspaces of $V^{\otimes\Nq} \otimes
A^{\otimes\ng}$ play a special role. They are examples of color
singlets in 
$(V\otimes\overline{V})^{\otimes \Nq} \otimes A^{\otimes2\ng}$. 
If we chose them to be mutually transversal,
\begin{equation}
\label{eq:transversal_projectors}
  \Proj_j \Proj_k = 0 \quad \forall\ j \neq k \, , 
\end{equation}
i.e.\ the image of each projector is contained in the kernel of all
the others, then hermiticity implies that they project onto mutually
orthogonal subspaces, and that the projectors are themselves
orthogonal with respect to the scalar product (\ref{eq:sp_trace}),
\begin{equation}
  \langle \Proj_j | \Proj_k \rangle 
  = \tr(\Proj_j^\dag \Proj_k)
  = \tr(\Proj_j \Proj_k) = 0 
  \quad \forall\ j \neq k \, .
  \label{eq:PidotPj}
\end{equation}
These projectors can therefore be used in the construction of
orthogonal bases.  Denoting by $d_j$ the dimension of the image of
$\Proj_j$ we also find
\begin{equation}
  \|\Proj_j\|^2 
  = \tr(\Proj_j^\dag \Proj_j) 
  = \tr(\Proj_j^2) 
  = \tr(\Proj_j)
  = d_j \, , 
\label{eq:HermProjSquare}
\end{equation}
and thus $\Proj_j/\sqrt{d_j}$ is normalized with respect to the scalar
product (\ref{eq:scalar_product}).

In the example of $qq\to qq$ above, the color space was spanned by these
projectors alone. In general this is not the case. If the same
multiplet appears several times in the decomposition of $V^{\otimes\Nq}
\otimes A^{\otimes\ng}$, then also operators describing transitions
from one instance of a multipet to a different instance of the same
multiplet constitute linearly independent vectors in color space.

Hermitian projectors onto $\SU(\Nc)$-invariant
irreducible subspaces of $A^{\otimes\ng}$ will be our starting point
for the construction of orthogonal bases of the color space within
$A^{\otimes 2\ng}$. Then we will see that these projectors also enable
the construction of orthogonal bases for the color space for
$A^{\otimes\ng} \to A^{\otimes(\ng+1)}$, i.e.\ for the color singlet
space within $A^{\otimes(2\ng+1)}$.  Finally, when there are external
quarks, we take advantage of $V \otimes \overline{V} = \bullet \oplus
A$, where $\bullet$ denotes the singlet, i.e.\ a subspace transforming
under the trivial representation. This implies
\begin{equation}
\label{eq:VVbar_binomi}
  (V \otimes \overline{V})^{\otimes\Nq} 
  = (\bullet \oplus A)^{\otimes\Nq} 
  = \bigoplus_{\nu=0}^{n_q} 
    \left( \begin{smallmatrix} \displaystyle n_q \\[1ex] 
                               \displaystyle \nu \end{smallmatrix} \right)
     A^{\otimes\nu} \, , 
\end{equation}
where $A^{\otimes0} = \bullet$, and thus, we are able to construct
orthogonal bases for the color spaces within $(V \otimes
\overline{V})^{\otimes\Nq} \otimes A^{\otimes 2\ng}$ or $(V \otimes
\overline{V})^{\otimes\Nq} \otimes A^{\otimes(2\ng+1)}$, 
as soon as we have constructed the projectors for
$A^{\otimes\nu}\ \forall\ \nu=1,\hdots,\Nq+\ng$.

\subsection{Trace bases}
\label{sec:TraceBasis}

For tree level processes involving only gluons, the most popular way
to keep track of the color structure is probably to use a basis
consisting of traces over $\SU(3)$ generators
\cite{Paton:1969je,Dittner:1972hm,Cvi76,Cvitanovic:1980bu,
  Mangano:1987xk,Mangano:1988kk,Nagy:2007ty,Sjodahl:2009wx}.  A
general amplitude $\mathcal{A}$ can then be written as 
\begin{equation}
  \mathcal{A} = \sum_{\sigma \in S_{\Ng-1}} \mathcal{A}_{\sigma}
      \tr[t^1 t^{\sigma(2)} \cdots t^{\sigma(\Ng)}],
  \label{eq:A}
\end{equation}
where $\sigma$ denotes a permutation, i.e.\ $\mathcal{A}$ is a sum over (color)
scalar subamplitudes $\mathcal{A}_{\sigma}$ (also referred to as color ordered,
dual or partial amplitudes) multiplying the color structures given by
the traces. 
Note that fixing the position of the first generator does not
impose any restriction due to the cyclicity of the trace. 
For tree-level
gluon-only processes there are thus $(\Ng-1)!$ basis vectors. In
diagrammatic notation these traces are quark loops with $\Ng$ gluon
lines attached. That every tree-level gluon amplitude can be written in
this way can be seen as follows \cite{Cvi76}. Consider any tree level
diagram and first rewrite any four-gluon vertex in terms of three
gluon vertices. Then replace the triple gluon vertices using
\begin{equation}
\label{eq:f}
  \ui f_{abc} = \frac{1}{\TR} \big[ \tr(t^a t^b t^c)-\tr(t^b t^a t^c) \big]  
  \quad \Leftrightarrow \quad 
  \parbox{1.5cm}{\epsfig{file=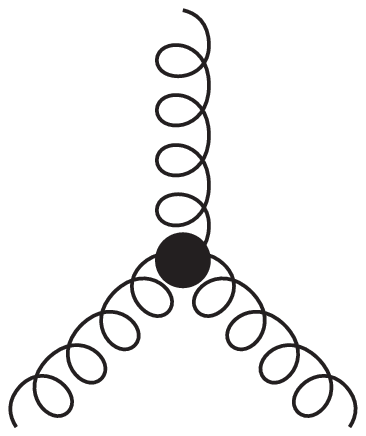,width=1.5cm}}
  = \frac{1}{\TR} \left[ 
    \parbox{1.5cm}{\epsfig{file=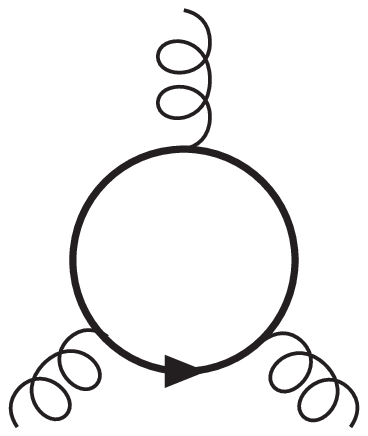,width=1.5cm}}
    - \parbox{1.5cm}{\epsfig{file=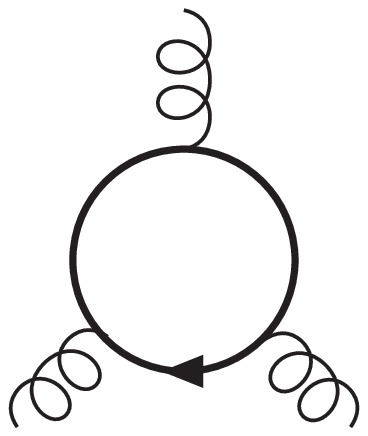,width=1.5cm}} \right] , 
\end{equation}
where the arbitrary normalization constant $\TR$ is defined by 
\begin{equation}
  \tr[t^a t^b] = \TR \, \delta_{ab}
  \quad \Leftrightarrow \quad 
  \parbox{2.4cm}{\epsfig{file=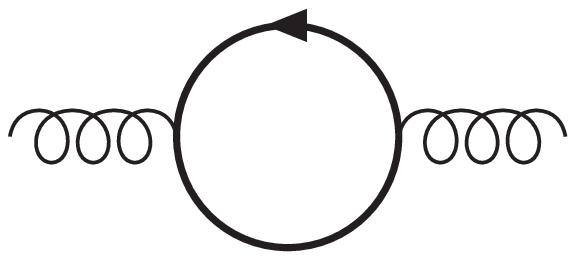,width=2.4cm}} 
  = \TR \parbox{1.2cm}{\epsfig{file=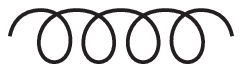,width=1.2cm}} \, .
\end{equation}
Finally, remove every internal gluon propagator using the Fierz-type identity
\begin{equation}
  (t^a)^i_j(t^a)^k_l = \TR 
  \Big[ \delta^i_l\delta^k_j-\frac{1}{\Nc}\delta^i_j\delta^k_l \Big] 
  \quad \Leftrightarrow \quad 
  \parbox{2cm}{\epsfig{file=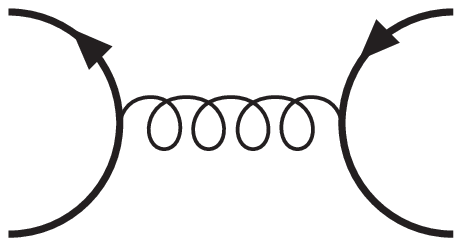,width=2cm}}
  = \TR \left[ 
    \parbox{1.5cm}{\epsfig{file=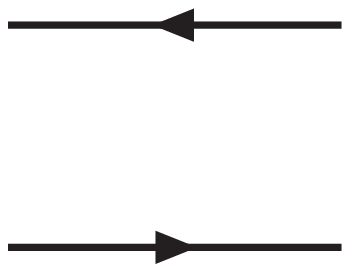,width=1.5cm}}
    - \frac{1}{\Nc} 
      \parbox{1.5cm}{\epsfig{file=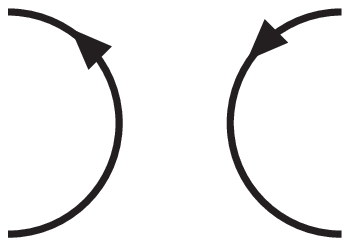,width=1.5cm}} \right] .
  \label{eq:Fierz}
\end{equation}
Noting that the color suppressed terms drop out, see
e.g.\ \cite[sec. 9.14]{Cvi08}, the final result is a sum of traces of
the form given in \eqref{eq:A}. At loop level it is necessary to also
incorporate basis vectors which are products of traces. In general,
considering processes to order $N_{\mathrm{loop}}$, it is necessary to
include states which are direct products of up to $N_{\mathrm{loop}}$
different traces. As $\tr[t^a]=0$, the basis vectors for calculations
to arbitrary order in the coupling constant have at most $\Ng/2$
traces since each trace has to contain at least two generators.
Considering all ways of partitioning $\Ng$ gluons into traces does
thus always give a basis which can be used to any order in
perturbation theory.  This basis is complete for $\Ng \leq \Nc$, but
it is overcomplete for $\Ng > \Nc$
\cite{Dittner:1972hm,Cvi76}. Moreover, it is not orthogonal.  This is
a significant drawback due to the rapid growth of the number of basis
vectors with the number of external gluons (partons in general).
Counting the number of basis vectors can be reduced to the problem of
mapping $\Ng$ units to $\Ng$ units without mapping a single one to
itself (no generator can stand alone inside a trace). There are thus
\begin{equation}
  \mbox{subfactorial}(\Ng) 
  = \Ng! \sum_{\nu=0}^{\Ng} \frac{(-1)^\nu}{\nu!}
  \approx \Ng!/e 
  \label{eq:Nv}
\end{equation}
basis vectors, giving rise to $\approx (\Ng!/e)^2$ terms when
calculating scalar products.

For processes involving quarks the basis may be constructed similarly,
by starting with connecting all $\Nq$ quark ends to the $\Nq$
anti-quark ends, and then attaching the gluons in all possible ways to
these open quark lines. Again, at loop level, new color structures
have to be considered. For calculations up to $N_{\mathrm{loop}}$ we,
in general, also have to include color structures which, in addition
to the $\Nq$ open quark lines, also have up to $N_{\mathrm{loop}}$
closed quarks lines, i.e.\ traces of subsets of generators. Again the
basis vectors will be non-orthogonal, and the number of basis vectors
will grow roughly like a factorial.  The exact number of basis vectors
for a total of $\Ng$ gluons and $\Nq$ $\qqbar$-pairs can be found
using the recursion relation
\begin{equation}
  N_{\mbox{\tiny vec}}[\Nq, \Ng]
  = N_{\mbox{\tiny vec}}[\Nq, \Ng-1](\Ng-1+\Nq)
    + N_{\mbox{\tiny vec}}[\Nq, \Ng-2](\Ng-1) \, , 
\label{eq:Nvqg}
\end{equation}
with 
\begin{eqnarray}
  N_{\mbox{\tiny vec}}[\Nq, 0]&=&\Nq!\;,\;\;\;  
  N_{\mbox{\tiny vec}}[\Nq, 1]=\Nq \Nq!\;,
  \label{eq:start_cond}
\end{eqnarray}
or, alternatively, by using an exponential generating function
\cite{Cvi76,WolfdieterLang}.  The first term in \eqref{eq:Nvqg} comes from
attaching the new gluon line to any of the existing (open or closed)
quark lines, whereas the last term comes from basis vectors in which
the generator for the new gluon stands inside the same trace as {\it
  one} of the $\Ng-1$ other gluons.

For special cases the number of degrees of freedom for the sub-amplitudes
have been seen to reduce significantly,
and powerful recursion relations have been derived.
Especially, this is the case 
for tree-level pure Yang-Mills theory, as in \eqref{eq:A},
\cite{Parke:1986gb,Mangano:1987xk,Kleiss:1988ne,Berends:1987me,Cachazo:2004kj,Britto:2004ap,Bern:2008qj,BjerrumBohr:2010ta,BjerrumBohr:2010zb}.
While these
strategies may significantly reduce the computational effort in the
situations they are tailored for, we here pursue a general approach.
We aim for minimal orthogonal bases, which can be used for any number
and kind of partons, and to any order in perturbation theory. We
demonstrate that such bases can be constructed using hermitian transversal
projectors onto different irreducible representations.  
The resulting bases are
orthogonal, and can easily be chosen minimal for any finite $\Nc$, such
as $\Nc=3$.

\subsection{Illustration: $gg\to gg$}
\label{sec:illustration_gg_to_gg}

Our method will be based on first constructing hermitian projectors
which decompose $A^{\otimes \ng}$ into irreducible subspaces invariant
under $\SU(\Nc)$. We will then show how these can be used for
constructing complete orthogonal bases, for processes involving up to
$2\ng+1$ gluons, and processes where a subset of the gluons has been
replaced by $\qqbar$-pairs. Let us sketch this procedure for $\ng=2$.

The $\SU(\Nc)$ irreducible representations involved in the decomposition of
$A^{\otimes\ng}$ can, e.g., be obtained, by multiplying the
corresponding Young diagrams \cite{Ham62},
\begin{eqnarray}
\Yboxdim{10pt}
  \begin{array}{ccccccccccccccccc}
    \upsmall{\Nc-1} \csep \upsmall{1} 
    & & \upsmall{\Nc-1} \csep \upsmall{1} 
    & & \upsmall{\Nc}
    & & \upsmall{\Nc-1} \csep \upsmall{1} 
    & & \upsmall{\Nc-1} \csep \upsmall{1} 
    & & \upsmall{\Nc-2} \csep \upsmall{1} \csep \upsmall{1}
    & & \upsmall{\Nc-1} \csep \upsmall{\Nc-1} \csep \upsmall{2}
    & & \upsmall{\Nc-1} \csep \upsmall{\Nc-1} \csep 
        \upsmall{1} \csep \upsmall{1}
    & & \upsmall{\Nc-2} \csep \upsmall{2}
    \\
    \yng(2,1)   
    & \otimes 
    &\yng(2,1)  
    & \ = \ 
    & \bullet 
    & \oplus 
    & \yng(2,1) 
    & \oplus 
    & \yng(2,1) 
    & \oplus 
    & \yng(3) 
    & \oplus 
    &  \yng(3,3)
    & \oplus 
    &  \yng(4,2)
    & \oplus 
    & \circ 
    \\[1ex]
    8
    & & 8 
    & & 1 
    & & 8\,
    & & 8 \,
    & & 10
    & & \overline{10} 
    & & 27
    & & 0 
  \end{array}
  \, .
\label{eq:SU388}
\end{eqnarray}
Here and in the following we represent irreducible representations in
several ways: On the first line we uniquely specify the multiplets in
terms of the lengths of the {\it columns} of the corresponding Young
diagrams. On the second line we specialize to $\Nc=3$ displaying
actual Young diagrams. There we denote by $\circ$ any irreducible representation
that does not exist for $\Nc=3$ but only for sufficiently large $\Nc$. 
Also recall that
we denote by $\bullet$ the trivial rep, i.e. $\bullet =
\Yboxdim{5pt}\yng(1,1,1)$ for $\Nc=3$. Finally, on the third line we
give the dimensions of the respective $\SU(3)$ multiplets.
 
Hermitian projectors corresponding to \eqref{eq:SU388} have been given
in several places. The earliest reference known to us is
\cite{MacFarlane:1968vc}, where they are given for $\Nc=3$. 
A derivation for arbitrary $\Nc$ in terms of birdtracks is given by
Cvitanovi\'c in \cite[sec.~6.D \& tab.~6.3]{Cvi84}, see also
\cite[sec.~9.12 \& tab.~9.4]{Cvi08}. Cvitanovi\'c employs
characteristic equations for invariant matrices in order to construct
the projectors. Our approach described in \secref{sec:Gluons}, which
is inspired by \cite{Cvi08}, will avoid factorizing characteristic
equations but instead provide an algorithm for directly writing down
the projectors. A slightly different diagrammatic derivation, also for
arbitrary $\Nc$, is given by Dokshitzer and Marchesini
\cite{Dokshitzer:2005ig}. Our construction for a certain class of
projectors in \secref{sec:new_multiplets} is a generalization of their
method. For the moment we list the two gluon projectors without
derivation,

\begin{eqnarray}
  \Proj^{1}_{g_1\, g_2\, g_3\,g_4}
  &=& \frac{1}{\Nc^2-1}\delta _{{g_1\,g_2}} \delta _{{g_3\,g_4}} \, ,
  \nonumber \\
  \Proj^{8s}_{g_1\, g_2\, g_3\,g_4}
  &=& \frac{\Nc}{2 T_R (\Nc^2-4)} d_{{g_1 g_2 i_1}} d_{{i_1 g_3 g_4}} \, ,
  \nonumber\\
  \Proj^{8a}_{g_1\, g_2\, g_3\,g_4}
  &=& \frac{-1}{2 \Nc T_R} \ui f_{g_1\,g_2\,i_1} \ui f_{i_1\,g_3\,g_4} \, ,
  \nonumber\\
  \Proj^{10}_{g_1\, g_2\, g_3\,g_4}
  &=& \frac{1}{4} (\delta_{g_1\,i_1} \delta_{g_2\,i_2} 
                   - \delta_{g_1\,i_2} \delta_{g_2\,i_1})
      \left[ \delta_{i_1\,g_3}\delta_{i_2\,g_4} 
             + \frac{1}{T_R^2} \tr(t^{i_1} t^{g_4} t^{i_2} t^{g_3}) \right]
      - \frac{1}{2} \, \Proj^{8a}_{g_1\, g_2\, g_3\,g_4} \, ,
  \nonumber\\
  \Proj^{\overline{10}}_{g_1\, g_2\, g_3\,g_4}
  &=& \frac{1}{4} (\delta_{g_1\,i_1} \delta_{g_2\,i_2} 
                   - \delta_{g_1\,i_2} \delta_{g_2\,i_1})
      \left[ \delta_{i_1\,g_3}\delta_{i_2\,g_4} 
             - \frac{1}{T_R^2} \tr(t^{i_1} t^{g_4} t^{i_2} t^{g_3}) \right]
      - \frac{1}{2} \, \Proj^{8a}_{g_1\, g_2\, g_3\,g_4} \, ,
  \nonumber\\
  \Proj^{27}_{g_1\, g_2\, g_3\,g_4}
  &=& \frac{1}{4} (\delta_{g_1\,i_1} \delta_{g_2\,i_2} 
                   + \delta_{g_1\,i_2} \delta_{g_2\,i_1})
      \left[ \delta_{i_1\,g_3}\delta_{i_2\,g_4} 
             + \frac{1}{T_R^2} \tr(t^{i_1} t^{g_4} t^{i_2} t^{g_3}) \right]
      \nonumber\\ &&
      - \frac{\Nc-2}{2\Nc} \, \Proj^{8s}_{g_1\, g_2\, g_3\,g_4} 
      - \frac{\Nc-1}{2\Nc} \, \Proj^{1}_{g_1\, g_2\, g_3\,g_4} \, ,
  \nonumber\\
  \Proj^{0}_{g_1\, g_2\, g_3\,g_4}
  &=& \frac{1}{4} (\delta_{g_1\,i_1} \delta_{g_2\,i_2} 
                   + \delta_{g_1\,i_2} \delta_{g_2\,i_1})
      \left[ \delta_{i_1\,g_3}\delta_{i_2\,g_4} 
             - \frac{1}{T_R^2} \tr(t^{i_1} t^{g_4} t^{i_2} t^{g_3}) \right]
      \nonumber\\ &&
      - \frac{\Nc+2}{2\Nc} \, \Proj^{8s}_{g_1\, g_2\, g_3\,g_4} 
      - \frac{\Nc+1}{2\Nc} \, \Proj^{1}_{g_1\, g_2\, g_3\,g_4} \, ,
  \label{eq:ggBasis}
\end{eqnarray}
where we have introduced the totally symmetric tensor
\begin{equation}
\label{eq:d_def}
  d_{abc} 
  := \frac{1}{T_R} \, \big[ \tr(t^a t^b t^c) + \tr(t^b t^a t^c) \big]
  =  \parbox{2.5cm}{\epsfig{file=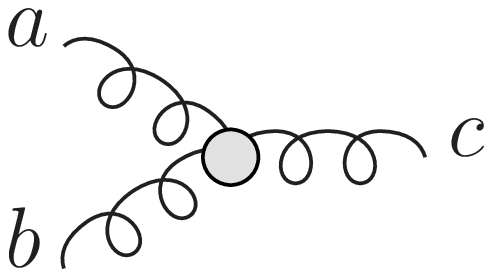,width=2.5cm}}. 
\end{equation}
In \eqref{eq:ggBasis},
and often in the following, we label projection operators by the
dimensions of $\SU(3)$ multiplets, although our construction is for
arbitrary $\Nc$, and for $\Nc \ne 3$ the dimensions differ. If a
multiplet appears several times we add some additional label, as for
the octets above. Since $\Proj^0$ vanishes for $\Nc=3$, we also have
$\Proj^{27} = \Proj^{27} + \Proj^{0}$ in this case, which allows to
write $\Proj^{27}$ in a simpler form,
\begin{equation}
  \Proj^{27}_{g_1\,g_2\,g_3\,g_4} 
  \underset{\Nc=3}{=}\
  \frac{1}{2} (\delta_{g_1\,g_3} \delta_{g_2\,g_4} 
               + \delta_{g_1\,g_4} \delta_{g_2\,g_3})
  - \Proj^{8s}_{g_1\, g_2\, g_3\,g_4} - \Proj^{1}_{g_1\, g_2\, g_3\,g_4}\;.
\end{equation}
This is the way in which $\Proj^{27}$ is given in
\cite{MacFarlane:1968vc}. As gluons transform in a real
representation, for processes involving only gluons, the decuplet
projectors occur only in the real combination
\cite{MacFarlane:1968vc,Kidonakis:1998nf,Oderda:1999kr,Sjodahl:2008fz,Sjodahl:2009wx}
\begin{equation} 
  \left(\Proj^{10}+\Proj^{\overline{10}}\right)_{g_1\, g_2\, g_3\,g_4}
  = \frac{1}{2} (\delta_{g_1\,g_3} \delta_{g_2\,g_4} 
                 - \delta_{g_1\,g_4} \delta_{g_2\,g_3})
    - \Proj^{8a}_{g_1\, g_2\, g_3\,g_4} \, .
\end{equation}
However, for processes involving quarks $\Proj^{10}$ and
$\Proj^{\overline{10}}$ can appear independently. In birdtrack
notation \cite{Cvi76,Cvi08} \eqref{eq:ggBasis} reads
\begin{align}
  \label{eq:ggBasis_birdtracks}
  \nonumber
  \Proj^1 &= \frac{1}{\Nc^2-1} 
    \parbox{1.7cm}{\epsfig{file=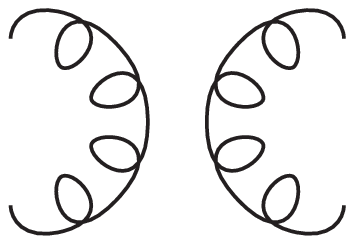,width=1.7cm}} 
  \\ \nonumber
  \Proj^{8s} &= \frac{\Nc}{2 T_R (\Nc^2-4)}
    \parbox{2.2cm}{\epsfig{file=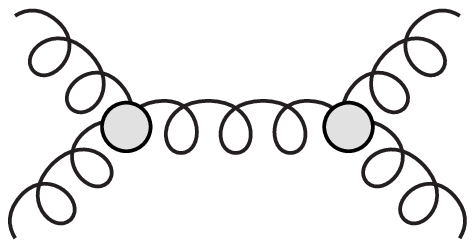,width=2.2cm}} 
  \\ \nonumber
  \Proj^{8a} &= \frac{1}{2 \Nc T_R}
    \parbox{2.2cm}{\epsfig{file=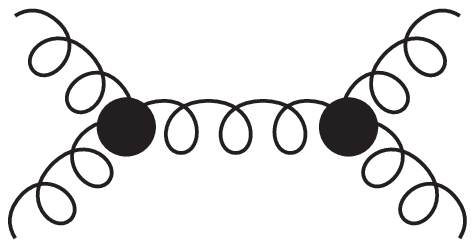,width=2.2cm}} 
  \\
  \Proj^{10} 
  &= \frac{1}{2} \parbox{2.2cm}{\epsfig{file=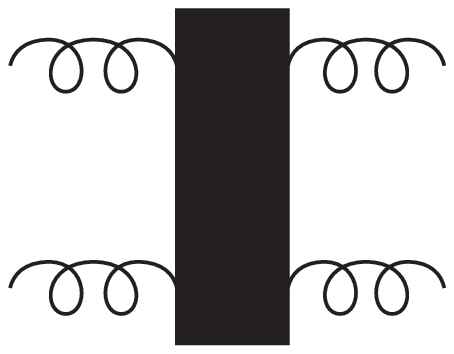,width=2.2cm}}
     + \frac{1}{2T_R^2} 
       \parbox{3.5cm}{\epsfig{file=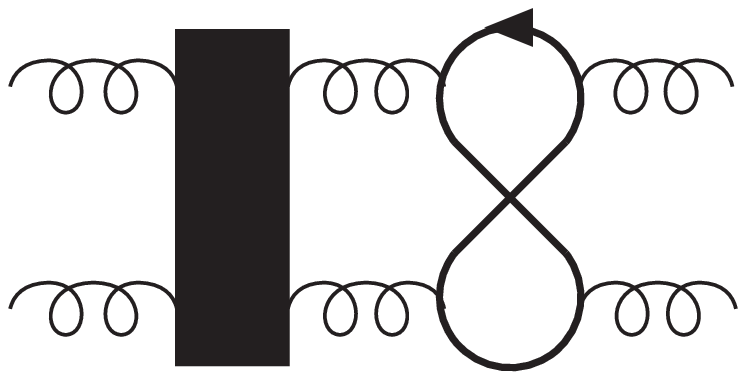,width=3.5cm}} 
     - \frac{1}{2} \, \Proj^{8a}
  \\ \nonumber
  \Proj^{\overline{10}} 
  &= \frac{1}{2} \parbox{2.2cm}{\epsfig{file=Figures/A.eps,width=2.2cm}}
     - \frac{1}{2T_R^2} 
       \parbox{3.5cm}{\epsfig{file=Figures/A8.eps,width=3.5cm}} 
     - \frac{1}{2} \, \Proj^{8a}
  \\ \nonumber
  \Proj^{27} 
  &= \frac{1}{2} \parbox{2.2cm}{\epsfig{file=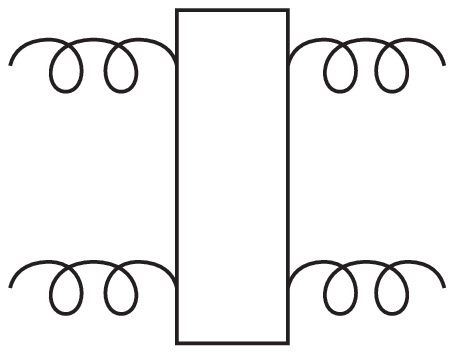,width=2.2cm}}
     + \frac{1}{2T_R^2} 
       \parbox{3.5cm}{\epsfig{file=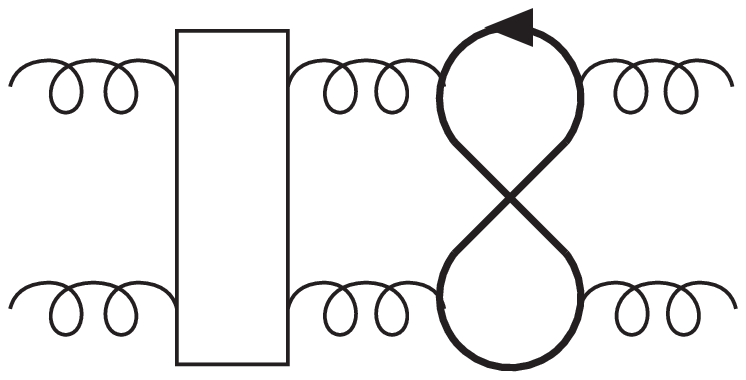,width=3.5cm}}
     - \frac{\Nc-2}{2\Nc} \, \Proj^{8s}
     - \frac{\Nc-1}{2\Nc} \, \Proj^{1}
  \\ \nonumber
  \Proj^{0} 
  &= \frac{1}{2} \parbox{2.2cm}{\epsfig{file=Figures/S.eps,width=2.2cm}}
     - \frac{1}{2T_R^2} 
       \parbox{3.5cm}{\epsfig{file=Figures/S8.eps,width=3.5cm}} 
     - \frac{\Nc+2}{2\Nc} \, \Proj^{8s}
     - \frac{\Nc+1}{2\Nc} \, \Proj^{1}.
\end{align}
The black and white bars denote anti-symmetrization and
symmetrization, respectively, see \appref{sec:birdtracks}. One can
easily verify that

\begin{equation}
  \parbox{3.5cm}{\epsfig{file=Figures/A8.eps,width=3.5cm}} 
  = \parbox{4.8cm}{\epsfig{file=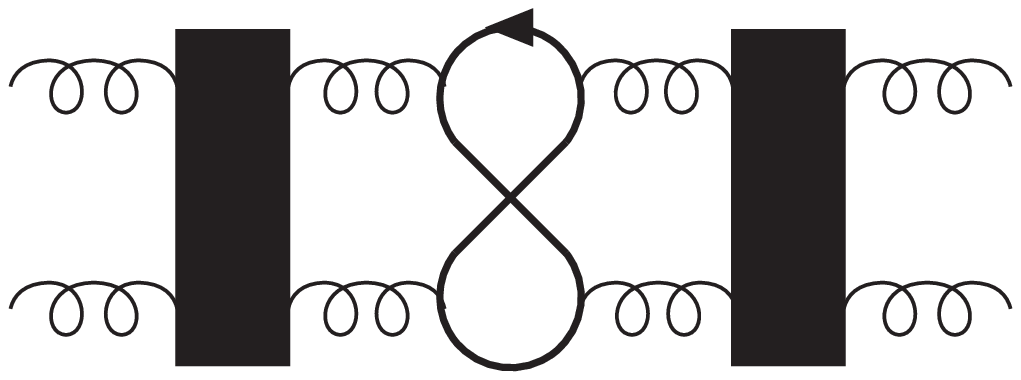,width=4.8cm}} \, , 
\end{equation}
and similarly for the symmetrized expression, making the hermiticity
of these projectors manifest.

From these projection operators orthogonal bases have been constructed
for processes involving up to five gluons \cite{Sjodahl:2008fz}. 
In general, knowing the projection operators for up to $\ng$ gluons it is
possible to construct orthogonal bases for QCD processes involving
up to $2\ng+1$ gluons (where we assume for the moment that there are
no quarks). The basis for $2\ng+1$ gluons can be constructed by
considering, e.g., $\ng\to\ng+1$.  The incoming gluons may then be
projected onto a multiplet $M$ using the projection operators for
$\ng$ gluons.  If the incoming $\ng$ gluons are in $M$ the outgoing
$\ng+1$ gluons must, due to color conservation, 
be in the same multiplet, see \appref{sec:invariant_tensors}. However, the multiplet $M$ may appear
more than once in $A^{\otimes\ng}$ or $A^{\otimes(\ng+1)}$ or
both. For example, there are six 27-plets in $A^{\otimes3}$, and one
in $A^{\otimes2}$. For $2g \to3g$ there are thus one (from the
incoming side) $\times$ six (from the outgoing side) possibilities for
the gluons to be in matching 27-plets. The 27-plets corresponding to
the case that two of the gluons in the outgoing $A^{\otimes 3}$ are in
a decuplet and an anti-decuplet, do, however, only appear in
combination.

\section{Hermitian quark projectors}
\label{sec:QuarkProjectors}

In this section we discuss projection operators for $\Nq$
quarks. Later, in \secref{sec:QuarkBasis} we use the hermitian
versions of these projection operators in order to construct an
orthogonal basis of the color space for $\Nq$ $\qqbar$ pairs.

A standard method for constructing projection operators onto
irreducible subspaces invariant under $\SU(\Nc)$ is to symmetrize and
anti-symmetrize according to the corresponding Young tableaux, and --
in the case of five or more quarks -- successively project out already
constructed projectors for Young tableaux of equal shape, see
e.g.\ \cite[sec. 5.4]{Littlewood}.  In this way, a complete set of
projection operators can be constructed for any number of quarks.
These projection operators are, however, {\it not} hermitian, see
\figref{fig:68and3bar8}, which implies that they are not suited for
constructing an orthogonal basis of the color space for $\Nq$ $\qqbar$
pairs, as we cannot use \eqref{eq:PidotPj}.

\FIGURE[t]{
  $\Proj_Y^{6,8}= \dfrac{4}{3}$
  \parbox{3cm}{\epsfig{file=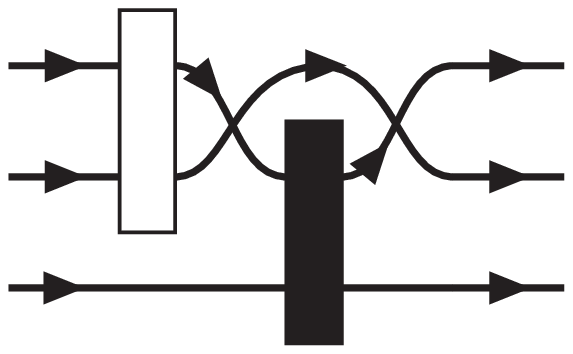,width=3cm}}\, ,\hspace{12mm}
  $\Proj^{6,8}= \dfrac{4}{3}$
  \parbox{3cm}{\epsfig{file=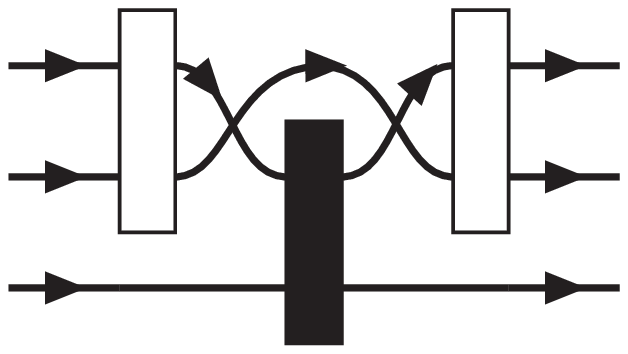,width=3cm}}\\[2ex]
  \hspace*{-13mm}
  $\Proj_Y^{\overline{3},8}= \dfrac{4}{3}$
  \parbox{3cm}{\epsfig{file=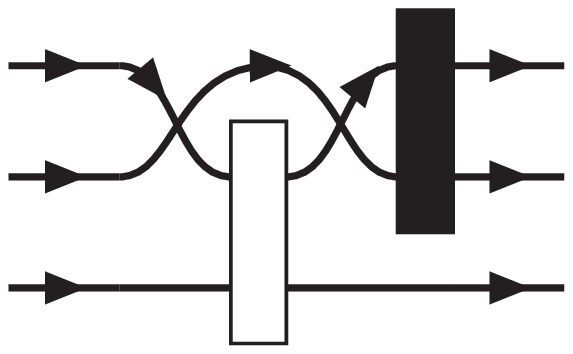,width=3cm}}\, ,\hspace{12mm}
  $\Proj^{\overline{3},8}= \dfrac{4}{3}$
  \parbox{3cm}{\epsfig{file=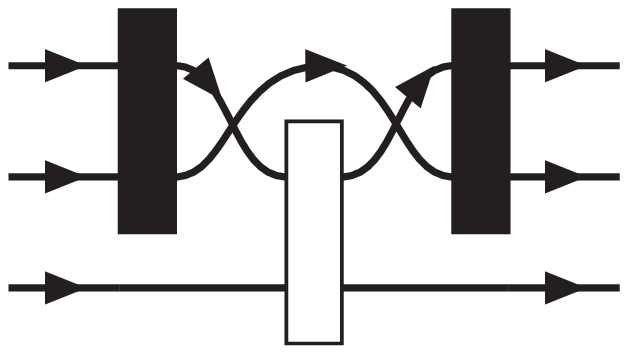,width=3cm}}
  \caption{\label{fig:68and3bar8} The standard Young projection
    operators $\Proj_Y^{6,8}$ and $\Proj_Y^{\overline{3},8}$ compared
    to their hermitian versions $\Proj^{6,8}$ and
    $\Proj^{\overline{3},8}$ from \eqref{eq:3qProjectors}.
    Clearly $\Proj^{6,8
      \dagger}\Proj^{\overline{3},8}=\Proj^{6,8}\Proj^{\overline{3},8}=0$.
    However, as can be seen from the symmetries, $\Proj_Y^{6,8\dagger}
    \Proj_Y^{\overline{3},8}\ne0$.
}}

Diagrammatically speaking, these operators have been constructed such
that products of distinct projectors vanish when contracting the {\it
outgoing} indices of the first projector with the {\it incoming}
indices of the second one; however, when calculating scalar products
in the color space of $\Nq$ $\qqbar$-pairs, the outgoing indices of
the first vector are contracted with the {\it outgoing} indices of the
second vector, cf.~\eqref{eq:scalar_product}.  Therefore, standard
Young projection operators are not orthogonal in the sense of
\eqref{eq:scalar_product}. By utilizing hermitian projection operators
this problem can be circumvented.

Hermitian Young projectors for three quarks were given in
\cite{Cvitanovic:1980bu}. In \cite{Cvi08} a general method for
constructing hermitian Young projectors is developed. This method is
based on solving certain characteristic equations. An alternative
approach for directly writing down hermitian Young projectors will be
presented elsewhere \cite{ProjectorPaper}.

The projectors can be expressed in terms of symmetrization and
anti-symmetrization operators, cf.~\eqref{eq:(anti-)symmetrizer}. Here
and in the following we label projection operators by the multiplets
built up successively when multiplying the partons,
i.e.\ 
\begin{equation}
  \Proj^{M_{2}, M_{3}}
\end{equation}
denotes a projector onto states where parton 1 and parton 2 are in a
multiplet $M_{2}$, and together with parton 3 form a multiplet
$M_{3}$. In this notation we have
\begin{align}\nonumber
  \Proj^{6,10}
  &=
  \parbox{2.5cm}{\epsfig{file=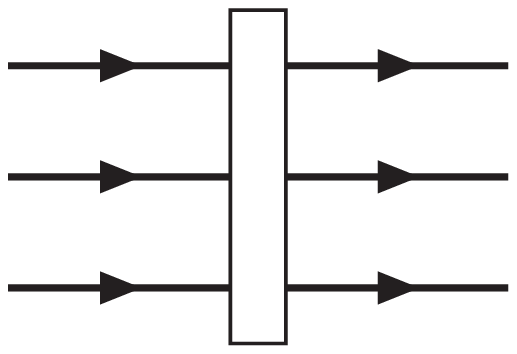,width=2.5cm}} \, , 
  &
  \Proj^{6,8}
  &= \frac{4}{3}
  \parbox{3cm}{\epsfig{file=Figures/68ProjH_no_prefac.eps,width=3cm}} \, , 
  \\[2ex]
  \Proj^{\overline{3},8}
  &=
  \frac{4}{3}
  \parbox{3cm}{\epsfig{file=Figures/3bar8ProjH_no_prefac.eps,width=3cm}} \, , 
  &
  \Proj^{\overline{3},1}
  &=
  \parbox{2.5cm}{\epsfig{file=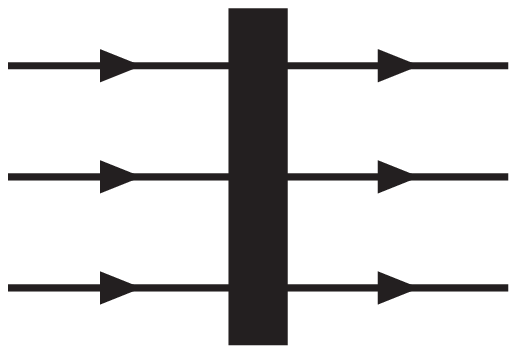,width=2.5cm}} \, .
\label{eq:3qProjectors}
\end{align}
In index notation, and written out as sums over permutations, these
projection operators read
\begin{align}
  \nonumber
  \Proj^{6,10}_{q_1\,q_2\,q_3\,q_4\,q_5\,q_6}
  &=\frac{1}{6}\big(
  \delta^{q_1}_{\,q_4}\delta^{q_2}_{\,q_5}\delta^{q_3}_{\,q_6}
  + \delta^{q_1}_{\,q_4}\delta^{q_2}_{\,q_6}\delta^{q_3}_{\,q_5}
  + \delta^{q_1}_{\,q_5}\delta^{q_2}_{\,q_4}\delta^{q_3}_{\,q_6}
  \\ \nonumber & \qquad\quad 
  + \delta^{q_1}_{\,q_5}\delta^{q_2}_{\,q_6}\delta^{q_3}_{\,q_4}
  + \delta^{q_1}_{\,q_6}\delta^{q_2}_{\,q_4}\delta^{q_3}_{\,q_5}
  + \delta^{q_1}_{\,q_6}\delta^{q_2}_{\,q_5}\delta^{q_3}_{\,q_4} 
  \big) \, ,  
  \\ \nonumber
  \Proj^{6,8}_{q_1\,q_2\,q_3\,q_5\,q_4\,q_6}
  &=\frac{1}{6}\big(
  2\delta^{q_1}_{\,q_5}\delta^{q_2}_{\,q_4}\delta^{q_3}_{\,q_6}
  -\delta^{q_1}_{\,q_5}\delta^{q_2}_{\,q_6}\delta^{q_3}_{\,q_4}
  +2\delta^{q_1}_{\,q_4}\delta^{q_2}_{\,q_5}\delta^{q_3}_{\,q_6}
  \\ \nonumber & \qquad\quad 
  -\delta^{q_1}_{\,q_4}\delta^{q_2}_{\,q_6}\delta^{q_3}_{\,q_5}
  -\delta^{q_1}_{\,q_6}\delta^{q_2}_{\,q_5}\delta^{q_3}_{\,q_4}
  -\delta^{q_1}_{\,q_6}\delta^{q_2}_{\,q_4}\delta^{q_3}_{\,q_5} 
  \big) \, ,  
  \\  
  \Proj^{\overline{3},8}_{q_1\,q_2\,q_3\,q_5\,q_4\,q_6}
  &=\frac{1}{6}\big(
  -2\delta^{q_1}_{\,q_5}\delta^{q_2}_{\,q_4}\delta^{q_3}_{\,q_6}
  -\delta^{q_1}_{\,q_5}\delta^{q_2}_{\,q_6}\delta^{q_3}_{\,q_4}
  +2\delta^{q_1}_{\,q_4}\delta^{q_2}_{\,q_5}\delta^{q_3}_{\,q_6}
  \\ \nonumber & \qquad\quad 
  +\delta^{q_1}_{\,q_4}\delta^{q_2}_{\,q_6}\delta^{q_3}_{\,q_5}
  +\delta^{q_1}_{\,q_6}\delta^{q_2}_{\,q_5}\delta^{q_3}_{\,q_4}
  -\delta^{q_1}_{\,q_6}\delta^{q_2}_{\,q_4}\delta^{q_3}_{\,q_5} 
  \big) \, ,  
  \\ \nonumber
  \Proj^{\overline{3},1}_{q_1\,q_2\,q_3\,q_4\,q_5\,q_6}
  &=\frac{1}{6}\big(
  \delta^{q_1}_{\,q_4}\delta^{q_2}_{\,q_5}\delta^{q_3}_{\,q_6}
  -\delta^{q_1}_{\,q_4}\delta^{q_2}_{\,q_6}\delta^{q_3}_{\,q_5}
  -\delta^{q_1}_{\,q_5}\delta^{q_2}_{\,q_4}\delta^{q_3}_{\,q_6}
  \\ \nonumber & \qquad\quad 
  +\delta^{q_1}_{\,q_5}\delta^{q_2}_{\,q_6}\delta^{q_3}_{\,q_4}
  +\delta^{q_1}_{\,q_6}\delta^{q_2}_{\,q_4}\delta^{q_3}_{\,q_5}
  -\delta^{q_1}_{\,q_6}\delta^{q_2}_{\,q_5}\delta^{q_3}_{\,q_4} 
  \big) \, .
\end{align}
As these projectors are hermitian, i.e.\ their birdtrack diagrams
(\ref{eq:3qProjectors}) are invariant under reflection about a
vertical line and simultaneous inversion of all arrows,
cf.~\appref{sec:birdtracks}, they are not only mutually transversal,
cf.~\eqref{eq:transversal_projectors}, but also orthogonal with
respect to the scalar product (\ref{eq:scalar_product}).

\section{Quark bases from hermitian quark projectors}
\label{sec:QuarkBasis}

When viewed as vectors in the color space for $(V \otimes
\overline{V})^{\otimes 3}$ the projectors in \eqref{eq:3qProjectors}
do not span the full space, since operators describing transitions
from one instance of a multiplet to any other instance of that
multiplet also transform as singlets under $\SU(\Nc)$. A basis of the
color space for three $\qqbar$ pairs thus contains four different
vectors derived from the octets. Normalized orthogonal basis
vectors can be chosen as follows, \cite[Fig.~21]{Cvitanovic:1980bu},
\begin{align}
  \nonumber
  \Vec^{6,10;6,10}
  &= \sqrt{\frac{6}{\Nc(\Nc^2+3\Nc+2)} } \; \Proj^{6,10} \, , 
  &
  \Vec^{6,8;6,8}
  &= \sqrt{\frac{3}{\Nc (\Nc^2-1)} } \; \Proj^{6,8} \, ,   
  \\ \nonumber
  \Vec^{6,8;\overline{3},8}
  &= \frac{2}{\sqrt{\Nc (\Nc^2-1)} } \;
  \raisebox{1mm}{\parbox{2.7cm}{
                 \epsfig{file=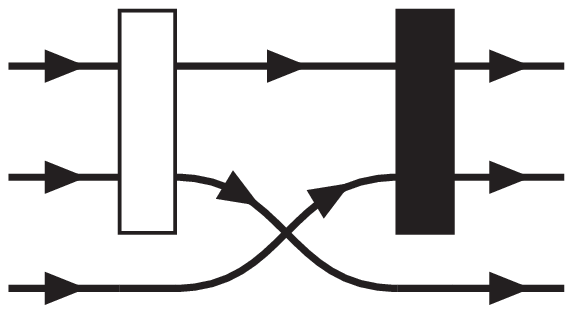,width=2.7cm}}} \, , 
  &
  \Vec^{\overline{3},8;6,8}
  &= \frac{2}{\sqrt{\Nc (\Nc^2-1)} } \;
  \raisebox{1mm}{\parbox{2.7cm}{
                 \epsfig{file=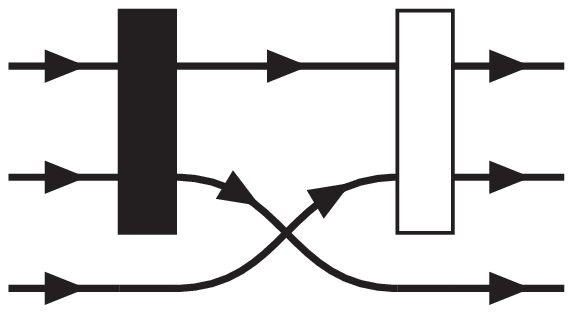,width=2.7cm}}} \, , 
  \\[1ex]
  \Vec^{\overline{3},8;\overline{3},8}
  &= \sqrt{ \frac{3}{\Nc (\Nc^2-1)}} \; \Proj^{\overline{3},8} \, , 
  &
  \Vec^{\overline{3},1;\overline{3},1}
  &= \sqrt{\frac{6}{\Nc(\Nc^2-3\Nc+2)} } \; \Proj^{\overline{3},1} \, ,
\label{eq:3qqbarBasisv1}
\end{align}
where each basis vector is denoted by the construction history on the
incoming and outgoing side in sequence, and the normalization is
consistent with \eqref{eq:HermProjSquare}. In index notation the two
vectors describing transitions between the octets can be written as 
\begin{equation}
\begin{split}
  \Vec^{6,8;\overline{3},8}_{q_1\,q_2\,q_3\,q_4\,q_5\,q_6}
  &= \frac{1}{2\sqrt{\Nc (\Nc^2-1)} } \, \big(
   \delta^{q_1}_{\,q_4}\delta^{q_2}_{\,q_6}\delta^{q_3}_{\,q_5}
  -\delta^{q_1}_{\,q_5}\delta^{q_2}_{\,q_6}\delta^{q_3}_{\,q_4}
  +\delta^{q_1}_{\,q_6}\delta^{q_2}_{\,q_4}\delta^{q_3}_{\,q_5}
  -\delta^{q_1}_{\,q_6}\delta^{q_2}_{\,q_5}\delta^{q_3}_{\,q_4}
  \big) \, ,
  \\
  \Vec^{\overline{3},8;6,8}_{q_1\,q_2\,q_3\,q_4\,q_5\,q_6}
  &= \frac{1}{2\sqrt{\Nc (\Nc^2-1)} } \, \big(
   \delta^{q_1}_{\,q_4}\delta^{q_2}_{\,q_6}\delta^{q_3}_{\,q_5}
  +\delta^{q_1}_{\,q_5}\delta^{q_2}_{\,q_6}\delta^{q_3}_{\,q_4}
  -\delta^{q_1}_{\,q_6}\delta^{q_2}_{\,q_4}\delta^{q_3}_{\,q_5}
  -\delta^{q_1}_{\,q_6}\delta^{q_2}_{\,q_5}\delta^{q_3}_{\,q_4}
  \big) \, . 
\end{split}
\end{equation}
These two basis vectors can be constructed as follows. In order to
find a vector describing a transition from $\overline{3},8$ to $6,8$
write down the birdtrack expression for $\Proj^{6,8}$ on the left and
that for $\Proj^{\overline{3},8}$ on the right; now one has to find a
non-vanishing way for connecting these diagrams. To this end, note
that there is a symmetrizer to the very left in $\Proj^{6,8}$
and an anti-symmetrizer to the very right in $\Proj^{\overline{3},8}$. 
If both lines leaving the white bar enter the black
bar then the whole expression vanishes, i.e.\ one of the lines leaving
the white bar has to be connected to the third line on the right. One
such choice is displayed in the diagram for
$\Vec^{6,8;\overline{3},8}$ above. Any other non-vanishing choice
yields the same vector up to a factor. 
Finally, the result has to be normalized
using the scalar product (\ref{eq:sp_trace}). The remaining vector
$\Vec^{\overline{3},8;6,8}$ can either be constructed in the same way,
or by taking the hermitian conjugate of $\Vec^{6,8;\overline{3},8}$.

Knowing the hermitian projection operators it is possible to similarly
construct the orthogonal basis vectors for processes involving more
$\qqbar$ pairs. The orthogonality can be seen by noting that
contracting the incoming or the outgoing indices gives 0.  We also
note that there are six basis vectors, in agreement with what is
obtained from the $\Nq!$ ways of connecting quark and anti-quark lines
in \eqref{eq:start_cond}.  In this case there are thus equally many
vectors for $\Nc=3$ as for $\Nc \to \infty$.

\section{Hermitian gluon projectors}
\label{sec:Gluons}

In this section we outline a general algorithm for constructing hermitian
projectors for all multiplets appearing in $A^{\otimes\ng}$ for
arbitrary $\ng$. The construction is recursive, i.e.\ the projectors
for the decomposition of $A^{\otimes\nu}$ with $\nu\leq\ng-1$ along
with their properties are used when constructing the projectors onto
multiplets within $A^{\otimes\ng}$. As an illustrating example we
treat the case $\ng=3$ along with the outline of the general construction.

For our algorithm it is important to keep track of for which $n$ a given
multiplet $M$ appears for the first time in the sequence $A^{\otimes
  n}$, $n=0,1,2,\hdots$. We denote this number by $\Nf(M)$ and call it
that multiplet's {\it first occurrence}. For instance the singlet has
$\Nf(\bullet)=0$ and the adjoint representation has first occurrence one. 
Some more examples, labeled by $\SU(3)$ Young diagrams, are
listed in \tabref{tab:first_occurrences}.

\TABLE[t]{
\Yboxdim{7pt}
\centering 
\begin{tabular}{c||c|c|c|c}
$\Nf$ & $0$ & $1$ & $2$ & $3$\\[1ex]
\hline&&&&\\[-2ex]
$\SU(3)$ & $\bullet=\yng(1,1,1)$ & $\yng(2,1)$ & $\yng(3)$ & $\yng(5,1)$ \\
Young diagrams & & & $\yng(3,3)$ & $\yng(5,4)$ \\[2mm]
& & & $\yng(4,2)$ & $\yng(6,3)$ 
\end{tabular}
\caption{\label{tab:first_occurrences}Examples of $\SU(3)$ Young
  diagrams sorted according to their first occurrence $\Nf$.}  }

In order to make sure that projectors onto all invariant subspaces are 
constructed we first decompose $A^{\otimes\ng}$ into multiplets, 
$A^{\otimes\ng} = \bigoplus_{j} M_j$, by multiplying Young diagrams.
The $\ng=2$ decomposition has already been preformed in \eqref{eq:SU388}.
For $\ng=3$ we have to multiply the r.h.s. of
\eqref{eq:SU388} term by term with another gluon. Multiplication of
the singlet trivially yields an octet,
\begin{equation}
  \label{eq:1x8}
\Yboxdim{10pt}
  \begin{array}{ccccc}
  \upsmall{\Nc}  
  && \upsmall{\Nc-1} \csep \upsmall{1} 
  && \upsmall{\Nc-1} \csep \upsmall{1}\\
  \bullet & \otimes & \yng(2,1) & \ = \ & \yng(2,1)\\[1ex]
  1 && 8 && 8 
  \end{array} \ .
\end{equation}
The product of two octets is already displayed in \eqref{eq:SU388}.
When multiplying the decuplet with an octet we have
\begin{equation}
\label{eq:10x8}
\Yboxdim{10pt}
  \begin{array}{ccccccccccccccc}
  \upsmall{\Nc-2} \csep \upsmall{1} \csep \upsmall{1} 
  && \upsmall{\Nc-1} \csep \upsmall{1} 
  && \upsmall{\Nc-1} \csep \upsmall{1} 
  && \upsmall{\Nc-2} \csep \upsmall{1} \csep \upsmall{1} 
  && \upsmall{\Nc-2} \csep \upsmall{1} \csep \upsmall{1} 
  && \upsmall{\Nc-1} \csep \upsmall{\Nc-1} \csep \upsmall{1} \csep \upsmall{1} 
  && \upsmall{\Nc-2} \csep \upsmall{2} 
  && \upsmall{\Nc-1} \csep \upsmall{\Nc-2} 
     \csep \upsmall{1} \csep \upsmall{1} \csep \upsmall{1}
  \\
  \yng(3) & \otimes 
  & \yng(2,1) & \ = \ 
  & \yng(2,1) & \oplus
  & \yng(3) & \oplus
  & \circ & \oplus
  & \yng(4,2) & \oplus
  & \circ & \oplus 
  & \yng(5,1) 
  \\[1ex]
  10 && 8 && 8 && 10 && (10) && 27 && 0 && 35
  \\[1.5ex]
  && &&
  && \upsmall{\Nc-1} \csep \upsmall{\Nc-2} \csep \upsmall{2} \csep \upsmall{1} 
  && \upsmall{\Nc-3} \csep \upsmall{1} \csep \upsmall{1} \csep \upsmall{1} 
  && \upsmall{\Nc-3} \csep \upsmall{2} \csep \upsmall{1}
  \\
  && &&& \oplus
  & \circ & \oplus
  & \circ & \oplus
  & \circ &.
  \\[1ex]
  && && && 0 && 0 && 0  
\end{array} \ 
\end{equation}
As above, we in general denote multiplets that do not appear for $\Nc=3$, but 
only for sufficiently large $\Nc$, by $\circ$. 
While the second $(\Nc-2,1,1)$-multiplet, has a Young tableaux shape
which is admissible for SU(3), it can be seen from Young tableaux multiplication
that it cannot appear. For such multiplets -- which are forbidden only 
by the construction -- we display the corresponding 
SU(3)-dimension in brackets.  
Similarly, for the anti-decuplet we get
\begin{equation}
\Yboxdim{10pt}
\label{eq:10barx8}
  \begin{array}{ccccccccccccccc}
  \upsmall{\Nc-1} \csep \upsmall{\Nc-1} \csep \upsmall{2}  
  && \upsmall{\Nc-1} \csep \upsmall{1} 
  && \upsmall{\Nc-1} \csep \upsmall{1} 
  && \upsmall{\Nc-1} \csep \upsmall{\Nc-1} \csep \upsmall{2} 
  && \upsmall{\Nc-1} \csep \upsmall{\Nc-1} \csep \upsmall{2} 
  && \upsmall{\Nc-1} \csep \upsmall{\Nc-1} \csep \upsmall{1} \csep \upsmall{1} 
  && \upsmall{\Nc-2} \csep \upsmall{2} 
  && \upsmall{\Nc-1} \csep \upsmall{\Nc-1} \csep \upsmall{\Nc-1} 
     \csep \upsmall{2} \csep \upsmall{1} 
  \\
  \yng(3,3) & \otimes 
  & \yng(2,1) & \ = \ 
  & \yng(2,1) & \oplus
  & \yng(3,3) & \oplus
  & \circ & \oplus
  & \yng(4,2) & \oplus
  & \circ & \oplus 
  & \yng(5,4) 
  \\[1ex]
  \overline{10} && 8 && 8 && \overline{10} && (\overline{10}) 
  && 27 && 0 && \overline{35}
  \\[1.5ex]
  && &&
  && \upsmall{\Nc-1} \csep \upsmall{\Nc-2} \csep \upsmall{2} \csep \upsmall{1} 
  && \upsmall{\Nc-1} \csep \upsmall{\Nc-1} \csep \upsmall{\Nc-1} 
     \csep \upsmall{3} 
  && \upsmall{\Nc-1} \csep \upsmall{\Nc-2} \csep \upsmall{3} 
  \\
  && &&& \oplus
  & \circ & \oplus
  & \circ & \oplus
  & \circ   &.
  \\[1ex]
  && && && 0 && 0 && 0  
\end{array} \ 
\end{equation}
Finally, for the products with the remaining two multiplets, $0$ and $27$, 
we obtain
\begin{equation}
\label{eq:27x8}
\Yboxdim{10pt}
  \begin{array}{cccccccccccccc}
  \upsmall{\Nc-1} \csep \upsmall{\Nc-1} \csep \upsmall{1} \csep \upsmall{1}  
  && \upsmall{\Nc-1} \csep \upsmall{1} 
  && \upsmall{\Nc-1} \csep \upsmall{1} 
  && \upsmall{\Nc-2} \csep \upsmall{1} \csep \upsmall{1} 
  && \upsmall{\Nc-1} \csep \upsmall{\Nc-1} \csep \upsmall{2} 
  && \upsmall{\Nc-1} \csep \upsmall{\Nc-1} \csep \upsmall{1} \csep \upsmall{1} 
  && \upsmall{\Nc-1} \csep \upsmall{\Nc-1} \csep \upsmall{1} \csep \upsmall{1} 
  \\
  \yng(4,2) & \otimes 
  & \yng(2,1) & \ = \ 
  & \yng(2,1) & \oplus
  & \yng(3) & \oplus
  & \yng(3,3) & \oplus
  & \yng(4,2) & \oplus
  & \yng(4,2) 
  \\[1ex]
  27 && 8 && 8 && 10 && \overline{10} && 27 && 27 
  \\[1.5ex]
  && &&
  && \upsmall{\Nc-1} \csep \upsmall{\Nc-1} \csep \upsmall{\Nc-1} 
     \csep \upsmall{2} \csep \upsmall{1} 
  && \upsmall{\Nc-1} \csep \upsmall{\Nc-2} \csep \upsmall{1} 
     \csep \upsmall{1} \csep \upsmall{1} 
  && \upsmall{\Nc-1} \csep \upsmall{\Nc-1} \csep \upsmall{\Nc-1} 
     \csep \upsmall{1} \csep \upsmall{1} \csep \upsmall{1} 
  && \upsmall{\Nc-1} \csep \upsmall{\Nc-2} \csep \upsmall{2} \csep \upsmall{1} 
  \\
  && &&& \oplus 
  & \yng(5,4) & \oplus
  & \yng(5,1) & \oplus
  & \yng(6,3) & \oplus
  & \circ     &,
  \\[1ex]
  && && && \overline{35} && 35 && 64 && 0  
\end{array} \ 
\end{equation}
\begin{equation}
\label{eq:0x8}
\Yboxdim{10pt}
  \begin{array}{ccccccccccccccccccccc}
  \upsmall{\Nc-2} \csep \upsmall{2}
  && \upsmall{\Nc-1} \csep \upsmall{1} 
  && \upsmall{\Nc-1} \csep \upsmall{1} 
  && \upsmall{\Nc-2} \csep \upsmall{1} \csep \upsmall{1} 
  && \upsmall{\Nc-1} \csep \upsmall{\Nc-1} \csep \upsmall{2} 
  && \upsmall{\Nc-2} \csep \upsmall{2}
  && \upsmall{\Nc-2} \csep \upsmall{2}
  && \upsmall{\Nc-1} \csep \upsmall{\Nc-2} \csep \upsmall{2} 
     \csep \upsmall{1}
  && \upsmall{\Nc-3} \csep \upsmall{2} \csep \upsmall{1} 
  && \upsmall{\Nc-1} \csep \upsmall{\Nc-2} \csep \upsmall{3}
  && \upsmall{\Nc-3} \csep \upsmall{3}
  \\
  \circ & \otimes 
  & \yng(2,1) & \ = \ 
  & \circ & \oplus
  & \circ & \oplus
  & \circ & \oplus
  & \circ & \oplus
  & \circ & \oplus 
  & \circ & \oplus
  & \circ & \oplus
  & \circ & \oplus
  & \circ 
  \\[1ex]
  0 && 8 && (8) && (10) && (\overline{10}) && 0 && 0 && 0 && 0 && 0 && 0 
\end{array} \ .
\end{equation}
Here the first three multiplets on the r.h.s. of the last
equation, $(\Nc-1,1)$, $(\Nc-2,1,1)$ and $(\Nc-1,\Nc-1,2)$, would be
allowed Young diagrams for $\Nc=3$. However, we denote them by $\circ$
and set the dimensions in brackets since they were obtained by
multiplication of $(\Nc-2,2)$, a multiplet that does not exist for
$\Nc=3$.

Looking at these decompositions of tensor products one can make two
observations: 
\begin{enumerate}
\item A multiplet $M' \subseteq M \otimes A$ always has first
  occurrence
  \begin{equation}
    \Nf(M') = \Nf(M)-1 \, , \ \Nf(M) \text{ or } \Nf(M)+1 \, . 
  \end{equation}
  In particular, there are no singlets in eqs.~(\ref{eq:10x8} --
  \ref{eq:0x8}).
\item The only multiplet which can show up several times in $M \otimes
  A$ is $M$ itself, all other multiplets appear at most once. In fact,
  $M$ can appear up to $\Nc-1$ times.
\end{enumerate}
Both observations are true in general and we prove them in
\appref{sec:first_occurrence}.

Below we outline the construction of 
the corresponding projectors $\Proj^{M_j}$ having the following properties:
\begin{itemize}
\item[(i)] $\Proj^{M_j} \Proj^{M_k} = \delta_{jk} \Proj^{M_j}$. We
  call this property {\it transversality},
  cf.~\eqref{eq:transversal_projectors}.
\item[(ii)] $\Proj^M=\CG^M\CG^{M\dag}$ with $\CG^M: A^{\otimes\Nf} \to
  A^{\otimes\ng}$, where $\Nf$ is the first occurrence of
  $M$. Choosing suitable bases in $A^{\otimes\ng}$ and $A^{\otimes\Nf}$
  the matrix elements of $\CG^M$ are Clebsch-Gordan coefficients. In
  birdtrack notation this means that there is always an intermediate
  section with $\Nf$ gluon lines in the middle of the diagram for each projector,
  \begin{equation}
  \label{eq:P=CCdag}
  \begin{split}
  \Proj^M \ &= \ 
  \parbox{5.5cm}{\epsfig{file=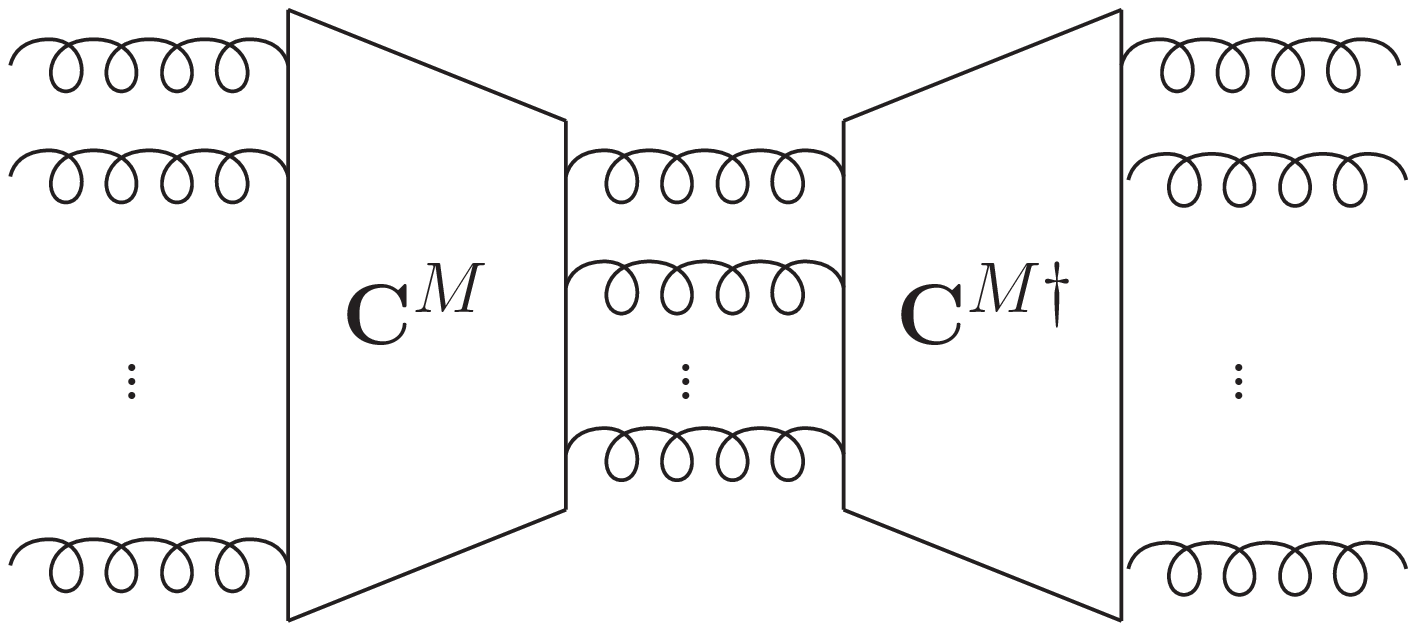,width=5.5cm}} \ .\\
  & \qquad \ng\text{ lines} \qquad \Nf\text{ lines} \qquad \ng\text{ lines} 
  \end{split} 
  \end{equation}
\item[(iii)] A projector $\Proj^{M'}$ onto a multiplet $M' \subseteq
  A^{\otimes\ng}$ appearing in the decomposition of 
  $M \otimes A^{\otimes (\ng-\nu)}$,
  i.e.\ $M \otimes A^{\otimes (\ng-\nu)} = M' \oplus \hdots\,$, satisfies
  \begin{equation} 
    (\Proj^M \otimes \eins_{A^{\otimes(\ng-\nu)}}) \Proj^{M'}
    = \Proj^{M'} (\Proj^M \otimes \eins_{A^{\otimes(\ng-\nu)}}) 
    = \Proj^{M'} \, , 
  \end{equation}
  where $\eins_{A^{\otimes(\ng-\nu)}} : A^{\otimes(\ng-\nu)} \to
  A^{\otimes(\ng-\nu)}$ denotes the identity operator. In terms of
  birdtracks this is written
  \begin{equation}
  \parbox{5cm}{\epsfig{file=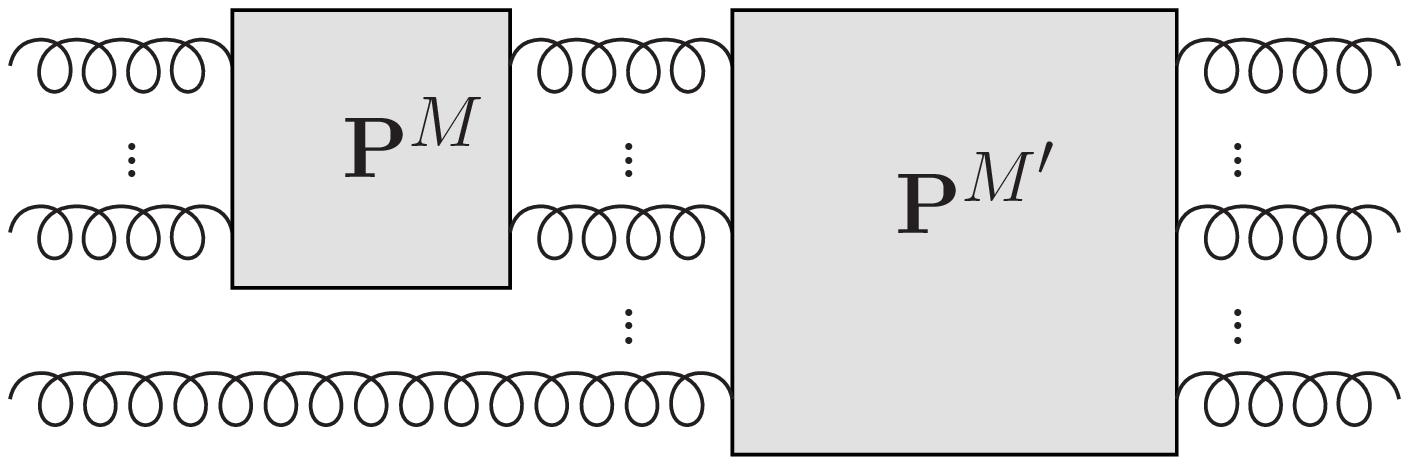,width=5cm}}
  = \parbox{5cm}{\epsfig{file=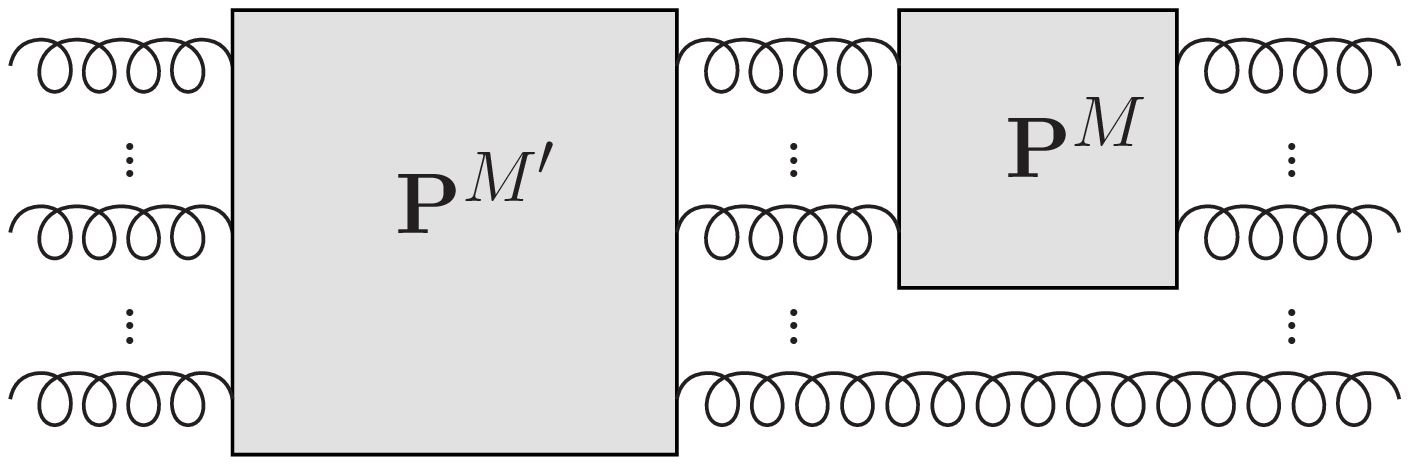,width=5cm}}
  = \parbox{3cm}{\epsfig{file=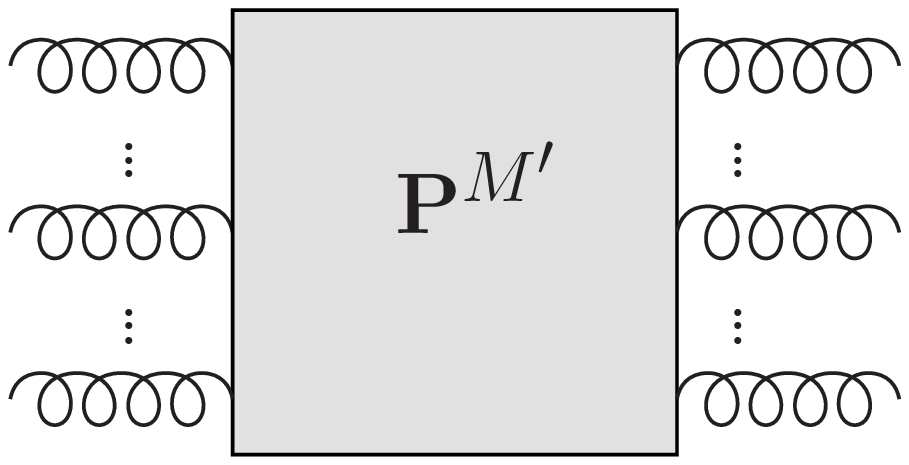,width=3cm}} \, , 
  \end{equation}
  i.e.\ the first $\nu$ gluons are in multiplet $M$ and together
  with the remaining gluons they form an overall multiplet $M'$.
\end{itemize}

The hermiticity of $\Proj^M$ is obvious from (ii).  Also note its
birdtrack manifestation in \eqref{eq:P=CCdag}: The diagram is
invariant under simultaneous mirroring about a vertical line through
the $\Nf$ gluon lines and reversing all arrows (which may appear
inside $\CG^M$), cf.\ \appref{sec:birdtracks}. Together with
transversality (i) hermiticity ensures that the projectors project
onto mutually orthogonal subspaces and are themselves mutually
orthogonal with respect to the scalar product (\ref{eq:sp_trace}),
cf.\ the discussion in \secref{sec:ColorSpace}.

From (ii) one can infer that by multiplying the Clebsch-Gordan
matrices in reverse order we obtain
\begin{equation}
\label{eq:CdagC}
\begin{split}
  \CG^{M\dag} \CG^M 
  \ = \ \parbox{5.5cm}{\epsfig{file=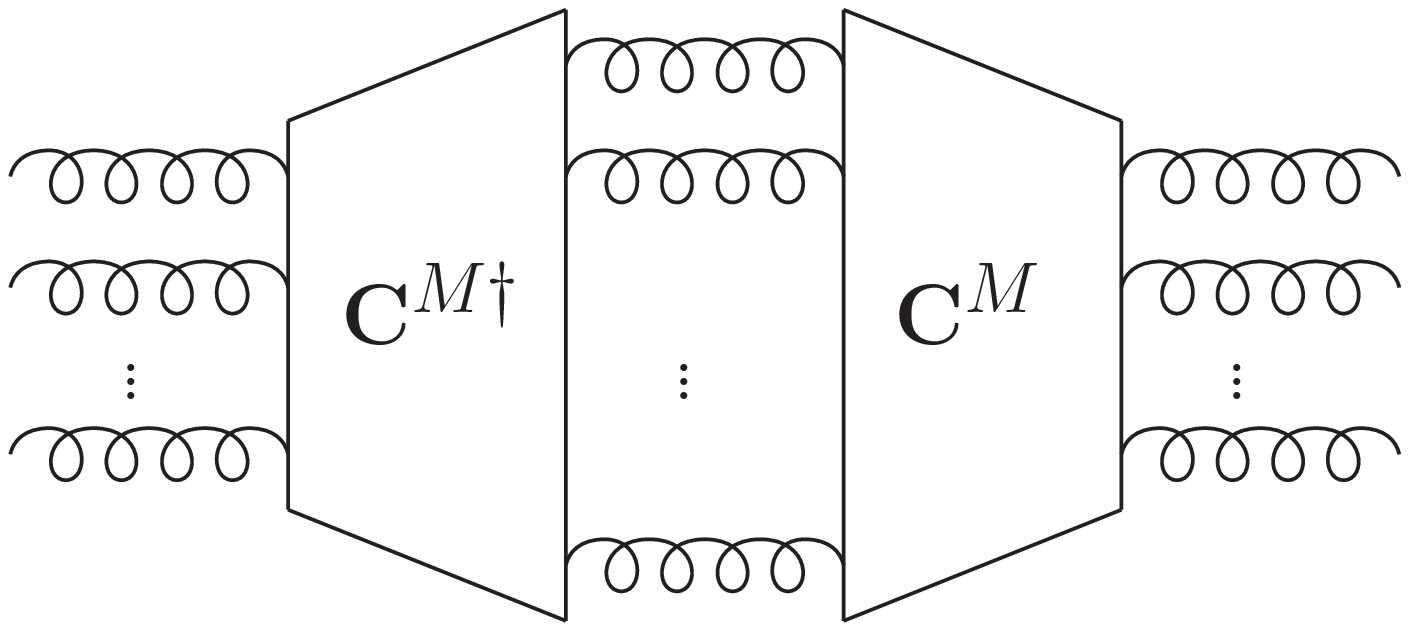,width=5.5cm}}
  \ &= \ \Proj^{M_f} \, , \\
  \Nf\text{ lines} \qquad \ng\text{ lines} \qquad \Nf\text{ lines} 
  \ & 
\end{split}
\end{equation}
where $M_f \subset A^{\otimes\Nf}$ carries the same irreducible
representation as $M$. For $M \neq M'$ we have 
\begin{equation} 
\label{eq:CMdagCM'=0}
  \CG^{M\dag} \CG^{M'} = 0 \, .
\end{equation}
We include proofs of eqs.~(\ref{eq:CdagC}) and (\ref{eq:CMdagCM'=0})
in \appref{sec:ClebschGordans}.

According to property (iii) a projector $A^{\otimes\ng} \to
A^{\otimes\ng}$ not only projects onto a definite multiplet, but also
ensures that the first $\nu$ gluons are in multiplet $M_\nu$, $\nu =
2,\hdots,\ng$. We refer to the sequence $M_2,M_3,\hdots,M_{\ng}$ as
the projector's {\it construction history}. It is convenient to label
projectors by their construction histories,
\begin{equation}
\label{eq:const-hist_P}
  \Proj^{M_2,M_3,\hdots,M_{\ng}} 
  = \parbox{12cm}{\epsfig{file=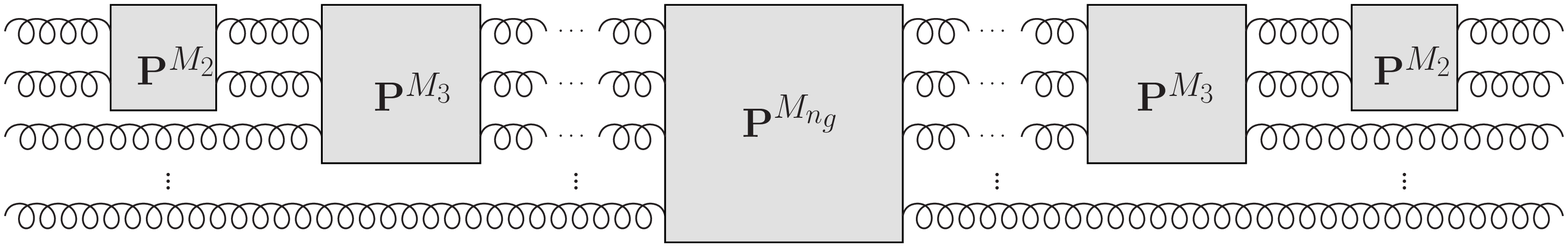,width=12cm}}
  \, . 
\end{equation}

We prove our algorithm by induction in $n_g$. We can start from either
$\ng=0$ or $\ng=1$, for which all properties are satisfied
trivially. However, it is instructive to revisit the $\ng=2$
projectors, which were given in eqs.~(\ref{eq:ggBasis}) and
(\ref{eq:ggBasis_birdtracks}), and verify that they also satisfy the
properties (i)--(iii). The only property which may not be immediately
obvious is (ii) for the $\Nf=2$ projectors $\Proj^{10},
\Proj^{\overline{10}}, \Proj^{27}$ and $\Proj^0$. Note, however,
that for $\ng=\Nf$ property (ii) is satisfied trivially with
$\CG^M = \Proj^{M} = \CG^{M\dag}$. 

Below we outline the recursive
construction of projectors for the decomposition of $A^{\otimes\ng}$
from the projectors for the decompositions of $A^{\otimes\nu}$,
$\nu\leq\ng-1$. Making sure that the properties (i)--(iii) are
retained by this algorithm will establish the induction step.
In order to keep track of which projectors have to be constructed in step
$\ng$ we proceed as follows. For each multiplet $M \subseteq
A^{\otimes(\ng-1)}$ we decompose $M \otimes A$ by multiplying the
corresponding Young diagrams, as done in \eqref{eq:SU388} and 
eqs.~(\ref{eq:10x8}--\ref{eq:0x8}) above.

Multiplets $M' \subseteq A^{\otimes\ng}$ with first occurrence
$\Nf(M')=\ng$ we refer to as {\it new~multiplets}. For all other multiplets
$M' \subseteq A^{\otimes\ng}$ we have $\Nf(M') < \ng$ and,
correspondingly, they are referred to as {\it old~multiplets}. Multiplets
$M' \subseteq M \otimes A$ with $\Nf(M')=\Nf(M)-1$ or $\Nf(M')=\Nf(M)$
are necessarily old multiplets. Multiplets with $\Nf(M')=\Nf(M)+1$ can
be either old or new depending on whether $M$ was old or new within
$A^{\otimes(\ng-1)}$. Our general strategy for obtaining all
projectors, is to first construct all projectors onto old multiplets, 
and then to use these projectors in the subsequent
construction of projectors onto new multiplets.  

Projectors $\Proj^{\hdots M,M'}$ onto old multiplets $M' \subseteq M
\otimes A$ can always be constructed as follows. Consider the
corresponding Clebsch-Gordan matrix $\CG^{\hdots M,M'}$. In order to
satisfy property (iii), there has to be a $\CG^{\hdots M} \otimes
\eins_A$ at the left end, whereas property (ii) requires a
$\Proj^{M'_f}$ at the right end,
\begin{equation}
  \CG^{\hdots M,M'} 
  = \parbox{8cm}{\epsfig{file=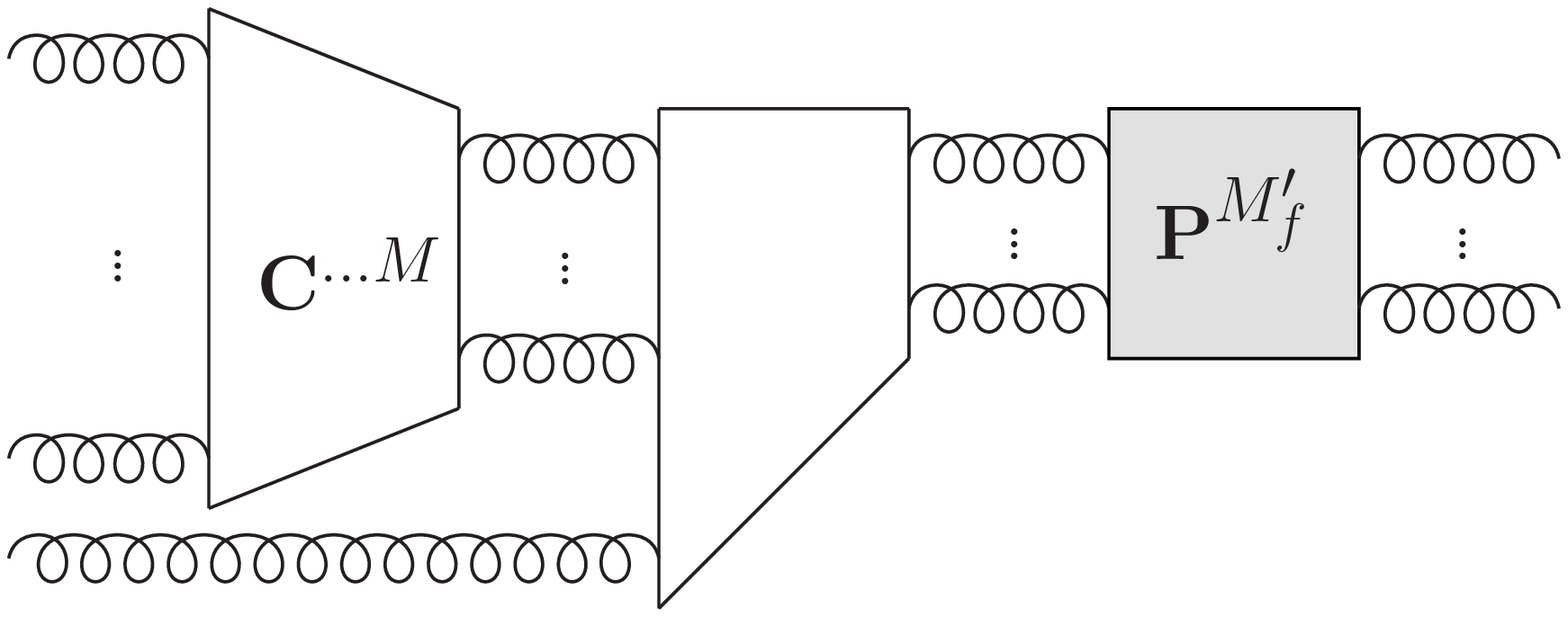,width=8cm}} \, .
  \label{eq:any_old}
\end{equation}
In the middle one has to connect the $\Nf(M'_f)$ gluon lines on the right to
the $\Nf(M)+1$ gluon lines on the left in such a way that the whole
expression does not vanish. Then, after appropriate normalization,
$\Proj^{\hdots M,M'} = \CG^{\hdots M,M'} \CG^{\hdots M,M'\dag}$ is the
desired projector. 

In principle, a non-vanishing connection can always be found be
  splitting all gluon lines entering the trapezoid into
  $\qqbar$-pairs, and then considering all ways of attaching the
  quark- and anti-quark ends. At least one such connection has to be
  non-zero. As this procedure may be tedious, in particular for many
  gluons, we provide more explicit recipes for the construction of
  projectors onto old multiplets in
  sections \ref{sec:old_starting_multiplet}--\ref{sec:new_to_same}. These
  recipes cover most of the frequently occurring cases. In particular,
  they directly yield the full set of 3-gluon projectors onto old
  multiplets. Projectors onto new multiplets require an independent
  construction which we develop in \secref{sec:new_multiplets}.

\subsection{Starting from an old multiplet, $\Nf(M) < \ng-1$}
\label{sec:old_starting_multiplet}

We begin with $M' \subseteq M \otimes A$ where the starting multiplet
for $\ng-1$ gluons was old, i.e.\ $\Nf(M) < \ng-1$. 
In this case we write down $\Proj^{...,M}$
as depicted in property (ii), add a gluon line below and draw a
projector $\Proj^{M'}$ over the $\Nf(M)+1$ gluon lines in the middle, 
\begin{equation}
  \Proj^{\hdots,M,M'} 
  = \parbox{6cm}{\epsfig{file=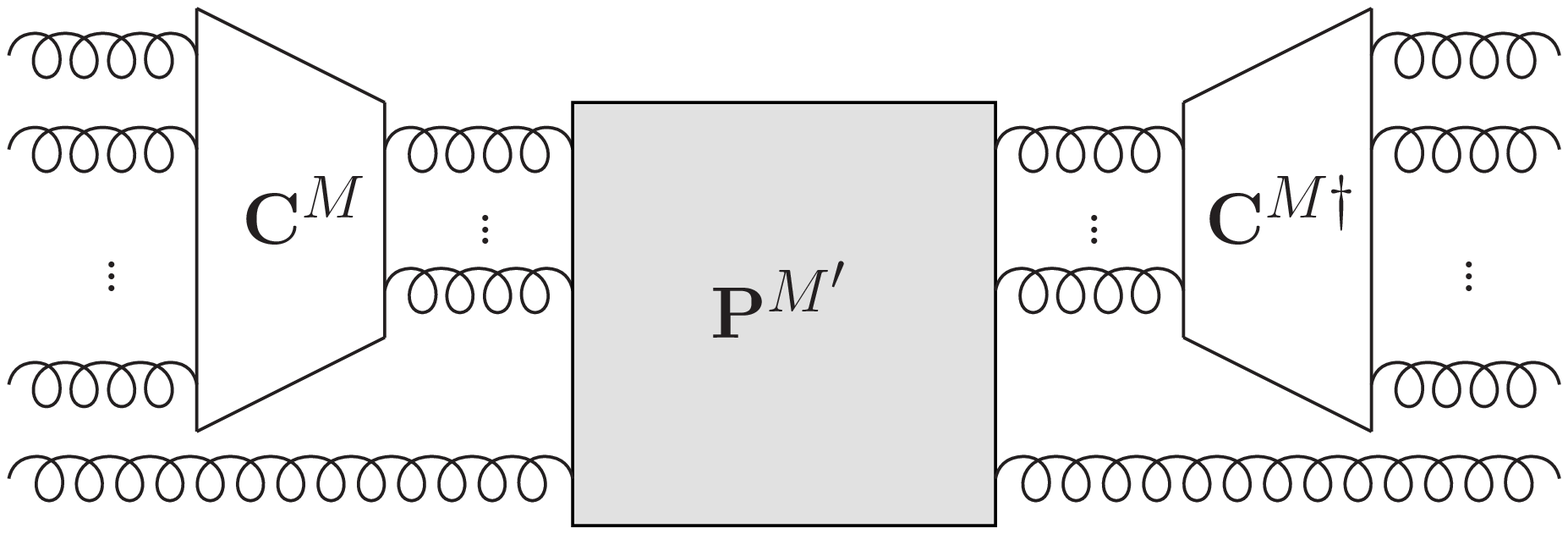,width=6cm}}
  \, .
\end{equation}
The projector $\Proj^{M'}$ has already been constructed in an earlier
step according to the induction hypothesis. The operator
$\Proj^{\hdots,M,M'}$ satisfies (ii) since $\Proj^{M'}$ does. 
To make it explicit we could insert another copy of $\Proj^{M'}$
in the middle of the r.h.s. above.
$\Proj^{\hdots,M,M'}$ also
satisfies (iii) since, $\Proj^{...,M}$ does and since
\begin{align}
  \nonumber
  \Proj^{\hdots,M,M'} \, (\Proj^{\hdots,M} \otimes \eins) 
  &=\parbox{10cm}{\epsfig{file=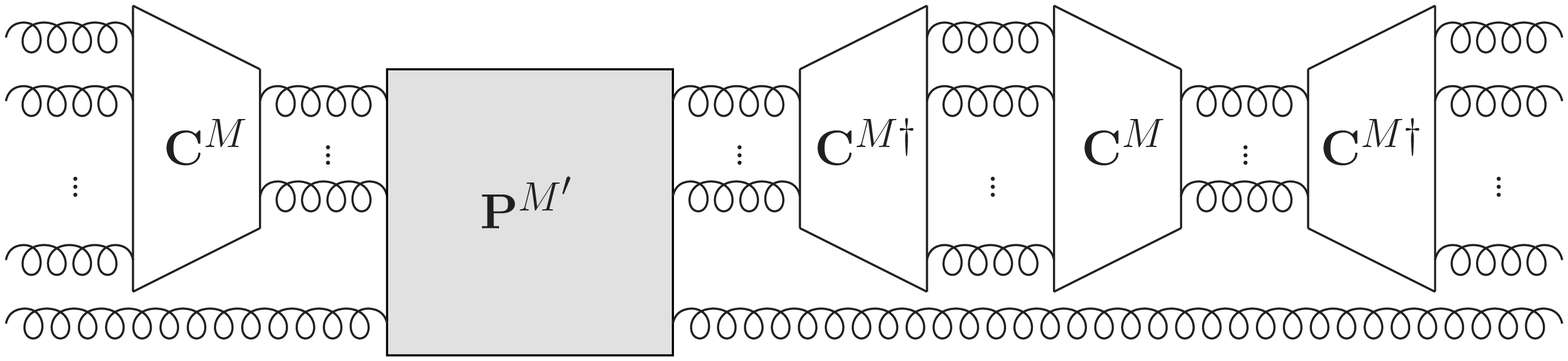,width=10cm}}
  \\ \nonumber
  &=\parbox{8.5cm}{\epsfig{file=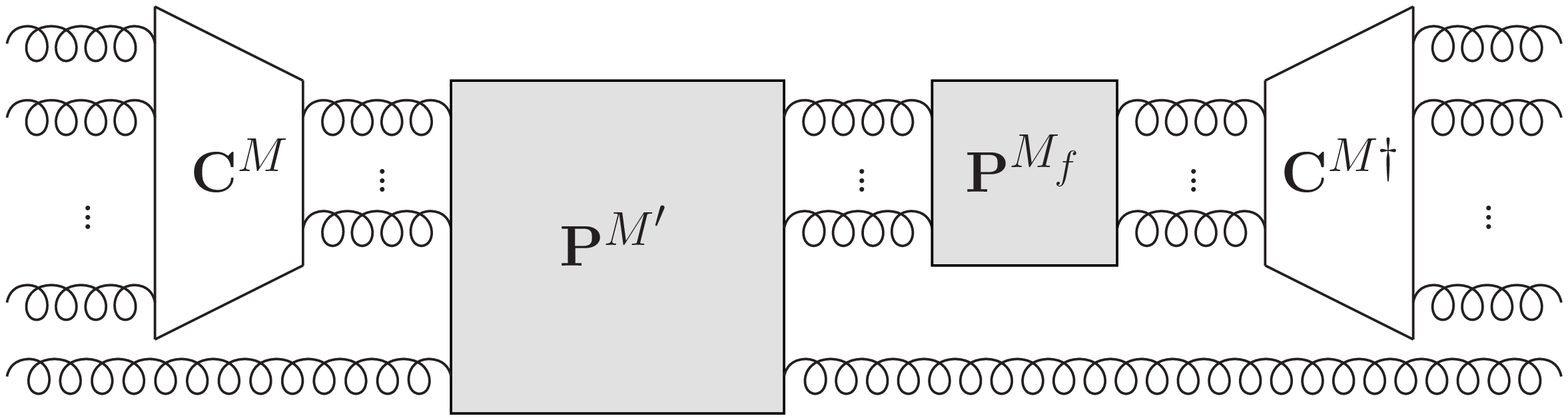,width=8.5cm}}
  \\ 
  &=\parbox{6.7cm}{\epsfig{file=Figures/old_starting_multiplet.eps,width=6.7cm}}
  \ = \ \Proj^{\hdots,M,M'} \, .
\end{align}
Here the second line holds since $\Proj^{\hdots,M}$ fulfills (ii)
which implies \eqref{eq:CdagC}. The first occurrence projector
$\Proj^{M_f}$ appearing in this way, however, also has to appear in
the construction history of $\Proj^{M'}$. Since $\Proj^{M'}$ in turn
satisfies (iii) we can readily omit $\Proj^{M_f}$ from the equation.
By the same arguments we establish that $\Proj^{\hdots,M,M'}$ actually
is a projector, i.e.\ that
$(\Proj^{\hdots,M,M'})^2=\Proj^{\hdots,M,M'}$. Writing
$(\Proj^{\hdots,M,M'})^2$ in terms of birdtracks
\begin{equation}
  \Big(\Proj^{\hdots,M,M'}\Big)^2 
  =\parbox{11cm}{\epsfig{file=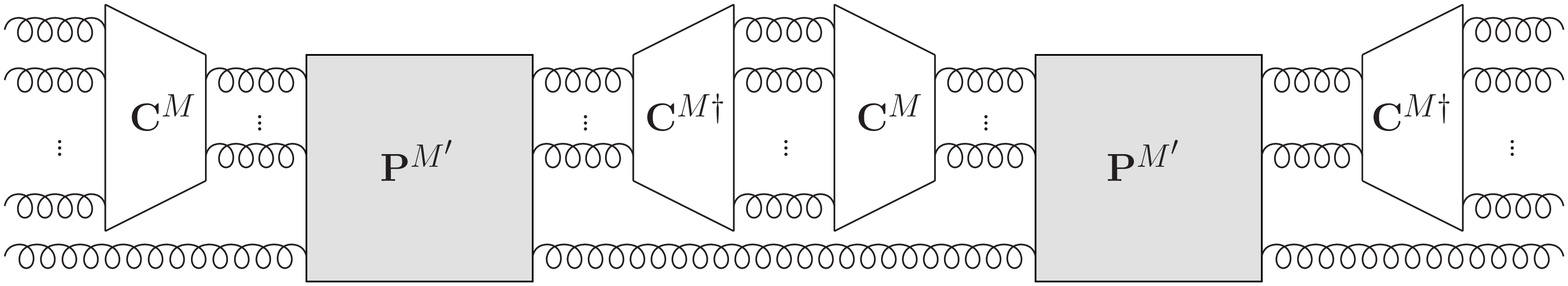,width=11cm}} \, ,
\end{equation}
the $\CG^{M\,\dagger}\CG^M$ again produces a $\Proj^{M_f}$ in the 
middle which can be absorbed into one of the $\Proj^{M'}$, and, since the 
latter is a projector, we have $(\Proj^{M'})^2=\Proj^{M'}$ and thus get back
$\Proj^{\hdots,M,M'}$.

The transversality property (i) for projectors constructed from
different starting multiplets $M$ is obvious due to (iii). 
For two different projectors constructed in the above described way,
starting from the same multiplet $M$, transversality follows by repeating the last
calculation with the second $\Proj^{M'}$ replaced by $\Proj^{M''}$
which then cancel.

The $\ng=3$ projectors which can be constructed in this way are
$\Proj^{1,8}$, $\Proj^{8s,M'}$ and $\Proj^{8a,M'}$ with
$M'=1,\,8s,\,8a,\,10,\,\overline{10},\,27,\,0$. As an example we note that 
$\Proj^{8s,27}$ may be written
\begin{equation}
  \Proj^{8s,27}_{g_1\,g_2\,g_3\,g_4\,g_5\,g_6}
  = \frac{1}{\TR}
  \frac{\Nc}{2(\Nc^2-4)}
  d_{g_1\,g_2\,i_1}\Proj^{27}_{i_1\,g_3\, i_2\,g_6 }d_{i_2\,g_4\,g_5}
\end{equation}
where the normalization derives from $\Proj^{8s}$ in \eqref{eq:ggBasis}.
The other projectors are stated in \appref{sec:3gProjectors}.

\subsection{Starting from a new multiplet and going back 
to one with lower $\Nf$}
\label{sec:new_to_older}

When constructing multiplets $M' \subseteq M \otimes A$ where $M$ was
new in the step before, i.e.\ $\Nf(M)=\ng-1$, we have to distinguish
the three cases $\Nf(M')=\Nf(M)-1,\Nf(M)$ and $\Nf(M)+1$.

We first treat the case $\Nf(M')=\Nf(M)-1=\ng-2$. If $M'$ also appeared in the
construction history immediately before $M$, then we find
\begin{equation}
 \Proj^{\hdots,M,M'} 
 = \frac{\dim M'}{\dim M} \
 \parbox{5cm}{\epsfig{file=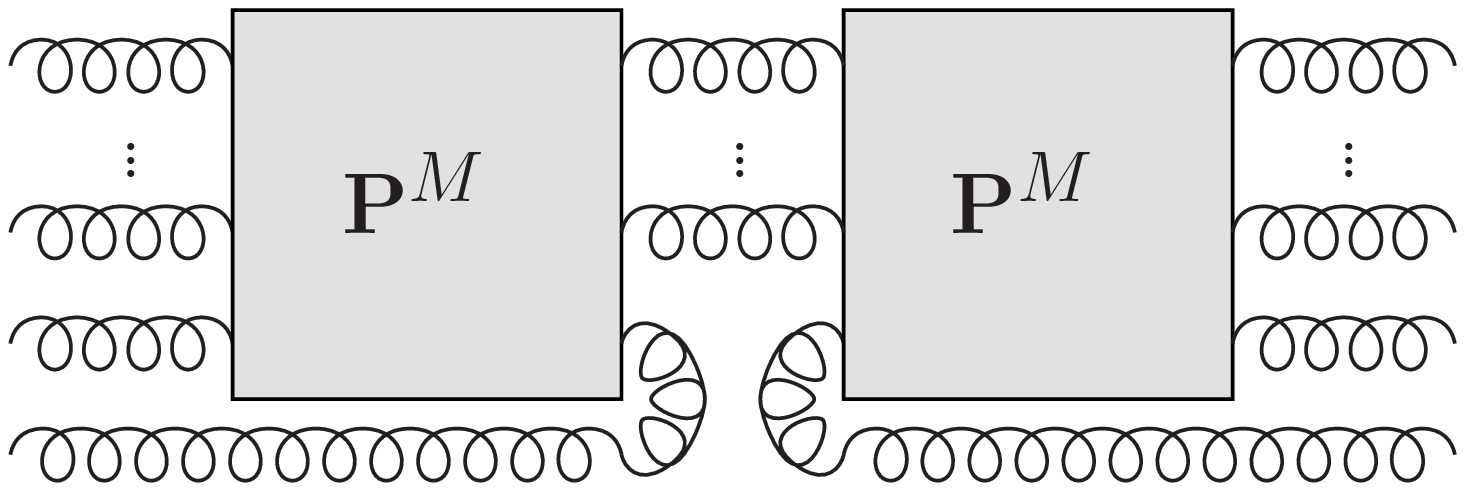,width=5cm}} \, .
\label{eq:back_in_hist}
\end{equation}
Clearly, $\Proj^{...,M,M'}$ satisfies properties (i), (ii) and
(iii). Due to the construction history the $\ng-2$ lines in the middle
carry the desired irreducible representation.  In
\appref{sec:proj_prop} we prove that $\Proj^{...,M,M'}$ as given
in \eqref{eq:back_in_hist} is a projector, and calculate the normalization.
For the three gluon case, the construction in \eqref{eq:back_in_hist}
always works as only the octet can precede the first occurrence two
multiplet $M$.  The projectors which can be constructed in this way
are $\Proj^{10,8}$, $\Proj^{\overline{10},8}$, $\Proj^{27,8}$ and
$\Proj^{0,8}$.
As an example we note that $\Proj^{27,8}$ can be written
\begin{equation}
    \Proj^{27,8}_{g_1\,g_2\,g_3\,g_4\,g_5\,g_6}
    =\frac{4 (\Nc+1)}{\Nc^2 (\Nc+3)} 
    \Proj^{27}_ {g_ 1\,g_ 2\,i_1\,g_3} \Proj^{27}_ {i_1\,g_ 6\,g_ 4\,g_ 5}
\end{equation}
where the prefactor is the ratio of the general $\Nc$ dimensions of the
``octet'' and the ``27''-plet.

For more than three gluons it may happen that $M'$ differs from the
multiplet preceding $M$ in the construction history. The above method
would then give 0. In this case we resort to \eqref{eq:any_old} in
order to find the corresponding projectors.

\subsection{Starting from a new multiplet and going to one with same $\Nf$}
\label{sec:new_to_same}

We now turn to the case $\Nf(M')=\Nf(M)=\ng-1$.  
If the multiplicity of $M'$ in $M \otimes A$ is one, we construct the 
corresponding projector by attaching a gluon to one of the internal gluons,
\begin{equation}
 \Proj^{\hdots,M,M'} 
 = \frac{\mbox{dim}(M')}{B(M,M')}
 \parbox{7cm}{\epsfig{file=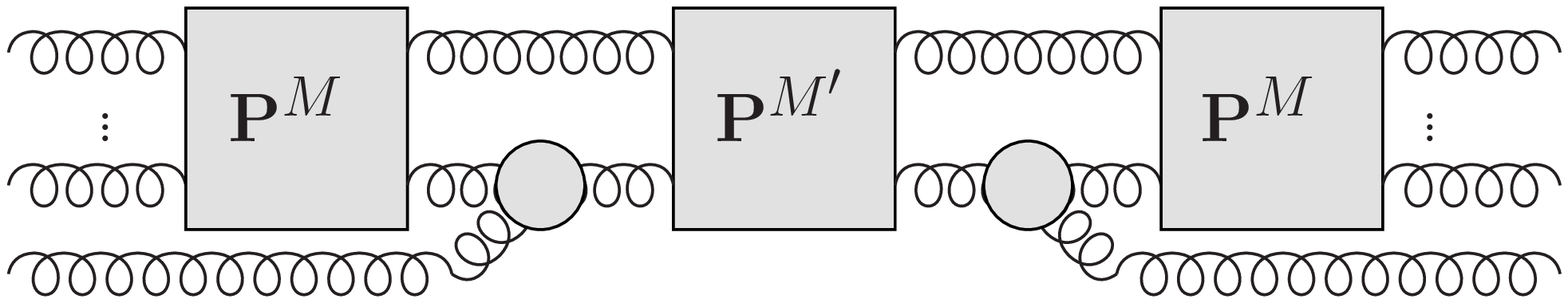,width=7cm}} \, .
\label{eq:new_to_same_nf}
\end{equation}
where the big gray circles can be any of $\ui f$ or $d$ (same on both
sides of $\Proj^{M'}$), and $B$ is the normalization factor from
\eqref{eq:new_to_same_nfN4}.  The projection
operators $\Proj^{27,10},$ $\Proj^{27,\overline{10}},$ $\Proj^{0,10},$
$\Proj^{0,\overline{10}}$, $\Proj^{10,27},$
$\Proj^{\overline{10},27},$ $\Proj^{10,0}$, and
$\Proj^{\overline{10},0}$ are constructed as indicated above. 

As an example we note that, after calculating the normalization, the
projector $\Proj^{27,10}$ may be written
\begin{equation}
 \Proj^{27,10}_{g_1\,g_2\,g_3\,g_4\,g_5\,g_6}
  = 
  \frac{1}{\TR}\frac{2 (\Nc+2)}{\Nc (\Nc+3)} 
  \Proj^{27}_{g_ 1\,g_ 2\,i_ 1\,i_ 2} d_ {i_ 2\,g_ 3\,i_ 3} \Proj^{10}_ {i_1\,i_3\,i_4\,i_6}   d_ {i_6\,g_6\,i_5} \Proj^{27}_ {i_4\,i_5\,g_4\,g_5}.
\end{equation}
For $\ng > 3$ it can happen that the above construction does not
work. This is the case if there is no instance of $M'$ in $m \otimes
A$, where $m$ is the multiplet preceding $M$ in the construction
history. In this case we refer to the general strategy from
\eqref{eq:any_old}.

For the multiplet $M$ appearing in many instances in $M \otimes A$, it
has to be guaranteed that projectors corresponding to all instances
are constructed and that the operators are hermitian and transversal.
We start by defining two projectors of the form
(\ref{eq:new_to_same_nf}), $\Proj^{\hdots,M,Md}$, for which both big
gray circles represent $d$ and $\Proj^{\hdots,M,Mf}$, where both
circles represent $\ui f$. For $\ng=3$ and $M=0$ or $M=27$ these two
projectors are transversal and we are done. 
We note in passing that the two-gluon octet-projectors are also constructed 
in this way, cf.~\eqref{eq:ggBasis_birdtracks}.

For the decuplets, i.e.\ for $M=10$ and $M=\overline{10}$, within
$A^{\otimes 3}$ the situation is slightly more complicated. It turns
out that 
\begin{equation}
  \Proj^{\hdots,M,Mf} \neq \Proj^{\hdots,M,Md}
  \qquad \text{but} \qquad 
   \Proj^{\hdots,M,Mf} \, \Proj^{\hdots,M,Md} \neq 0 \, , 
\end{equation}
i.e.\ the projectors are different but not transversal.  In this case
we keep one of them, $\Proj^{\hdots,M,Mf}$, for our final list. Then
we construct another transversal operator by projecting
$\Proj^{\hdots,M,Md}$ onto the orthogonal complement of the image of
$\Proj^{\hdots,M,Mf}$,
\begin{equation}
\label{eq:TMMfd}
  \Tens^{\hdots,M,Mfd} 
  := \big( \eins_{A^{\otimes 3}} - \Proj^{\hdots,M,Mf} \big) \, 
     \Proj^{\hdots,M,Md} \, 
      \big( \eins_{A^{\otimes 3}} - \Proj^{\hdots,M,Mf} \big) \, .
\end{equation}
The resulting tensor is proportional to the desired projector,
$\Proj^{\hdots,M,Mfd} = \alpha \Tens^{\hdots,M,Mfd}$, and the
normalization is determined by taking the trace and solving for
$\alpha$. This yields the second projector
\begin{equation}
\label{eq:PMMfd}
  \Proj^{\hdots,M,Mfd} 
  = \frac{\dim M}{\tr \Tens^{\hdots,M,Mfd}} \, \Tens^{\hdots,M,Mfd} \, ,
\end{equation}
which is transversal to $\Proj^{\hdots,M,Mf}$. In
\appref{sec:proj_prop} we show that $\Proj^{\hdots,M,Mfd}$ can be
written in the form of \eqref{eq:new_to_same_nf} where the big
gray circles represent a linear combination of $d$ and $\ui f$,
and in \appref{sec:3gProjectors} we give the corresponding projection
operators in this form.

For $\Nc>3$ and $\ng>3$, there may be more than two instances of $M$
in $M \otimes A$, 
allowing for the
definition of more than two projectors, which -- if not already transversal --
can be made so by recursively projecting onto orthogonal complements
as above. In this case the original set of projectors can always be found 
by applying the method outlined in \eqref{eq:any_old}.

\subsection{Projection operators onto new multiplets}
\label{sec:new_multiplets}

In order to be able to reach a new multiplet, i.e.\ one with first
occurrence $\Nf(M')=\ng$, we have to start with a multiplet $M$ which
was new for $\ng-1$ gluons, i.e. $\Nf(M)=\ng-1$.

For constructing the projection operators we split the gluons into 
$q\bar{q}$~pairs, such that symmetrization and anti-symmetrization 
can be done in the quark and anti-quark indices separately.
We start by constructing tensors
\begin{equation}
  \Tens^{\hdots M,M'}
  =  \parbox{9cm}{\epsfig{file=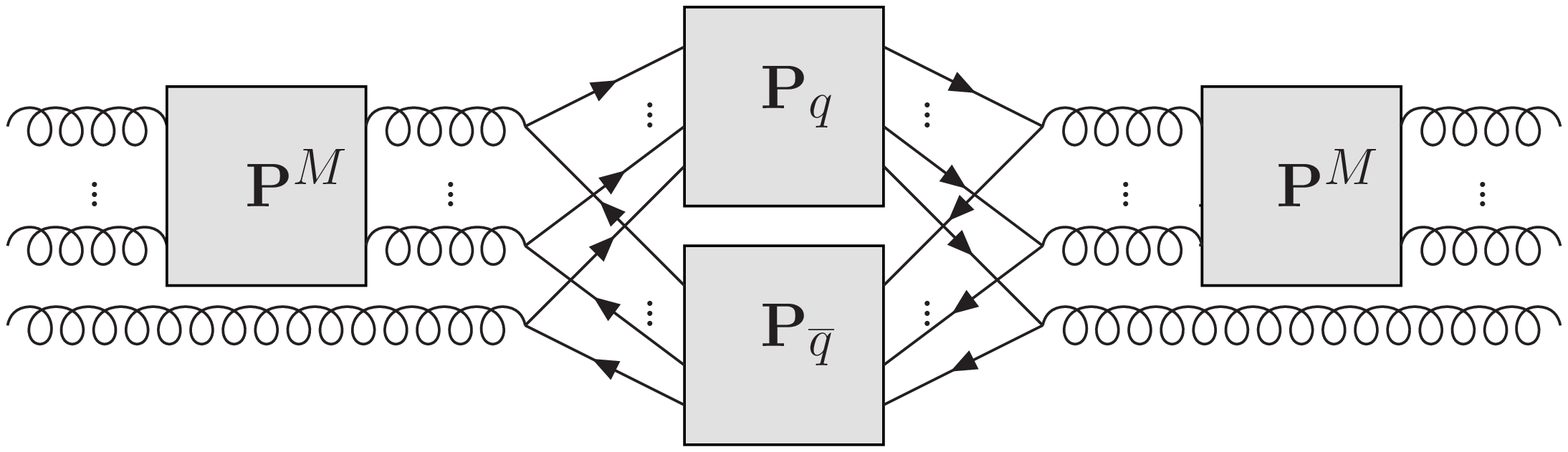,width=9cm}} \, , 
  \label{eq:new_projector}
\end{equation}
where $\Proj_{q}$ projects onto a multiplet $M_q \subset
V^{\otimes\ng}$ and $\Proj_{\qbar}\,$ onto $\overline{M_{\qbar}} \subset
\overline{V}^{\otimes\ng}$, (in the notation of
\appref{sec:first_occurrence}). The motivation for the definition
$\Tens^{\hdots M,M'}$ is best understood by reading the expression
from the center outwards. In the middle section we project onto $M_q
\otimes \overline{M_{\qbar}}$, which contains at most one (and for
sufficiently large $\Nc$ exactly one) new multiplet, as we show in \appref{sec:first_occurrence}.

The projector onto $M_q \otimes \overline{M_{\qbar}}\,$ is sandwiched
between generators projecting $(V\otimes\overline{V})^{\otimes\ng} \to
A^{\otimes\ng}$. This removes projections onto some (not all) old
multiplets, but leaves the projection onto the new multiplet
untouched, as the difference between
$(V\otimes\overline{V})^{\otimes\ng} = (A\oplus\bullet)^{\otimes\ng}$
and $A^{\otimes\ng}$ is $\bigoplus_{\nu=0}^{n_q-1}
\left( \begin{smallmatrix} \displaystyle n_q \\ \displaystyle
  \nu \end{smallmatrix} \right) A^{\otimes\nu}$,
cf.~\eqref{eq:VVbar_binomi}, i.e.\ contains only multiplets of lower
first occurrence.

Finally we sandwich between $\Proj^M \otimes \eins_A$, thus projecting
onto $M \otimes A$ and making sure that property (iii), i.e.\ the
construction history, is satisfied. 
Since there is only one new multiplet within $M_q \otimes \overline{M_{\qbar}}$ 
and only one within $M \otimes A$,
and since $A^{\ng} \subseteq (V\otimes \overline{V})^{\ng}$ we
can always choose the quark and anti-quark multiplets in such a way
that the resulting $\Tens^{\hdots M,M'}$ is non-zero and contains a
part which is proportional to the projector onto $M'$ which we want to
construct. The details of how to choose these quark and anti-quark 
projectors are discussed towards the end of this section.

The tensor $\Tens^{\hdots M,M'}$, seen as linear operator from $M
\otimes A$ to $M \otimes A$ contains a part mapping the new multiplet
$M'$ to itself.  However, there are also pieces mapping other
multiplets $m \subseteq M \otimes A$ to (sometimes a different
instance of) $m$.  These pieces have to be projected out. 
To this end we define the hermitian operator
\begin{equation}
\label{eq:define_Q}
  \mathbf{Q} = \eins_{A^{\otimes\ng}} 
  - \underset{\Nf(m) < \ng}
  {\sum_{m \subseteq A^{\otimes\ng} }} 
    \Proj^m \, , 
\end{equation}
which projects onto the new multiplets within $A^{\otimes\ng}$, and
sandwich $\Tens^{\hdots M,M'}$ between it,
\begin{equation}
\label{eq:define_Ttilde}
  \widetilde{\Tens}^{\hdots M,M'} 
  := \mathbf{Q} \, \Tens^{\hdots M,M'} \, \mathbf{Q} \, .
\end{equation}
Note that for practical purposes in \eqref{eq:define_Q} it is
sufficient to sum over $m\subseteq M \otimes A$ with $\Nf(m) < \ng$,
as all other terms vanish in \eqref{eq:define_Ttilde}. 
Using 
\begin{equation}
  \Tens^{...,M,M'}=\sum_{m,m' \in M \otimes A} t_{m,m'}\CG^{m}\CG^{m'\dagger}\;,
\end{equation}
and rewriting all projection operators in terms of Clebsch-Gordan matrices, 
it is not hard to prove that \eqref{eq:define_Ttilde} can be further 
simplified to
\begin{equation}
  \widetilde{\Tens}^{...,M,M'} 
  = \Tens^{...,M,M'} 
    - \underset{\Nf(m) < \ng}{\sum_{m \subseteq M \otimes A}} 
      \Proj^m \, \Tens^{...,M,M'} \, .
\end{equation}
This is the way in which the three gluon projection operators are constructed.
The desired projector is proportional to $\widetilde{\Tens}^{\hdots M,M'}$
and the normalization is found by taking the trace in 
 $\Proj^{\hdots M,M'} = \alpha \widetilde{\Tens}^{\hdots M,M'}$, 
yielding
\begin{equation}
\label{eq:NewProj}
  \Proj^{\hdots M,M'} 
  = \frac{\dim M'}{\tr\big(\widetilde{\Tens}^{\hdots M,M'}\big)} \, 
    \widetilde{\Tens}^{\hdots M,M'} \, .
\end{equation}
It remains to show that the resulting projector is hermitian. Using
hermitian quark projectors, see~\secref{sec:QuarkProjectors},
the hermiticity of $\Proj^{\hdots M,M'}$ is obvious from inserting
\eqref{eq:new_projector} into \eqref{eq:define_Ttilde}. 

Interestingly $\Proj^{\hdots M,M'}$ is also hermitian when using
conventional, i.e.\ non-hermitian, Young projectors. It follows from
Schur's lemma, see \appref{sec:invariant_tensors}, that invariant
projectors onto multiplets can only be non-hermitian if the multiplet
is not unique within the space from which one projects. Within
$A^{\otimes 3}$, e.g., the $64$-plet is unique and thus an invariant
projector onto it is automatically hermitian, whereas the $35$-plet
appears within both, $10 \otimes 8$, 
and $27 \otimes 8$, see eqs.~(\ref{eq:10x8}) and (\ref{eq:27x8}). 
In the latter case, of a multiplet which is not unique within $A^{\otimes \ng}$, 
it can (and actually does) happen that, when constructing the projector onto a 
particular instance of $M'=35$, the central part of \eqref{eq:new_projector} 
-- without the $\Proj^M$-projectors -- contains terms projecting onto 
an instance of $35$ having a different construction history
{\it and} terms which map one of the instances to the
other. Of these only the latter terms can be non-hermitian. However,
since $M'$ is unique within $M \otimes A$ these terms are removed by
the $\Proj^M$-projectors in \eqref{eq:new_projector}, and, after
removing the lower first occurrence parts in \eqref{eq:define_Ttilde},
the resulting $\widetilde{\Tens}^{\hdots M,M'}$ is hermitian -- 
irrespectively of whether the $P_q$ and $P_{\qbar}$ are hermitian or not.

Finally, we note that property (ii) follows trivially from the
hermiticity using
$\CG^{...M,M'}=\CG^{\dagger...M,M'}=\Proj^{...M,M'}$, and that the
construction history property is manifest. The projectors thus fulfill
all of (i)-(iii).

As an example consider the projector onto $\overline{35} \subset
A^{\otimes 3}$ coming from $27 \subset A \otimes A$. From the Young
tableau \tyoung{12,3} for the quarks, and $\overline{\tyoung{123}}$
for the anti-quarks we get
\begin{equation}
\Yboxdim{10pt}
 \label{eq:3535Barv1}
 \begin{array}{ccc}
   \upsmall{\Nc-1} \csep \upsmall{\Nc-1}\csep \upsmall{\Nc-1}& &
     \csep \upsmall{2} \csep \upsmall{1} 
  \\
  \yng(3,3) & \otimes &  \young(12,3)
  \\[1ex]
  \\[1.5ex]
\end{array} 
 =
  \begin{array}{c}
   \upsmall{\Nc-1} \csep \upsmall{\Nc-1} \csep \upsmall{\Nc-1} 
     \csep \upsmall{2} \csep \upsmall{1} 
  \\
  \yng(5,4) 
  \\[1ex]
  \overline{35}
  \\[1.5ex]
\end{array} \  \oplus \mbox{old multiplets} \, .
\end{equation}
As \tyoung{12,3} is symmetric in the first two indices, and since the
27-plet is contained in $\tyoung{12} \otimes \overline{\tyoung{12}}$,
these $q$- and $\qbar$-diagrams are chosen such that the projection
onto the $27$-plet in the first two indices is non-vanishing,
\begin{equation}
  \Tens^{27,\overline{35}}
  =  \parbox{9cm}{\epsfig{file=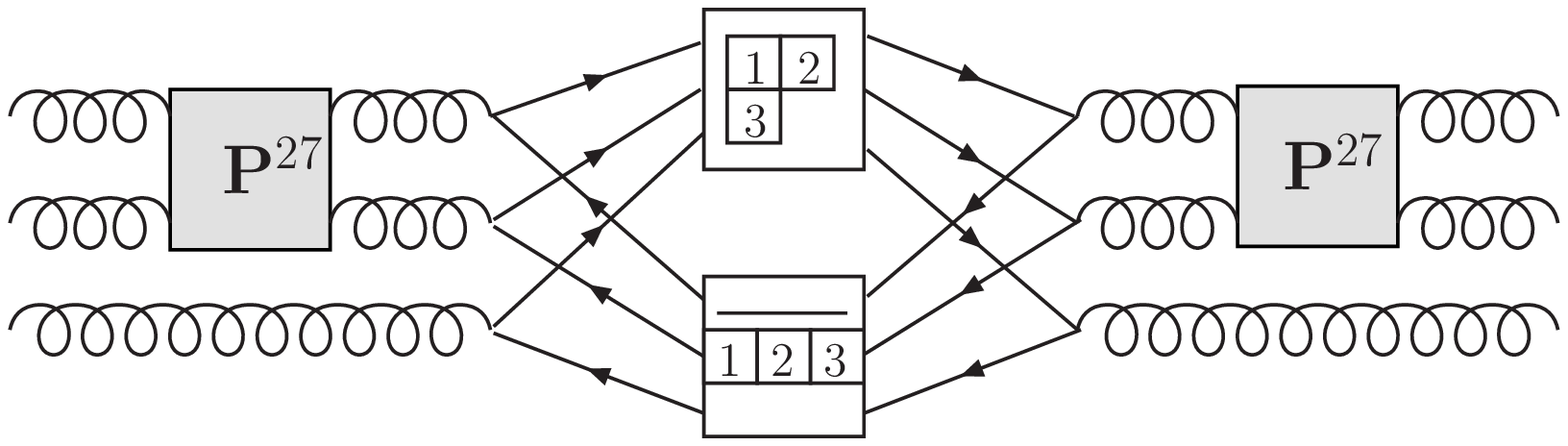,width=9cm}} \neq 0 \,.
\label{eq:2735}
\end{equation}
In this way, by considering the symmetry of the quarks and the
anti-quarks in the $\ng-1$ projector, and using the $\ng$ quark and 
anti-quark projectors where the $\ng^{th}$ quark and anti-quark are added to their respective
Young diagrams, in such a way that the Young diagrams have the right
shapes to guarantee $M' \subseteq M_q \otimes \overline{M_{\qbar}}$,
it is always possible to find
suitable $q$- and $\qbar$-diagrams. This  allows for a unique
construction of all instances of new multiplets, i.e. each new multiplet
within $A^{\otimes \ng}$ has a corresponding new multiplet within 
$(V \otimes \overline{V})^{\ng}$, defined in this way.

For the three gluon case the projection operators are written down in
\appref{sec:3gProjectors}. It has been checked that these projection
operators sum to unity, i.e.
\begin{equation}
  \sum_{I \in A^{\otimes 3}} \Proj^I_{g_1\,g_2\,g_3\,g_4\,g_5\,g_6}=\delta_{g_1\,g_3}\delta_{g_2\,g_4}\delta_{g_3\,g_6},
\end{equation}
and that they are hermitian
\begin{equation}
  \Proj_{g_1\,g_2\,g_3\,g_4\,g_5\,g_6}=\Proj^{*}_{g_4\,g_5\,g_6\, g_1\,g_2\,g_3}.
\end{equation}
As a consequence of the hermiticity it follows that the real
projection operators, and the real linear combinations are symmetric,
i.e.
\begin{equation}
  \Proj_{g_1\,g_2\,g_3\,g_4\,g_5\,g_6}=\Proj_{\,g_4\,g_5\,g_6\, g_1\,g_2\,g_3}\quad
  \mbox{for real combinations}.
\end{equation}

\subsection{Further remarks}
\label{sec:general_remarks}
We conclude this section with some general remarks.  First we note
that the fact that the multiplets can be uniquely constructed from the
quark and anti-quark projection operators also introduces an
alternative $\Nc$-independent description of all multiplets.  
They can be labeled using first the lengths of all quark columns, and
then the lengths of all anti-quark columns, e.g. $12$ and $111$ in the
above case of $\overline{35}$.
We use this notation for the new
multiplets arising for three gluons, which have dimension 0 for
$\SU(3)$. As an example we note that $\Proj^{27,\overline{35}}$
alternatively could have been written $\Proj^{c11c11,c21c111}$.  Using
this notation also translates straightforwardly to the representations of
Young diagrams used in eqs. (\ref{eq:1x8})-(\ref{eq:0x8}).  The length
of the first column is $\Nc$ minus the length of the last anti-quark
column, the length of the second column is $\Nc$ minus the length of
the second last anti-quark column, etc. After this follow columns
with lengths given by the quark columns. The notation also immediately
reveals the first occurrence of a multiplet; it is simply the sum of
the quark-column lengths (which equals the sum of the anti-quark
column lengths -- before conjugating).  To distinguish the
different {\it instances} of the new multiplets we note that they could
alternatively have been labeled using Young tableaux with quark and
anti-quark numbers filled in as in \eqref{eq:2735}.

In this context we also remark that the fact that $\Nc$ is small in QCD
leads to a significant reduction of the number of projection
operators in the $\ng$ gluon space, partly since many new projection
operators vanish for small $\Nc$, and partly because projectors may be
forbidden by construction, such as $\Proj^{0,8}$.  Similarly, there is
a reduction in the number of basis vectors in the space of $2\ng$
gluons.  In fact, as is argued in \appref{sec:NumberOfProjectors}, the
number of basis vectors grows only exponentially, as opposed to
factorially, c.f. \eqref{eq:Nv}, for finite $\Nc$.  As can be seen in
\tabref{table:states}, for more than a few gluons, the reduction in
the number of projectors and in the dimension of the vector space is significant.

\TABLE[t]{
\centering 
\begin{tabular}{c c c c c} 
\hline\hline 
Case & Projectors $\Nc=3$ & Projectors $\Nc=\infty$ & Vectors $\Nc=3$ & Vectors $\Nc=\infty$ \\ [0.5ex] 
\hline 
2g $\to$ 2g & 6 & 7 & 8 & 9 \\ 
3g $\to$ 3g & 29 & 51 & 145 & 265  \\
4g $\to$ 4g & 166 & 513 & 3 598 & 14 833\\
5g $\to$ 5g & 1 002 & 6 345 & 107 160 & 1 334 961\\[1ex] 
\hline 
\end{tabular}
\label{table:states}
\caption{Number of projectors and basis vectors for $\ng \to \ng$ gluons 
({\it without} imposing projection operators and vectors to appear in 
self-conjugate combinations). In the $\Nc\rightarrow \infty$ limit 
the number of vectors for a total of $\Ng$, incoming plus outgoing, gluons is
given by subfactorial($\Ng$) as in \eqref{eq:Nv}.
For $\Nc=3$ the number of projectors are found by counting all irreducible 
representations occurring in $A^{\otimes \ng}$, and the number of 
vectors are obtained by considering all possible transitions between
multiplets of same type on the incoming and outgoing side.}
}

We finally remark that as the gluon transforms under a real representation,
only projection operators and basis vectors which are invariant
under charge conjugation may appear
for purely gluonic processes
\cite{Kidonakis:1998nf,Oderda:1999kr,Sjodahl:2008fz,Sjodahl:2009wx}.
Non-invariant projectors may 
only appear together with their charge conjugated versions.
Thus, for example $\Proj^{10,35}$ may only occur together with 
$\Proj^{\overline{10},\overline{35}}$. This reduces the number of 
projection operators for $\ng=3$ gluons from 51 to 36 in the $\Nc \to \infty$ limit,
and from 29 to 21 for $\Nc=3$.

\section{Multiplet based gluon bases}
\label{sec:GluonBasis}
In order to construct the vectors in the color space for $A^{\otimes 6}$ 
we may group the gluons as $g_1\,g_2\,g_3\to g_4\,g_5\,g_6$.
We can (naively)
combine any instance of an incoming multiplet with any (other) instance of the 
same outgoing multiplet as there is no conserved quantum number
which forbids the transition from one instance of a multiplet to another
instance of the same multiplet.
However, as long as only gluons are involved, all vectors have to be invariant under 
charge conjugation. The construction of the basis for the six gluon 
space, is discussed in detail in \appref{sec:6gBasis}.

Alternatively we could have grouped the gluons as 
$g_1 g_2\to g_3 g_4 g_5 g_6$ in which case only multiplets with first 
occurrence up to two could have appeared on the left hand side, and therefore 
also on the right hand side. On the incoming side we then have the multiplets 
as enumerated in \eqref{eq:SU388}.
For keeping track of the multiplets on
the outgoing side we would need to find all the multiplets with first 
occurrence up to two, when Young multiplying four gluons.

The multiplets with first occurrence up to two, arising when multiplying
a first occurrence two multiplet with an octet can be read off from
eqs. (\ref{eq:1x8})-(\ref{eq:0x8}). 
However, we also have multiplets with first 
occurrence two, when multiplying the three gluon multiplets with first 
occurrence
three with a gluon. To enumerate these multiplets we need to Young multiply
the first occurrence three states in 
eqs. (\ref{eq:1x8})-(\ref{eq:0x8}) with a gluon.
After having performed this task we can, however, construct the basis vectors 
using projection operators with first occurrence up to three. 

As the first occurrence cannot change
by more then one unit when multiplying with a gluon, it is in general the 
case that we never need projection operators with first occurrence 
larger than $\ng$ when treating up to $2\,\ng+1$ gluons. 
This is true independently of how the gluons are grouped.
In particular the $\ng=3$ projection operators are also sufficient to 
construct orthonormal bases for up to seven gluons.

\section{General construction of multiplet bases}
\label{sec:QuarksAndGluons}

To treat the general case involving both quarks and gluons we note that 
for each quark (outgoing quark or incoming anti-quark) there is an
anti-quark (incoming quark or outgoing anti-quark). We can therefore always
start the process of sub-grouping by paring up each quark with an 
anti-quark. Each $\qqbar$ pair can either be in a singlet, reducing the 
basis construction to the corresponding problem without this $\qqbar$ pair, 
or in an octet. In the latter case the basis construction is equivalent to the
construction where the $\qqbar$ pair is traded for a gluon. Below we exemplify
the basis construction by constructing orthogonal bases for processes
involving six colored partons. Since, for each quark there
is a corresponding anti-quark, we may have 
three $\qqbar$-pairs,
two $\qqbar$-pairs and 2 gluons,
one $\qqbar$-pair and four gluons or 
six gluons, as treated above.
Mathematica .m-files containing the bases constructed in this way are 
attached electronically.

\subsection{Example: Three $\qqbar$ pairs}

This case is dealt with extensively in \secref{sec:QuarkBasis} where 
a basis is constructed using the hermitian quark projection operators
for $q_1 q_2 q_3 \to q_4 q_5 q_6$.
\TABLE[t]{
\centering 
\begin{tabular}{c c c c c c c c} 
\hline\hline 
$\SU(3)$ dimension/notation & 1 & 8 \\ [0.5ex] 
Multiplet, general notation & c0c0 & c1c1\\ [0.5ex] 
\hline 
In $\q_1 \qbar_2$ 
& $(12)^1$  
& $(12)^8$ \\
\hline
Out $\q_3 \qbar_4 \q_5\qbar_6 $ 
& $( (34)^1 (56)^1 )^1$
& $((34)^1 (56)^8)^8$, $((34)^8 (56)^1)^8$
\\
& $( (34)^8 (56)^8 )^1$  
& $((34)^8 (56)^8)^{8s}$, $((34)^8 (56)^8)^{8a}$\\
\hline 
\end{tabular}
\label{table:3qqbar}
\caption{The multiplets appearing in the construction of the bases
corresponding to 
$\q_1 \qbar_2 \to q_3 \qbar_4 q_5 \qbar_6$.
On the incoming side $q_1 \qbar_2$ may be in a singlet 
or in an octet. Due to color conservation the outgoing multiplet
must be the same. However, in this case, there are may ways to build
up the singlet or octet. To have an overall singlet the quarks 
$q_3$ and $\qbar_4$ may separately form a singlet if $q_5$ and $\qbar_6$
do too.
Alternatively $q_3$ and $\qbar_4$ may be in an octet which then combines
with another octet from $q_5$ and $\qbar_6$ to from a total singlet.
For the octets we have a total of four options. All in all, the multiplets
may be enumerated as above. The vector space 
has thus $1 \times 2$ dimensions for the singlets and 
$1 \times 4$ dimensions from the octets.
} }
In this section we note that we can equally well construct the basis using 
the gluon projection operators and grouping the partons as
$q_1 \qbar_2 \to q_3 \qbar_4 q_5 \qbar_6$. The multiplets on the 
ingoing and outgoing side may then be constructed as in 
\tabref{table:3qqbar}, by first grouping $\qqbar$-pairs to form
octets or singlets. Clearly the dimension of the basis must still be 
6 as in \secref{sec:QuarkBasis}.
Having enumerated all the basis vectors as in \tabref{table:3qqbar},
we may write down the basis using the (somewhat redundant) notation
$\Vec^{M_{12};M_{34},M_{56}, M_{3456}}$,
\begin{eqnarray}
  \Vec^{1;1,1,1}_{q_1\,q_2\,q_3\,q_4\,q_5\,q_6}&=& 
  \frac{1}{\sqrt{\Nc^3}}
  \delta^{q_1}_{q_2}\delta^{q_4}_{q_3}\delta^{q_6}_{q_5}\;,
  \nonumber \\
  \Vec^{1;8,8,1}_{q_1\,q_2\,q_3\,q_4\,q_5\,q_6}&=& 
  \frac{1}{\TR}
  \frac{1}{\sqrt{\Nc (\Nc^2-1)}}
  \delta^{q_1}_{q_2}(t^{i_1})^{q_4}_{q_3} (t^{i_1})^{q_6}_{q_5} \;,
  \nonumber \\
  \Vec^{8;1,8,8}_{q_1\,q_2\,q_3\,q_4\,q_5\,q_6}&=&
  \frac{1}{\TR}
  \frac{1}{\sqrt{\Nc (\Nc^2-1)}}(t^{i_1})^{q_1}_{q_2}\delta^{q_4}_{q_3} (t^{i_1})^{q_6}_{q_5}\;, 
  \nonumber \\
  \Vec^{8;8,1,8}_{q_1\,q_2\,q_3\,q_4\,q_5\,q_6}&=& 
  \frac{1}{\TR}
  \frac{1}{\sqrt{\Nc (\Nc^2-1)}}(t^{i_1})^{q_1}_{q_2}(t^{i_1})^{q_4}_{q_3} \delta^{q_6}_{q_5}\;, 
  \nonumber \\
  \Vec^{8;8,8,8s}_{q_1\,q_2\,q_3\,q_4\,q_5\,q_6}&=& 
  \frac{1}{\TR^2}
  \sqrt{\frac{\Nc}{2(\Nc^4-5 \Nc^2+4)}}
  (t^{i_1})^{q_1}_{q_2}d_{i_1\,i_2\,i_3}(t^{i_2})^{q_4}_{q_3} (t^{i_3})^{q_6}_{q_5} \;,
  \nonumber \\
  \Vec^{8;8,8,8a}_{q_1\,q_2\,q_3\,q_4\,q_5\,q_6}&=& 
  \frac{1}{\TR^2}
  \frac{1}{\sqrt{2\Nc(\Nc^2-1)}}(t^{i_1})^{q_1}_{q_2}\ui f_{i_1\,i_2\,i_3}(t^{i_2})^{q_4}_{q_3} (t^{i_3})^{q_6}_{q_5}\;,
  \label{eq:3qqbarBasisv2}
\end{eqnarray}
where the normalization has been fixed to get an orthonormal basis.
We note that in this case we did not have to make use of the two
or three gluon projection operators, which is immediately clear from 
the grouping of the partons, since on the left hand side we cannot get
a multiplet with first occurrence larger than one.
The basis in \eqref{eq:3qqbarBasisv2} is related to that in 
\eqref{eq:3qqbarBasisv1} by an orthogonal transformation.

\subsection{Example: Two $\qqbar$ pairs and two gluons}
\label{sec:2qqbar2g}

\TABLE[t]{
\centering 
\begin{tabular}{c c c c c c c c} 
\hline\hline\\[-2.5ex]  
$\SU(3)$ dim & 1 & 8 & 10 & $\overline{10}$ & 27  & 0 \\ [0.5ex] 
Multiplet & c0c0 & c1c1 & c11c2 & c2c11 & c11c11  & c2c2 \\ [0.5ex] 
\hline\\[-2.5ex]  
In $\q_1 \qbar_2 g_3 $ 
& $((12)^83)^1$  
& $((12)^13)^8$
& $((12)^83)^{10}$  
& $((12)^83)^{\overline{10}}$ 
& $((12)^83)^{27}$
& $((12)^83)^{0}$\\ 
& 
&$((12)^83)^{8s}$ 
& 
& 
& 
& \\
& 
&$((12)^83)^{8a}$ 
& 
& 
& 
& \\
\hline\\[-2.5ex]
\end{tabular}
\label{table:2qqbar2g} 
\caption{Table describing the multiplets used in the construction of the bases
corresponding to $\q_1 \qbar_2 g_3 \to q_4 \qbar_5 g_6$.
As the incoming and outgoing particle content is the same,
the possible multiplets for the outgoing partons are identical to those 
for the incoming.
For each basis vector, any instance of a multiplet $M$ on the incoming side 
can be combined with any instance of $M$ on the outgoing side.
For $\q_1 \qbar_2 g_3 \to q_4 \qbar_5 g_6$ we thus have 
$1+3^2+1+1+1+1=14$ basis vectors, reducing to $13$ for $\SU(3)$.
}
} 

In order to construct a basis for processes involving two $\qqbar$-pairs and two gluons
we may for example group the partons as $\q_1 \qbar_2 g_3 \to q_4 \qbar_5 g_6$
and use Young multiplication to arrive at the multiplet possibilities
in \tabref{table:2qqbar2g}.
(Alternatively we could use $g_3\,g_6 \to \q_1 \qbar_2 q_4 \qbar_5$
in which case we would not have any projection operators.)

The $q_1 \qbar_2$, can be either in a singlet or in an octet. 
If they are in a singlet, the singlet is combined with the gluon $g_3$
to an overall octet. On the other hand, if $q_1 \qbar_2$ are in an octet, 
when combined with $g_3$, the overall multiplet may be any of
$1, 8^s,8^a,10,\overline{10},27,0$.
On the outgoing side, the same method of subgrouping can be applied. 
From this it is immediately clear that we need gluon projectors with first
occurrence up to two, but not higher.

To construct the projectors, the states with corresponding construction
history on the incoming and outgoing side have to be joined, giving 
\begin{eqnarray}
  \Proj^{8,1}_{q_1\,q_2\,g_3\,q_4\,q_5\,g_6}&=&
  \frac{1}{\TR}
  (t^{i_1})^{q_1}_{q_2} \; \Proj^{1}_{g_3\,i_1\,g_6\,i_2} (t^{i_2})^{q_5}_{q_4}\;,
  \nonumber \\
  \Proj^{1,8}_{q_1\,q_2\,g_3\,q_4\,q_5\,g_6}&=&
  \frac{1}{\Nc}
  \delta^{q_1}_{q_2}\; \delta_{g_3\,g_6}\; \delta^{q_5}_{q_4} \;,\nonumber \\
  \Proj^{8,8s}_{q_1\,q_2\,g_3\,q_4\,q_5\,g_6}&=&
  \frac{1}{\TR}
  (t^{i_1})^{q_1}_{q_2} \; \Proj^{8s}_{g_3\,i_1\,g_6\,i_2} (t^{i_2})^{q_5}_{q_4}\;,
  \nonumber \\
  \Proj^{8,8a}_{q_1\,q_2\,g_3\,q_4\,q_5\,g_6}&=&
  \frac{1}{\TR}
  (t^{i_1})^{q_1}_{q_2} \; \Proj^{8a}_{g_3\,i_1\,g_6\,i_2} (t^{i_2})^{q_5}_{q_4}\;,
  \nonumber \\
  \Proj^{8,10}_{q_1\,q_2\,g_3\,q_4\,q_5\,g_6}&=&
  \frac{1}{\TR}
  (t^{i_1})^{q_1}_{q_2} \; \Proj^{10}_{g_3\,i_1\,g_6\,i_2} (t^{i_2})^{q_5}_{q_4}\;,\nonumber \\
  \Proj^{8,\overline{10}}_{q_1\,q_2\,g_3\,q_4\,q_5\,g_6}&=&
  \frac{1}{\TR}
  (t^{i_1})^{q_1}_{q_2} \; \Proj^{\overline{10}}_{g_3\,i_1\,g_6\,i_2} (t^{i_2})^{q_5}_{q_4}\;,\nonumber \\
  \Proj^{8,27}_{q_1\,q_2\,g_3\,q_4\,q_5\,g_6}&=&
  \frac{1}{\TR}
  (t^{i_1})^{q_1}_{q_2} \; \Proj^{27}_{g_3\,i_1\,g_6\,i_2} (t^{i_2})^{q_5}_{q_4}\;,\nonumber \\
  \Proj^{8,0}_{q_1\,q_2\,g_3\,q_4\,q_5\,g_6}&=&
  \frac{1}{\TR}
  (t^{i_1})^{q_1}_{q_2} \; \Proj^{0}_{g_3\,i_1\,g_6\,i_2} (t^{i_2})^{q_5}_{q_4}\;,
\end{eqnarray}
where the factor $1/\TR$ in the norm (when present) compensates for the factor 
$\TR$ coming from contraction of quarks to form gluon projectors when 
squaring the above projectors.

The basis vectors in the six parton space are given by allowing all 
instances of a given multiplet to go into any instance of the same 
multiplet, as enumerated in \tabref{table:2qqbar2g}. The vector space 
is thus 14 dimensional, reducing to 13 for $\SU(3)$. 
Their explicit forms are given in \appref{sec:2qqbar2gVectors}.

\subsection{Example: One $\qqbar$ pair and four gluons}

To find an orthonormal basis for processes involving one $\qqbar$ pair and 
four gluons, we again utilize the method of sub-grouping, sorting the
partons as $\q_1\, \qbar_2\, g_3 \to g_4\, g_5 \,g_6$, 
and finding the multiplets listed in \tabref{table:1qqbar4g}. 

We note that we could as well have sorted the partons (for example) as 
$\q_1\, \qbar_2\, \to g_3 \,g_4\, g_5 \,g_6$, in which case we would have 
had to perform the four gluon Young tableaux multiplication on the 
right hand side.

\TABLE[t]{\small
\centering 
\begin{tabular}{c c c c c c c c} 
\hline\hline\\[-2.5ex]  
$\SU(3)$ dim & 1 & 8 & 10 & $\overline{10}$ & 27  & 0 \\ [0.5ex] 
Multiplet & c0c0 & c1c1 & c11c2 & c2c11 & c11c11  & c2c2 \\ [0.5ex] 
\hline\\[-2.5ex] 
In $\q_1 \qbar_2 g_3 $ 
& $((12)^83)^1$  
& $((12)^13)^8$
& $((12)^83)^{10}$  
& $((12)^83)^{\overline{10}}$ 
& $((12)^83)^{27}$
& $((12)^83)^{0}$\\ 
& 
&$((12)^83)^{8s}$ 
& 
& 
& 
& \\
& 
&$((12)^83)^{8a}$ 
& 
& 
& 
& \\
\hline\\[-2.5ex] 
Out $g_4 g_5 g_6 $ 
& 
& $((45)^{1}6)^{8}$
& 
& 
& 
& \\ 
& $((45)^{8s}6)^1$  
& \quad $
((45)^{8s}6)^{8s/a}$
& $((45)^{8s}6)^{10}$  
& $((45)^{8s}6)^{\overline{10}}$ 
& $((45)^{8s}6)^{27}$ 
& $((45)^{8s}6)^{0}$\\ 
& $((45)^{8a}6)^1$  
& \quad $
((45)^{8a}6)^{8s/a}$
& $((45)^{8a}6)^{10}$  
& $((45)^{8a}6)^{\overline{10}}$ 
& $((45)^{8a}6)^{27}$ 
& $((45)^{8a}6)^{0}$\\ 
& 
& $((45)^{10}6)^8$
& $((45)^{10}6)^{10}$  
& $((45)^{\overline{10}}6)^{\overline{10}}$
& $((45)^{10}6)^{27}$ 
& $((45)^{10}6)^{0}$\\ 
& 
& $((45)^{\overline{10}}6)^8$
& $((45)^{10}6)^{10}$  
& $((45)^{\overline{10}}6)^{\overline{10}}$
& $((45)^{\overline{10}}6)^{27}$ 
& $((45)^{\overline{10}}6)^{0}$\\ 
& 
& $((45)^{27}6)^8$
& $((45)^{27}6)^{10}$  
& $((45)^{27}6)^{\overline{10}}$
& $((45)^{27}6)^{27}$
& $((45)^{0}6)^{0}$\\
&
& $((45)^{0}6)^8$
& $((45)^{0}6)^{10}$  
& $((45)^{0}6)^{\overline{10}}$
& $((45)^{27}6)^{27}$ 
& $((45)^{0}6)^{0}$\\ 
\hline 
\end{tabular}
\label{table:1qqbar4g} 
\caption{Table describing the multiplets used in the construction of the bases
corresponding to 
$\q_1 \qbar_2 g_3 \to g_4 g_5 g_6$. 
In each case basis vectors corresponding to any instance of a multiplet
$M$ on the incoming side can be combined with any instance of the same 
outgoing multiplet.
We thus have a 
$2+3\times 9+6+6+6+6=53$ dimensional vector space, reducing to 
$2+3\times 8+4+4+6+0=40$ for $\SU(3)$.
}} 

With the chosen treatment the 
incoming side is treated precisely as in the case of 
$\q_1 \qbar_2 g_3 \to q_4 \qbar_5 g_6$ above.
Due to color conservation, only states with first occurrence up to
two appearer on the outgoing side, meaning that we again only need the 
gluon projection operators with first occurrence up to two.

If $\q_1 \qbar_2$ are in an octet, the basis construction is similar to the case 
of $2g \to 3g$ \cite{Sjodahl:2008fz},
with the exception that also non-self conjugate states may appear.
This leads to a doubling of the number of basis vectors
for the part of the sub-space where $\q_1 \qbar_2$ are in an octet.

As the types of partons on the incoming and outgoing side are not the 
same, there are no projection operators, but there are 53 orthogonal basis 
vectors out of which 13 vanish for $\SU(3)$. Due to the size of the basis, we do
not display the basis vectors but only attach them electronically
to this publication.
A future publication of the Mathematica package used in the construction of
the projection operators and basis vectors is planned.

\section{Conclusions and outlook}
\label{sec:conclusions}

In this paper we outline a general algorithm for constructing orthogonal 
(normalized) multiplet bases for color summed calculations in 
QCD, for any number of quarks and gluons, any $\Nc$, 
and to any order in perturbation theory.

This is accomplished by first constructing gluon projection operators
projecting onto irreducible representations. We outline, how to
construct these projection operators recursively for an arbitrary
number of gluons, and illustrate the method by constructing a complete
set of orthogonal projection operators for three gluons.

A key idea for the construction is the splitting of gluons into
$\qqbar$-pairs, and the subsequent usage of hermitian (anti-) quark
projection operators, as illustrated in \eqref{eq:new_projector}. We
find that, in the $\Nc \to \infty$ limit, there is a one to one
correspondence between the quark and anti-quark symmetries, and the
gluon projection operators, cf. \appref{sec:first_occurrence}, whereas
for small $\Nc$ many projection operators vanish.  As a consequence of
this uniqueness, we note that the Young tableaux corresponding to
different $\Nc$ stand in a one to one -- or one to zero -- relation to
each other.  We also remark that choosing explicit indices we are able
to calculate the Clebsch-Gordan matrices $\CG^M$ in an $\Nc$
independent manner.

As an illustration we explicitly construct three gluon projection 
operators projecting onto mutually orthogonal subspaces, and the corresponding 
six gluon basis. Note, however, that the three gluon projectors can also be 
used for constructing orthogonal bases for up to seven gluons, or up to
$\ng+\Nq=7$ gluons and $\qqbar$-pairs in general.

Using this type of gluon projection operators, we note 
that we can easily construct complete sets of basis vectors for an 
arbitrary number of $\qqbar$-pairs and gluons.
The bases constructed in this way have the advantage of being orthogonal.
As the number of basis vectors (for $\Nc\to \infty$) 
scales roughly as a factorial in the number
of gluons and $\qqbar$-pairs, cf. \eqref{eq:Nv} and \eqref{eq:Nvqg},
this is a very strong advantage for processes involving many partons.
These bases can also trivially be made minimal for the $\Nc$ under 
consideration, by just omitting the vanishing basis vectors.
For many partons, this leads to a significant reduction in
the number of terms that have to be treated.
For example, for a total of 10 gluons, there are about one million
basis vectors for $\Nc\to\infty$, which in the standard, non-orthogonal,
trace-type basis would give rise to an unmanageable 
$\approx 10^{12}$ elements to calculate while squaring an amplitude. 
On the other hand, for the $\Nc=3$, in the
minimal multiplet basis, we instead have to treat about $10^5$ terms.
In fact we prove in \appref{sec:NumberOfProjectors} 
that the number of basis vectors scales at most as an exponential, 
rather than a factorial, in the number of gluons plus $\qqbar$-pairs.

The usage of these orthogonal SU(3) minimal bases therefore has the potential
to speed up exact multi-parton calculations significantly.
However, it should be remarked that in order to facilitate this, additional 
theoretical progress is advantageous. For the standard trace-type bases powerful
recursion relations in the number of external particles can be 
employed for special cases
\cite{Parke:1986gb,Mangano:1987xk,Kleiss:1988ne,Berends:1987me,Cachazo:2004kj,Britto:2004ap,Bern:2008qj,BjerrumBohr:2010ta,BjerrumBohr:2010zb}
and it remains to be explored how these would manifest themselves in 
a multiplet type basis.

In general (and especially for classes of processes where no efficient recursion 
relations can be found),
one would want efficient algorithms for sorting Feynman diagrams in the multiplet
basis. If this can be done at tree level, and the effect of gluon exchange
could be found, 
the color structure of higher order calculations could probably be 
dealt with efficiently.
Although the effect 
of gluon emission is simple in many situations, and although the soft 
anomalous dimension matrices 
\cite{Sotiropoulos:1993rd,Kidonakis:1998nf,Kyrieleis:2005dt,Dokshitzer:2005ig,Sjodahl:2008fz} describing the effect
of gluon exchange on the various basis vectors, have been found to 
be relatively sparse, a complete systematic treatment is still pending.

\section*{Acknowledgments}
We thank Johan Grönqvist, Jonas Lampart and Wolfdieter Lang for
helpful discussions. 
M.S. was supported by a Marie Curie Experienced Researcher fellowship of 
the MCnet Research Training network, contract MRTN-CT-2006-035606, 
by the Helmholtz Alliance "Physics at the Terascale" and by the 
Swedish Research Council, contract number 621-2010-3326.

\appendix

\section{Some birdtrack conventions}
\label{sec:birdtracks}

We introduced Cvitanovi\'c's birdtrack notation \cite{Cvi76,Cvi08} in
sections \ref{sec:ColorSpace} and \ref{sec:illustration_gg_to_gg}. Here
we collect some properties and conventions.

When translating the birdtrack diagram for a projector to index
notation we subgroup the indices on the l.h.s. and on the r.h.s.,
respectively, i.e.
\begin{equation}
\label{eq:a_projector}
  \parbox{5cm}{\epsfig{file=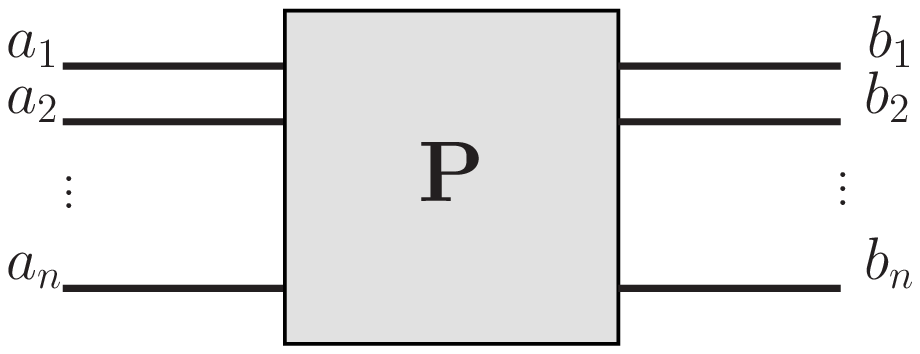,width=5cm}} 
  = \Proj_{a_1\,a_2\,\hdots\,a_n\,b_1\,b_2\,\hdots\,b_n}\;,
\end{equation}
where all lines could be
gluon \parbox{1.5cm}{\epsfig{file=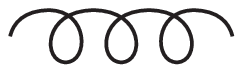,width=1.5cm}},
quark \parbox{1.5cm}{\epsfig{file=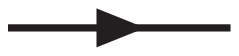,width=1.5cm}}
or
anti-quark \parbox{1.5cm}{\epsfig{file=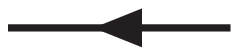,width=1.5cm}}
lines. With this convention we deviate from \cite[sec.~4.1]{Cvi08},
where all indices are read off in anti-clockwise order, in which case
the r.h.s. of \eqref{eq:a_projector} would read
$\Proj_{a_1\,\hdots\,a_n\,b_n\,\hdots\,b_1}$. However, for the
structures constants $f_{abc}$ and the totally symmetric tensor
$d_{abc}$ we do adopt the convention of assigning indices in an
anti-clockwise order, see eqs.~(\ref{eq:t_and_f_def}) and
(\ref{eq:d_def}).

Tensor products are taken by writing the birdtrack diagrams one below
the other, 
\begin{equation}
\label{eq:tensor_product}
  \raisebox{-.5mm}{\parbox{3cm}{\parbox{3cm}{
  \epsfig{file=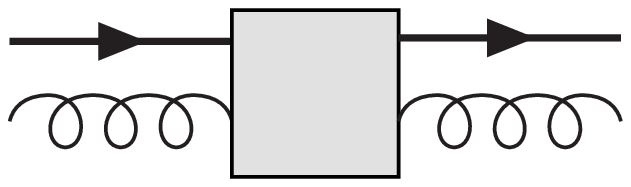,width=3cm}}}} \otimes
  \parbox{3cm}{\epsfig{file=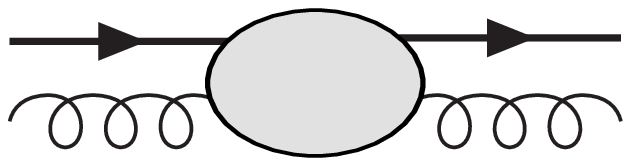,width=3cm}} 
  = \begin{matrix} \parbox{3cm}{\epsfig{file=Figures/square.eps,width=3cm}} 
    \\ \parbox{3cm}{\epsfig{file=Figures/blob.eps,width=3cm}} \end{matrix} \, , 
\end{equation}
and index contractions are achieved by joining lines, 
\begin{equation}
\label{eq:index_contraction}
  {}^{q_1}_{g_1}\!\raisebox{-1mm}{\parbox{3cm}{
  \epsfig{file=Figures/square.eps,width=3cm}}}{}^{q_2}_{g_2} \quad
  {}^{q_2}_{g_2}\!\raisebox{-.5mm}{\parbox{3cm}{
  \epsfig{file=Figures/blob.eps,width=3cm}}}{}^{q_3}_{g_3} 
  = {}^{q_1}_{g_1}\!\raisebox{-.5mm}{\parbox{4.6cm}{
    \epsfig{file=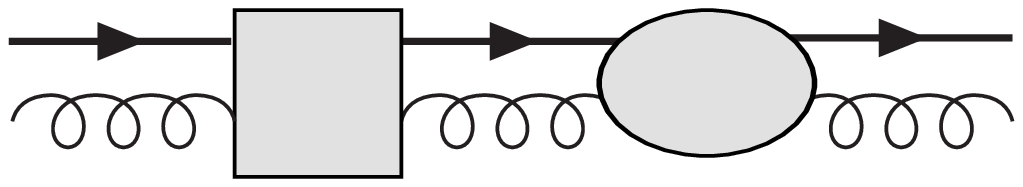,width=4.6cm}}}{}^{q_3}_{g_3} \, .
\end{equation}
In particular this implies that traces are taken by joining left and
right legs,
\begin{equation}
\label{eq:trace}
  \tr \left( 
      \parbox{3cm}{ \epsfig{file=Figures/square.eps,width=3cm}} \right) 
  = \raisebox{.5mm}{\parbox{1.7cm}{ 
    \epsfig{file=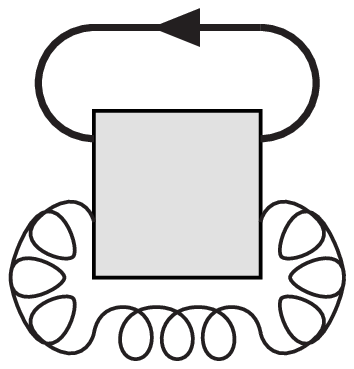,width=1.7cm}}} \, .
\end{equation}
The complex conjugate of a diagram is obtained by reversing all arrows
\begin{equation}
\label{eq:complex_conjugate}
  \parbox{4.6cm}{\epsfig{file=Figures/square_blob.eps,width=4.6cm}}^*
  = \parbox{4.6cm}{\epsfig{file=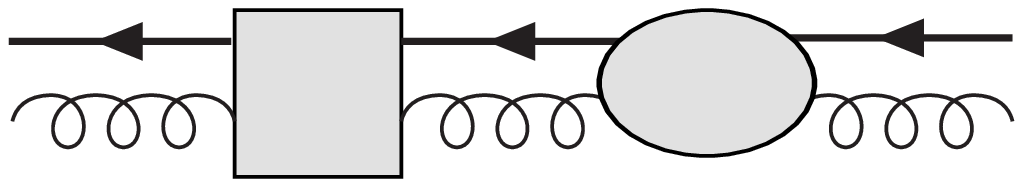,width=4.6cm}} \, , 
\end{equation}
and hermitian conjugation amounts to mirror the diagram across a
vertical line {\it and} to reverse all arrows, 
\begin{equation}
\label{eq:Hermitian_conjugate}
  \parbox{4.6cm}{\epsfig{file=Figures/square_blob.eps,width=4.6cm}}^\dag
  = \reflectbox{\parbox{4.6cm}{
    \epsfig{file=Figures/square_blob_star.eps,width=4.6cm}}}\, . 
\end{equation}
The latter is to be compared to 
\begin{equation}
  (\Proj^\dag)_{a_1\,a_2\,\hdots\,a_n\,b_1\,b_2\,\hdots\,b_n} 
  = (\Proj_{b_1\,b_2\,\hdots\,b_n\,a_1\,a_2\,\hdots\,a_n})^*
\end{equation}
in index notation.

We frequently have to symmetrize or anti-symmetrize over a set of
indices which in birdtrack notation is indicated by a white or a
black bar, respectively,
\begin{equation}
\label{eq:(anti-)symmetrizer}
  \parbox{3cm}{\epsfig{file=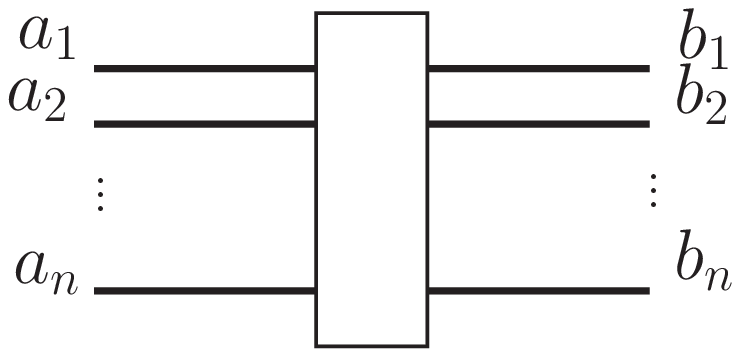,width=3cm}} 
  = \mathbf{S}_{a_1\,a_2\,\hdots\,a_n\,b_1\,b_2\,\hdots\,b_n} \, , \qquad 
    \parbox{3cm}{\epsfig{file=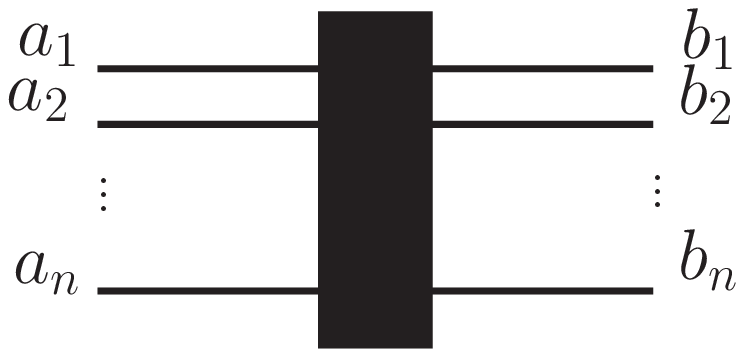,width=3cm}} 
  = \mathbf{A}_{a_1\,a_2\,\hdots\,a_n\,b_1\,b_2\,\hdots\,b_n} \, .
\end{equation}
Written out as sums over permutations these read
\begin{equation}
\begin{split}
  \mathbf{S}_{a_1\,a_2\,\hdots\,a_n\,b_1\,b_2\,\hdots\,b_n}
  &= \frac{1}{n!} \sum_{\sigma\in S_n} \delta_{a_1 b_{\sigma(1)}} \,
     \delta_{a_2 b_{\sigma(2)}} \cdots \, \delta_{a_n b_{\sigma(n)}} \, , 
  \\
  \mathbf{A}_{a_1\,a_2\,\hdots\,a_n\,b_1\,b_2\,\hdots\,b_n}
  &= \frac{1}{n!} \sum_{\sigma\in S_n} \mathrm{sign}(\sigma) \, 
     \delta_{a_1 b_{\sigma(1)}} \,
     \delta_{a_2 b_{\sigma(2)}} \cdots \, \delta_{a_n b_{\sigma(n)}} \, .
\end{split} 
\end{equation}
 
\section{First occurrence in Young multiplication}
\label{sec:first_occurrence}

In this appendix we analyze the first occurrence of multiplets in terms of Young
diagrams. In particular we show that there is at most one new
multiplet, having $\Nf=n$, within $\overline{M_{\qbar}} \otimes M_q$ with
$ M_{\qbar},M_q \subseteq V^{\otimes n}$.
We also derive a rule for determining
the first occurrence $\Nf(M)$ of a given multiplet $M$, and show that
$\Nf(M')-\Nf(M) \in \{ -1,0,1\}$ for $M' \subseteq M \otimes A$.  We
assume that the reader is familiar with the labeling of irreducible
representation of $\SU(\Nc)$ by Young diagrams as well as the rules
for conjugating and multiplying Young diagrams, see
e.g. \cite[secs. 7.12 \& 10]{Ham62} and \cite[sec. 9.8]{Cvi08}.

The first occurrence $\Nf(M)$ of a multiplet $M$, which is contained
in $A^{\otimes n}$ for some $n$, was defined in
\secref{sec:Gluons}. We now define the first occurrence $\NfV(M)$
within the sequence $(V \otimes \overline{V})^{\otimes n}$ in the same
way, as the smallest $n \geq 0$ for which $M \subseteq (V \otimes
\overline{V})^{\otimes n}$. Let us show that 
\begin{equation} 
\label{eq:Nf-equivalence}
  \Nf(M) = \NfV(M) 
\end{equation}
for any $M$ which appears in either one of the
sequences. On the one hand, note that due to $A^{\otimes n} \subseteq (V \otimes
  \overline{V})^{\otimes n}$ any $M \subseteq A^{\otimes n}$ is also
  contained in $(V \otimes \overline{V})^{\otimes n}$, and thus
  $\NfV(M) \leq \Nf(M)$. On the other hand, due to
  \eqref{eq:VVbar_binomi} any $M \subseteq (V \otimes
  \overline{V})^{\otimes n}$ also has to be contained in $A^{\otimes
    \nu}$ for some $\nu \leq n$, and therefore $\NfV(M) \geq
  \Nf(M)$. Together these two inequalities establish
  \eqref{eq:Nf-equivalence} and, consequently, we no longer
  distinguish between $\Nf$ and $\NfV$.

In order to analyze first occurrences we exploit that $(V \otimes
\overline{V})^{\otimes n}$ is isomorphic to $\overline{V}^{\otimes n}
\otimes V^{\otimes n}$, i.e.\ we have
\begin{equation}
  A^{\otimes n} 
  \subseteq (V \otimes \overline{V})^{\otimes n}
  \cong \overline{V}^{\otimes n} \otimes V^{\otimes n} \, .
\end{equation}
In terms of Young diagrams this means that for every instance of a
multiplet $M
\subseteq A^{\otimes n}$ there exist $n$-box Young diagrams
corresponding to multiplets $M_q$ and $M_{\qbar}$ such
that\footnote{By ``$n$-box diagram'' we only mean that $M_q, M_{\qbar}
  \subseteq V^{\otimes n}$, i.e.\ if the diagrams have one or more
  columns with $\Nc$ boxes it does not matter for our discussion
  whether we keep these columns or omit them -- as one is allowed to for
  $\SU(\Nc)$ Young diagrams.}
\begin{equation}
  M \subseteq \overline{M_{\qbar}} \otimes M_q \, .
\end{equation}
In the following we refer the Young diagrams for $M_{q}$ and
$\overline{M_{\qbar}}$ as quark diagram and anti-quark diagram,
respectively, or $q$- and $\qbar$-diagram for short. Recall that the
$\qbar$-diagram is obtained from the diagram for $M_{\qbar}$ by first
supplementing the latter with additional boxes at the bottom until all
columns have length $\Nc$, then rotating the resulting rectangular
diagram by $180^\circ$, and finally removing the original boxes. For
instance, for $\Nc=5$ we have
\begin{eqnarray}
  \overline{\tyng{4,2}}=
  \tyoung{\;\;\;\;,\;\;\;\;,\;\;\;\;,\;\;\bullet\bullet,\bullet\bullet\bullet\bullet} 
  \label{eq:QbarDiagram}
\end{eqnarray}
where $\tyoung{\bullet}$ denotes the {\it absence} of a box,
i.e.\ these are the boxes which were removed in the last step. We may
thus view the $\qbar$-diagram as a rectangular diagram with $\Nc$ rows
with $n$ boxes cut out.

Now consider $\overline{M_{\qbar}} \otimes M_q$ with $\Nc$ such that
the rightmost column of the $\qbar$-diagram has as many boxes as the
leftmost column of the $\q$-diagram. In this case we obtain
\begin{eqnarray}
  \tyoung{\;\;\;\;,\;\;\;\;,\;\;\;\;,\;\;\bullet\bullet,\bullet\bullet\bullet\bullet}
  \otimes 
  \tyoung{qqq,qq,q}
  =  \tyoung{\;\;\;\;qqq,\;\;\;\;qq,\;\;\;\;q,\;\;\bullet\bullet,\bullet\bullet\bullet\bullet}
&\oplus& 
\begin{array}{c}
\mbox{diagrams in which at least one quark box} \\
\mbox{occupies a cut out space}
\end{array} \, .
\label{eq:QQbarMultiply}
\end{eqnarray}
However, all diagrams in which a quark box occupies a cut out space, such as 
\begin{eqnarray}
  \tyoung{\;\;\;\;qq,\;\;\;\;qq,\;\;\;\;q,\;\;q\bullet,\bullet\bullet\bullet\bullet} \; ,
\end{eqnarray}
are already contained in $\overline{V}^{\otimes(n-1)} \otimes
V^{\otimes(n-1)}$, in the above case within
\begin{eqnarray}
  \tyoung{\;\;\;\;,\;\;\;\;,\;\;\;\;,\;\;\;\bullet,\bullet\bullet\bullet\bullet}
  \otimes
  \tyoung{qq,qq,q} \; .
\end{eqnarray}
Thus, the first occurrence of these diagrams is less than $n$, here $n-1$. 
The first diagram on the r.h.s.\ of \eqref{eq:QQbarMultiply}, however,
cannot be obtained for smaller $n$, as there would not be enough boxes
available. 
Hence, its first occurrence is $\Nf=n$, i.e.\ it
corresponds to a new multiplet in the terminology of
\secref{sec:Gluons}.
We also note that all Young diagrams of this kind, i.e. with $n$ cut
out spaces $\tyoung{\bullet}$ from the $\qbar$-diagram and $n$ boxes
$\tyoung{q}$ from the $q$-diagram, appear within $(V \otimes
\overline{V})^{\otimes n}$, and thus also within $A^{\otimes n}$.
Therefore, we can use the shapes of the quark and anti-quark diagrams
for determining the first occurrence $n_f$.

For larger $\Nc$, such that the rightmost column of the
$\qbar$-diagram is longer than the leftmost column of the
$\q$-diagram, the structure of \eqref{eq:QQbarMultiply} remains the same, e.g.,
for $\Nc=6$,
\begin{eqnarray}
  \tyoung{\;\;\;\;,\;\;\;\;,\;\;\;\;,\;\;\;\;,\;\;\bullet\bullet,\bullet\bullet\bullet\bullet}
  \otimes 
  \tyoung{qqq,qq,q}
  =  \tyoung{\;\;\;\;qqq,\;\;\;\;qq,\;\;\;\;q,\;\;\;\;,\;\;\bullet\bullet,\bullet\bullet\bullet\bullet}
&\oplus& 
\begin{array}{c}
\mbox{diagrams in which at least one quark box} \\
\mbox{occupies a cut out space}
\end{array} \, ,
\end{eqnarray}
as all other ways of appending all quark boxes to the $\qbar$-diagram
without occupying a cut out space are forbidden by the rules for Young
multiplication. Therefore, also in this case we obtain exactly one
multiplet with first occurrence $\Nf=n$. On the other hand, for
smaller $\Nc$, such that the rightmost column of the $\qbar$-diagram
is shorter than the leftmost column of the $\q$-diagram, it is
impossible to append all quark boxes to the $\qbar$-diagram without
occupying a cut out space. In this case we thus obtain only diagrams
with $\Nf < n$, i.e.\ diagrams corresponding to old multiplets.
We conclude that {\it within 
$\overline{M_{\qbar}} \otimes M_q$, with
$M_{\qbar}, M_q \subseteq V^{\otimes n}$, 
there is always at most
 one new multiplet}, and for sufficiently large $\Nc$ there is
exactly one new multiplet. Our construction of projectors onto new
multiplets builds on this. Another consequence is that new
multiplets can be labeled uniquely by the $q$- and
$\qbar$-diagrams. We introduce a corresponding notation in 
\secref{sec:general_remarks}.

The above discussion also provides us with a graphical rule for
determining the first occurrence $\Nf(M)$ of a given multiplet
$M$. First, draw the Young diagram corresponding to $M$ and 
mark empty spaces at the bottom of it by $\tyoung{\bullet}$ until
all columns have length $\Nc$. Then draw a vertical line through the
diagram such that the number of cut out spaces $\tyoung{\bullet}$ to
the left of the line is the same as the number of boxes $\tyoung{\;}$
to the right of it. This number is the first occurrence of $M$. As
examples we consider the octet, the decuplet and the $35$-plet for
$\SU(3)$,
\begin{align}
\nonumber \\ \label{eq:first_occurrence_examples}
\mbox{
  \begin{picture}(0,0)
  \put(11,-10) {\line(0,1) {40}}
  \end{picture}
  \tyoung{\;\;,\;\bullet,\bullet\bullet}}
  \; , && 
\mbox{
  \begin{picture}(0,0)
  \put(10.5,-10) {\line(0,1) {40}}
  \end{picture}
  \tyoung{\;\;\;,\bullet\bullet\bullet,\bullet\bullet\bullet}}
  \; , &&  
\mbox{
  \begin{picture}(0,0)
  \put(17.5,-10) {\line(0,1) {40}}
  \end{picture}
  \tyoung{\;\;\;\;\;,\;\bullet\bullet\bullet\bullet,\bullet\bullet\bullet\bullet\bullet}} \; , 
\\ \nonumber 
\end{align}
and obtain their first occurrences $1,2$ and $3$, respectively.

We now turn to the question of how the first occurrence changes under
multiplication with an additional gluon. More precisely, for $M'
\subseteq M \otimes A$ we want to know how $\Nf(M)$ and $\Nf(M')$ are
related. By definition we can obtain $\Nf(M')=\Nf(M)+1$ but not
higher, we will see that this also is in agreement with 
the first occurrence counting above. 
 Concerning the lowest possible value of $\Nf(M')$ one may
heuristically argue as follows. The additional gluon can form a
singlet with one of the other gluons, leaving the remaining gluons in
a multiplet with first occurrence $\Nf(M)-1$. (If they were in a state
with lower first occurrence, they could not have built up a state with
first occurrence $\Nf(M)$ before the multiplication with the
additional gluon.) Overall we thus expect
\begin{equation}
  \Nf(M)-1 \leq \Nf(M') \leq \Nf(M)+1 
  \quad \forall\ M' \subseteq M \otimes A \, .
\end {equation}
In order to prove this we carry out the Young multiplication with the
adjoint representation and use the rule for determining the first
occurrence, which we have derived above.

Recall that the Young diagram for the adjoint representation has two
columns, the first one with $\Nc-1$ boxes and the second column with
one box. For the following discussion it is convenient to define $k =
\Nc-1$. In order to calculate $M \otimes A$ we have to append the
labeled boxes of 
\begin{equation}
\label{eq:adjoint_rep_Young_diagram}
  \tyoung{11,2,\cdot,\cdot,k}
\end{equation}
in all allowed ways to the Young diagram for $M$. 
We label the way in which the boxes 
are appended 
by a pattern like
\begin{equation}
\label{eq:pattern_example}
  \tyoung{11,\bullet,2,\cdot,\cdot,k} \, , 
\end{equation}
which stands for appending the boxes $\tyoung{11}$ to row 1, nothing
to row 2, box $\tyoung{2}$ to row 3, etc., and box $\tyoung{k}$ to row
$\Nc$. For instance, if the initial multiplet was 
\begin{equation}
\label{eq:intial_example}
  \tyoung{\,\,\,\,\,\,,\,\,\,\,\,,\,\,\,,\,\,,\,\,} 
\end{equation}
then the pattern (\ref{eq:pattern_example}) is a shorthand notation for
the resulting Young diagram
\begin{equation}
\label{eq:final_example}
  \tyoung{\,\,\,\,\,\,11,\,\,\,\,\,\bullet,\,\,\,2,\,\,\cdot,\,\,\cdot,k} \, .
\end{equation}
In this way we can enumerate all multiplets which could possibly arise
after Young multiplication with the adjoint representation by
\begin{equation}
  \tyoung{11,\bullet,2,\cdot,\cdot,k},\;
  \tyoung{11,2,\bullet,\cdot,\cdot,k},\;...\;
  \tyoung{11,2,\cdot,\cdot,\bullet,k},\;
  \tyoung{11,2,\cdot,\cdot,k,\bullet},\;
  \;\;\;\; 
  \tyoung{\bullet,11,2,\cdot,\cdot,k},\;
  \tyoung{1,12,\bullet,\cdot,\cdot,k},\;...\;
  \tyoung{1,12,\cdot,\cdot,\bullet,k},\;
  \tyoung{1,12,\cdot,\cdot,k,\bullet},\;
  \;\;...\;\; 
  \tyoung{\bullet,1,2,\cdot,\cdot,1k},\;
  \tyoung{1,\bullet,2,\cdot,\cdot,1k},\;...\;
  \tyoung{1,2,\bullet,\cdot,\cdot,1k},\;
  \tyoung{1,2,\cdot,\cdot,\bullet,1k},\;
  \;\;\;\;
  \tyoung{1,1,2,\cdot,\cdot,k},\;
  \tyoung{1,2,1,\cdot,\cdot,k},\;...\;
  \tyoung{1,2,\cdot,\cdot,1,k},\;
  \tyoung{1,2,\cdot,\cdot,k,1}.
\label{ResultingMultiplets}
\end{equation}
For a given initial multiplet many of these are typically forbidden, but the
list is always exhaustive.

We first consider the last patterns, all of the same shape, with $\Nc$
unit length rows. They give rise to different instances of the initial
multiplet. There are thus up to $\Nc-1$ instances of the multiplet $M$
within $M\otimes A$, coming from the different placements of the
second $\tyoung{1}$. In these cases the first occurrence of the
resulting multiplet $M'$ is, trivially, equal to the first occurrence
of the initial multiplet $M$.

All other patterns in (\ref{ResultingMultiplets}) have different
shapes, and, accordingly, the corresponding resulting Young diagrams
(if they are allowed) also have different shapes. Thus, all of these
patterns correspond to unique multiplets in $M \otimes A$, which carry
non-equivalent irreducible representations of $\SU(\Nc)$. It remains
to determine their first occurrence.
 
Assume that we have calculated the first occurrence of the initial
multiplet $M$ according to the rules, described before the examples
(\ref{eq:first_occurrence_examples}). In particular we have
  placed the vertical line such that the number of boxes $\tyoung{\;}$
  to the right, and the number of cut out spaces $\tyoung{\bullet}$ to
  the left of it are the same. Keep this line in place while appending
  the boxes according to the pattern. Now each box which we append to
the Young diagram for $M$ either increases the number of boxes
$\tyoung{\;}$ to the right of the vertical line by one or decreases
the number of cut out spaces $\tyoung{\bullet}$ to the left of it by
one. Thus, the difference between the former and the latter number
always increases by $\Nc$.  In order to determine the first occurrence
of the resulting multiplet $M'$ we thus have to compensate this 
increase by moving the vertical line one column to the right, as by 
this, in each
row we either decrease the number of boxes $\tyoung{\;}$ to the right
of it by one or increase the number of cut out spaces
$\tyoung{\bullet}$ to the left of it by one.

Consider now again the patterns in (\ref{ResultingMultiplets})
which append two boxes to one row and nothing to one of the other
rows. Think of applying these according to the following two-step
procedure. First append one box to each row and move the vertical
line one box to the right as justified in the preceding
paragraph. Then move one box from the row in which the pattern has
the $\tyoung{\bullet}$ to the row in which the pattern has two
boxes.  We illustrate this for the example of
eqs.~(\ref{eq:pattern_example})--(\ref{eq:final_example}),
\begin{align}
\nonumber \\ \label{eq:moving_box_example}
\mbox{
  \begin{picture}(0,0)
  \put(24,-10) {\line(0,1) {60}}
  \end{picture}
  \tyoung{\,\,\,\,\,\,,\,\,\,\,\,\bullet,\,\,\,\bullet\bullet\bullet,\,\,\bullet\bullet\bullet\bullet,\,\,\bullet\bullet\bullet\bullet,\bullet\bullet\bullet\bullet\bullet\bullet}}
  \qquad \raisebox{6mm}{$\mapsto$} \quad 
\mbox{
  \begin{picture}(0,0)
  \put(31,-10) {\line(0,1) {60}}
  \end{picture}  
  \tyoung{\,\,\,\,\,\,\,,\,\,\,\,\,\,\bullet,\,\,\,\,\bullet\bullet\bullet,\,\,\,\bullet\bullet\bullet\bullet,\,\,\,\bullet\bullet\bullet\bullet,\,\bullet\bullet\bullet\bullet\bullet\bullet}} 
  \qquad \raisebox{6mm}{$\mapsto$} \quad 
\mbox{
  \begin{picture}(0,0)
  \put(31,-10) {\line(0,1) {60}}
  \end{picture}  
  \tyoung{\,\,\,\,\,\,\,\,,\,\,\,\,\,\bullet\bullet\bullet,\,\,\,\,\bullet\bullet\bullet\bullet,\,\,\,\bullet\bullet\bullet\bullet\bullet,\,\,\,\bullet\bullet\bullet\bullet\bullet,\,\bullet\bullet\bullet\bullet\bullet\bullet\bullet}} \; ,
\\ \nonumber
\end{align}
where in the second step the box is moved from row 2 to row 1. Note
once more that it does not matter whether we keep or omit the first
column of the resulting Young diagram.

The first occurrence of the resulting diagram now depends only on
whether the box, which is moved in the last step, crosses the vertical
line, and, if it does, in which direction. If it does not cross the
line, then the first occurrence does not change (but the shape of the
diagram does). This is the case in the example
(\ref{eq:moving_box_example}), and consequently the first occurrence
remains $5$. However, the first occurrence changes whenever the
moving box crosses the vertical line. If the box crosses the line from
left to right then $\Nf(M')=\Nf(M)+1$, e.g.\ if it is moved from row
3, 4, 5 or 6 to row 1 or 2 in the example. If the box crosses
the line from right to left, then $\Nf(M')=\Nf(M)-1$, e.g.\ if it is
moved from row 1 or 2 to row 4, 5 or 6 in the example. As only
one box moves, the first occurrence can change by at most 1.
From this reasoning we also see immediately from the Young diagram shapes
that there cannot be any ``$27$''-plet within ``$0$''$\otimes$``$8$''$\subset A^{\otimes3}$.

We summarize the results of what can happen under multiplication with
one additional gluon, i.e.\ for $M' \subseteq M \otimes A$:
\begin{enumerate}
\item[(i)] If $M'=M$ then it can appear up to $\Nc-1$ times in $M
  \otimes A$.
\item[(ii)] All other $M' \subseteq M \otimes A$ appear only once. In
  particular all new multiplets within $M \otimes A$ appear with
  multiplicity one.
\item[(iii)] The first occurrence of $M'$ differs from that of $M$ by
  at most 1,\\ i.e.\ $\Nf(M') \in \{ \Nf(M)-1,\Nf(M),\Nf(M)+1 \}$.
\end{enumerate}
The fact that $\Nf$ can change by at most one unit has a well-known
manifestation for $\SU(2)$ (spin), namely that when multiplying a spin-$j$
state with a spin-1 state then the resulting state's total spin is
either $j-1$, $j$ or $j+1$.
 
\section{Properties of the Clebsch-Gordan matrices $\CG^M$}
\label{sec:ClebschGordans}

In this appendix we prove some properties of the Clebsch-Gordan matrices 
$\CG^M :
A^{\otimes\Nf} \to A^{\otimes\ng}$
and projection operators introduced in \secref{sec:Gluons}.

We first consider \eqref{eq:CdagC} which states the following.  If
$\Proj^{M} = \CG^M \CG^{M\dag}$ projects onto a multiplet $M \subseteq
A^{\otimes\ng}$ then $\Proj^{M_f} = \CG^{M\dag} \CG^M$ is also
projector, projecting onto $M_f \subseteq A^{\otimes\Nf}$, a first
occurrence of the multiplet $M$. Notice first that by cyclic
permutation under the trace
\begin{equation}
  d:= \dim M = \tr \Proj^M = \tr \Proj^{M_f} = \dim M_f \, , 
\end{equation}
i.e.\ if $\Proj^{M_f}$ is a projector, then it projects onto a subspace
with the same dimension as $M$. It remains to show that (a) $\Proj^{M_f}$
actually is a projector and that (b) $M_f$ carries an irreducible
representation of $\SU(\Nc)$ equivalent to that carried by $M$.

(a) Recall that both $\CG^M\CG^{M\dag}$ and
  $\CG^{M\dag}\CG^M$ are diagonalizable since they are manifestly
  hermitian. All their non-zero eigenvalues are identical,
  including their multiplicities, for the following reason: If
$\lambda \neq 0$ is an eigenvalue of $\CG^M\CG^{M\dag}$ then there
exists a non-zero $v \in A^{\otimes\ng}$ such that $\CG^M\CG^{M\dag} v
= \lambda v$. Defining $u \in A^{\otimes\Nf}$ by $u=\CG^{M\dag} v$ the
eigenvalue equation can be rewritten as $\CG^M u = \lambda v$ and by
multiplication with $\CG^{M\dag}$ on the left we obtain
\begin{equation} 
  \CG^{M\dag} \CG^M u = \CG^{M\dag} \lambda v = \lambda u \, , 
\end{equation}
i.e.\ $\lambda$ is also an eigenvalue of $\CG^{M\dag}\CG^M$ with
eigenvector $u$. As $\Proj^M=\CG^M\CG^{M\dag}$ is a projector its only
non-zero eigenvalue is $1$ and by the above argument this is also true
for $\Proj^{M_f}=\CG^{M\dag}\CG^M$, i.e.\ $\Proj^{M_f}$ is also a
projector.

(b) In order to analyze whether $M_f$ is invariant under $\SU(\Nc)$ it
is necessary to distinguish carefully between a multiplet $M$ and the
irreducible representation which it carries. We denote by $\Ad$ the
adjoint representation of $\SU(\Nc)$, in particular, for every group
element $\gre\in\SU(\Nc)$, $\Ad(\gre)$ is a linear operator $A \to A$. The
space $A^{\otimes\ng}$ carries $\Ad^{\otimes\ng}$, the $\ng$-fold
tensor product of the adjoint representation. By $\Gamma_M$ we denote
the irreducible representation carried by $M \subseteq
A^{\otimes\ng}$.

Now we choose bases $\{v_j\}$ for $A^{\otimes\ng}$ and $\{u_j\}$ for
$A^{\otimes\Nf}$ such that $v_1,\hdots,v_{d}$ span $M$ and
$u_1,\hdots,u_{d}$ span $M_f$. In this basis $\Proj^M$ reads
\begin{equation}
\label{eq:PM_as_matrix}
  \Proj^M = \begin{pmatrix} \eins_d && 0 \\ 0 && 0 \end{pmatrix}
\end{equation}
where $\eins_d$ denotes the $d \times d$ unit
matrix. Similarly, $\Proj^{M_f} = \left( \begin{smallmatrix} \eins_d & 0
  \\ 0 & 0 \end{smallmatrix}\right)$. Recall, however, that in general
the zero blocks have different sizes for $\Proj^M$ and $\Proj^{M_f}$. 

Next we choose a vector $v \in M$ and define $u:=\CG^{M\dag} v$. As a
consequence of (a) we have $u \in M_f$ and $v=\CG^M u$. We know how $v$
transforms under $\SU(\Nc)$,
\begin{equation}
  \Ad^{\otimes\ng}(\gre) \, v 
  = \begin{pmatrix} \Gamma_M(\gre) && 0 \\ 0 && 0 \end{pmatrix} v \ 
  =: \tilde{v} \in \ M
  \qquad \forall\ \gre \in \SU(\Nc) \, .
\end{equation}
Again we have $\tilde{u} := \CG^{M\dag} \, \tilde{v} \in M_f$,
i.e.\ the transformation $\Ad^{\otimes\ng}(\gre)$ leads to a
transformation $\Gamma_{M_f}(\gre): M_f \to M_f$ defined by
\begin{equation}
\label{eq:GMf_def}
  \tilde{u} 
  =: \begin{pmatrix} \Gamma_{M_f}(\gre) && 0 \\ 0 && 0 \end{pmatrix} u
  = \CG^{M\dag} \begin{pmatrix} \Gamma_M(\gre) && 0 \\ 0 && 0 \end{pmatrix} 
    \CG^M \, u.
\end{equation}
It remains to show that $\Gamma_{M_f}$ is a representation of
$\SU(\Nc)$ and that it is equivalent to $\Gamma_M$. Denoting the upper
left $d \times d$ block of $\CG^M$ by $U$, i.e. 
\begin{equation}
  \CG^M = \begin{pmatrix} U && \cdots \\ \vdots && \ddots \end{pmatrix} \, , 
\end{equation}
we observe that
\begin{equation}
\label{eq:GMf=UdagGMU}
  \Gamma_{M_f}(\gre) = U^\dag \, \Gamma_M(\gre) \, U \, .
\end{equation}
Finally, we show that $U$ is unitary which concludes the proof. To
this end notice that $\Gamma_M$ maps the identity $\eins_{\Nc} \in
\SU(\Nc)$ to the $d\times d$ unit matrix, i.e.\ $\Gamma_M(\eins_{\Nc})
= \eins_d$, since $\Gamma_M$ is a representation. From
\eqref{eq:GMf_def} we infer
\begin{equation}
\label{eq:GMf(1)}
\begin{split}
  \begin{pmatrix} \Gamma_{M_f}(\eins_{\Nc}) && 0 \\ 0 && 0 \end{pmatrix} u
  &= \CG^{M\dag} \begin{pmatrix} \eins_d && 0 \\ 0 && 0 \end{pmatrix} 
    \CG^M \, u
  \underset{\mathrm{(\ref{eq:PM_as_matrix})}}{=} \CG^{M\dag} \Proj^M \CG^M \, u
  = \CG^{M\dag} \CG^M \CG^{M\dag} \CG^M \, u\\
  &= \big(\Proj^{M_f}\big)^2 \, u
  = \begin{pmatrix} \eins_d && 0 \\ 0 && 0 \end{pmatrix} u \, , 
\end{split}
\end{equation}
i.e.\ 
$\eins_d 
\underset{\mathrm{(\ref{eq:GMf(1)})}}{=} \Gamma_{M_f}(\eins_{\Nc}) 
\underset{\mathrm{(\ref{eq:GMf=UdagGMU})}}{=} U^\dag \, \Gamma_M(\eins_{\Nc}) \, U 
= U^\dag U$.

Having shown \eqref{eq:CdagC} we turn to \eqref{eq:CMdagCM'=0} which
is proved easily using $\CG^{M\dag} \CG^{M'} = 0 \ \Leftrightarrow \ \|
\CG^{M\dag} \CG^{M'} \| = 0$ and the norm (\ref{eq:sp_trace}):
\begin{equation}
  \| \CG^{M\dag} \CG^{M'} \|^2 
  = \tr( \CG^{M'\dag} \CG^M \CG^{M\dag} \CG^{M'}) 
  = \tr(\Proj^{M} \Proj^{M'}) 
  = 0 
  \quad \text{for } M \neq M' \, .
\end{equation}  

When constructing basis vectors from projectors in \appref{sec:6gBasis}
we are also interested in the norm of $\Vec :=
\CG^M \CG^{M'\dag}$, 
\begin{equation}
  \| \Vec \|^2
  = \tr( \CG^{M'} \CG^{M\dag} \CG^M \CG^{M'\dag}) 
  = \tr(\Proj^{M_f} \Proj^{M'_f}) 
  = \delta_{M_f M'_f} \tr(\Proj^{M_f}) 
  = \delta_{M_f M'_f} \dim M_f, 
  \label{eq:VecNorm}
\end{equation}
where $\delta_{M_f M'_f} $ indicates that the instances of the first
occurrence multiplets must be the same.

\section{Projectors for $ggg \to ggg$ in $\SU(\Nc)$}
\label{sec:3gProjectors}

For large enough $\Nc$ ($\Nc \geq 6$) there are 51 projection operators 
projecting onto irreducible subspaces. They are stated in 
\eqref{eq:3gProjectors}. Note, however, that for $\Nc=3$, this number
is reduced to 29, as many projectors correspond to multiplets whose Young 
diagrams are not admissible (as, for example, column 2 is longer than column 1),
or were constructed using a non-admissible intermediate state 
(as is the case for $\Proj^{0,8}$).

In addition, as the gluon transforms under a real representation, 
for processes with only gluons, the projection operators can only occur 
in real (i.e. symmetric) combinations, 
further reducing the number of physically independent projection
operators to 21.
The full set of 51 projection operators reads
\begin{eqnarray}
\label{eq:3gProjectors}
  \Proj^{8s,1}_{g_1\,g_2\,g_3\,g_4\,g_5\,g_6}
  &=&\frac{1}{\TR}
  \frac{\Nc}{2 \left(\Nc^4-5 \Nc^2+4\right)} d_ {g_1\,g_2\,g_3} d_ {g_4\,g_5\,g_6}\nonumber 
  \\ \nonumber
  \Proj^{8a,1}_{g_1\,g_2\,g_3\,g_4\,g_5\,g_6} 
  &=&\frac{1}{\TR}
  \frac{-1}{2 \Nc(\Nc^2- 1)} \ui f_ {g_1\,g_2\,g_3} \ui f_ {g_4\,g_5\,g_6}
  \\ \nonumber
  \Proj^{1,8}_{g_1\,g_2\,g_3\,g_4\,g_5\,g_6}
  &=&
  \frac{1}{\Nc^2-1} \delta _ {g_1\,g_2} \delta_{g_4\,g_5}  \delta _ {g_3\,g_6}
  \\ \nonumber
  \Proj^{8s,8s}_{g_1\,g_2\,g_3\,g_4\,g_5\,g_6}
  &=&\frac{1}{\TR^2}
  \frac{\Nc^2}{4 \left(\Nc^2-4\right)^2} d_ {g_1\,g_ 2\,i_ 1} d_{i_1\,g_3\,i_2} d_ {g_ 4\,g_ 5\,i_ 3} d_ {i_ 3\,g_ 6\,i_ 2}
   \\ \nonumber
  \Proj^{8s,8a}_{g_1\,g_2\,g_3\,g_4\,g_5\,g_6}
  &=&\frac{1}{\TR^2}
  \frac{-1}{4 (\Nc^2-4)} d_ {g_ 1\,g_ 2\,i_ 1} \ui f_ {i_1\,g_ 3\,i_ 2}  d_ {g_4\,g_5\,i_ 3} \ui f_ {i_ 3\,g_6\,i_ 2}
  \\ \nonumber
  \Proj^{8a,8s}_{g_1\,g_2\,g_3\,g_4\,g_5\,g_6}
  &=&\frac{1}{\TR^2}
  \frac{-1}{4 (\Nc^2-4)} \ui f_ {g_ 1\,g_ 2\,i_ 1} d_{i_1\,g_3 \,i_ 2}  i  f_ {g_4\,g_5\,i_3} d_ {i_ 3\,g_ 6\,i_2}
  \\ \nonumber
  \Proj^{8a,8a}_{g_1\,g_2\,g_3\,g_4\,g_5\,g_6}
  &=&\frac{1}{\TR^2}
  \frac{1}{4 \Nc^2} \ui f_{g_ 1\,g_ 2\,i_ 1} \ui f_ {i_1\,g_ 3\,i_ 2} \ui f_{g_4\,g_5\,i_3} \ui f_{i_3\,g_ 6\,i_ 2}
  \\ \nonumber
  \Proj^{10,8}_{g_1\,g_2\,g_3\,g_4\,g_5\,g_6}
  &=&
  \frac{4}{\Nc^2-4} \Proj^{10}_{g_ 1\,g_ 2\,i_1\,g_ 3} \Proj^{10}_{i_1\,g_ 6\,g_ 4\,g_ 5}
  \\ \nonumber
  \Proj^{\overline{10},8}_{g_1\,g_2\,g_3\,g_4\,g_5\,g_6}
  &=&
  \frac{4}{\Nc^2-4} \Proj^{\overline{10}}_ {g_1\,g_2\,i_1\,g_3} 
  \Proj^{\overline{10}}_ {i_1\,g_ 6\,g_ 4\,g_ 5}
  \\ \nonumber
  \Proj^{27,8}_{g_1\,g_2\,g_3\,g_4\,g_5\,g_6}
  &=&
 \frac{4 (\Nc+1)}{\Nc^2 (\Nc+3)} \Proj^{27}_ {g_ 1\,g_ 2\,i_1\,g_3} \Proj^{27}_ {i_1\,g_ 6\,g_ 4\,g_ 5}
  \\ \nonumber
  \Proj^{0,8}_{g_1\,g_2\,g_3\,g_4\,g_5\,g_6}
  &=& 
 \frac{4 (\Nc-1)}{(\Nc-3) \Nc^2} \Proj^{0}_ {g_1\,g_ 2\,i_1\,g_ 3} \Proj^{0}_ {i_1\,g_ 6\,g_ 4\,g_ 5}
  \\ \nonumber
  \Proj^{8s,10}_{g_1\,g_2\,g_3\,g_4\,g_5\,g_6}
  &=&\frac{1}{\TR}
\frac{\Nc}{2 \left(\Nc^2-4\right)} d_ {g_ 1\,g_ 2\,i_ 1} \Proj^{10}_{i_ 1\,g_ 3\,i_ 2\,g_ 6}d_ {i_2\,g_ 4\,g_ 5} 
  \\ \nonumber
  \Proj^{8a,10}_{g_1\,g_2\,g_3\,g_4\,g_5\,g_6}
  &=&-\frac{1}{\TR}
  \frac{1}{2 \Nc} \ui f_{g_1\,g_2\,i_1}  \Proj^{10}_ {i_ 1\,g_ 3\,i_ 2\,g_ 6} \ui f_ {i_2\,g_ 4\,g_ 5}
  \\ \nonumber
  \Proj^{10,10f}_{g_1\,g_2\,g_3\,g_4\,g_5\,g_6}
  &=& 
  \CG^{10f}_{g_1\,g_2\,g_3;\,i_1\,i_2} \CG^{10f\,\dagger}_{i_1\,i_2;\,g_4\,g_5\,g_6}
  \\ \nonumber
 \Proj^{10,10fd}_{g_1\,g_2\,g_3\,g_4\,g_5\,g_6}
  &=&  
  \CG^{10fd}_{g_1\,g_2\,g_3;\,i_1\,i_2} \CG^{10fd\,\dagger}_{i_1\,i_2;\,g_4\,g_5\,g_6}
  \\ \nonumber
  \Proj^{27,10}_{g_1\,g_2\,g_3\,g_4\,g_5\,g_6}
  &=& 
  \frac{1}{\TR}\frac{2 (\Nc+2)}{\Nc (\Nc+3)} 
  \Proj^{27}_{g_ 1\,g_ 2\,i_ 1\,i_ 2} d_ {i_ 2\,g_ 3\,i_ 3} \Proj^{10}_ {i_1\,i_3\,i_4\,i_6}   d_ {i_6\,g_6\,i_5} \Proj^{27}_ {i_4\,i_5\,g_4\,g_5}
  \\ \nonumber
  \Proj^{0,10}_{g_1\,g_2\,g_3\,g_4\,g_5\,g_6}
  &=&
  \frac{1}{\TR} \frac{2 (\Nc-2)}{\Nc(\Nc-3)} 
  \Proj^{0}_ {g_ 1\,g_ 2\,i_ 1\,i_ 2} d_ {i_ 2\,g_ 3\,i_ 3} \Proj^{10}_ {i_ 1\,i_ 3\,i_ 4\,i_ 6} d_ {i_6\,g_ 6\,i_5} \Proj^{0}_ {i_ 4\,i_ 5\,g_ 4\,g_ 5}  
  \\ \nonumber
  \Proj^{8s,\overline{10}}_{g_1\,g_2\,g_3\,g_4\,g_5\,g_6}
  &=&\frac{1}{\TR}
\frac{\Nc}{2 \left(\Nc^2-4\right)} d_ {g_ 1\,g_ 2\,i_ 1} \Proj^{\overline{10}}_{i_ 1\,g_ 3\,i_ 2\,g_ 6}d_ {i_2\,g_ 4\,g_ 5} 
  \\ \nonumber
  \Proj^{8a,\overline{10}}_{g_1\,g_2\,g_3\,g_4\,g_5\,g_6}
  &=&-\frac{1}{\TR}
  \frac{1}{2 \Nc} \ui f_{g_1\,g_2\,i_1}  \Proj^{\overline{10}}_ {i_ 1\,g_ 3\,i_ 2\,g_ 6} \ui f_ {i_2\,g_ 4\,g_ 5}
  \\ \nonumber
  \Proj^{\overline{10},\overline{10}f}_{g_1\,g_2\,g_3\,g_4\,g_5\,g_6}
  &=& 
  \CG^{\overline{10}f}_{g_1\,g_2\,g_3;\,i_1\,i_2} \CG^{\overline{10}f\,\dagger}_{i_1\,i_2;\,g_4\,g_5\,g_6}
    \\ \nonumber
    \Proj^{\overline{10},\overline{10}fd}_{g_1\,g_2\,g_3\,g_4\,g_5\,g_6}
    &=&    
    \CG^{\overline{10}fd}_{g_1\,g_2\,g_3;\,i_1\,i_2} \CG^{\overline{10}fd\,\dagger}_{i_1\,i_2;\,g_4\,g_5\,g_6}   
    \\ \nonumber
    \Proj^{27,\overline{10}}_{g_1\,g_2\,g_3\,g_4\,g_5\,g_6}
  &=&\frac{1}{\TR}
\frac{2 (\Nc+2)}{\Nc (\Nc+3)} \Proj^{27}_{g_ 1\,g_ 2\,i_ 1\,i_ 2} d_ {i_ 2\,g_ 3\,i_ 3} \Proj^{\overline{10}}_ {i_1\,i_3\,i_4\,i_6}   d_ {i_6\,g_6\,i_5} \Proj^{27}_ {i_4\,i_5\,g_4\,g_5}
  \\ \nonumber
  \Proj^{0,\overline{10}}_{g_1\,g_2\,g_3\,g_4\,g_5\,g_6}
  &=&\frac{1}{\TR}
  \frac{2 (\Nc-2)}{\Nc(\Nc-3)} \Proj^{0}_ {g_ 1\,g_ 2\,i_ 1\,i_ 2} d_ {i_ 2\,g_ 3\,i_ 3} \Proj^{\overline{10}}_ {i_ 1\,i_ 3\,i_ 4\,i_ 6} d_ {i_6\,g_6\,i_5} \Proj^{0}_ {i_ 4\,i_ 5\,g_ 4\,g_ 5}  
  \\ \nonumber
  \Proj^{8s,27}_{g_1\,g_2\,g_3\,g_4\,g_5\,g_6}
  &=& \frac{1}{\TR}
  \frac{\Nc}{2(\Nc^2-4)}
  d_{g_1\,g_2\,i_1}\Proj^{27}_{i_1\,g_3\, i_2\,g_6 }d_{i_2\,g_4\,g_5}
  \\ \nonumber
  \Proj^{8a,27}_{g_1\,g_2\,g_3\,g_4\,g_5\,g_6} 
  &=& \frac{1}{\TR}
  \frac{-1}{2\Nc}
  \ui f_{g_1\,g_2\,i_1}\Proj^{27}_{i_1\,g_3\, i_2\,g_6 }\ui f_{i_2\,g_4\,g_5}\nonumber
  \\ \nonumber
  \Proj^{10,27}_{g_1\,g_2\,g_3\,g_4\,g_5\,g_6}
  &=& \frac{1}{\TR}
  \frac{2\Nc}{\Nc^2-\Nc-2}
  \Proj^{10}_{g_1\,g_2\, i_1\, i_2 } d_{i_2\,g_3\,i_3} \Proj^{27}_{i_1\,i_3\, i_4\, i_6 }
  d_{i_6\,g_6\,i_5} \Proj^{10}_{i_4\,i_5\, g_4\, g_5 }
  \\ \nonumber
 \Proj^{\overline{10},27}_{g_1\,g_2\,g_3\,g_4\,g_5\,g_6}
  &=& \frac{1}{\TR}
  \frac{2\Nc}{\Nc^2-\Nc-2}
  \Proj^{\overline{10}}_{g_1\,g_2\, i_1\, i_2 } d_{i_2\,g_3\,i_3} 
  \Proj^{27}_{i_1\,i_3\, i_4\, i_6 }
  d_{i_6\,g_6\,i_5} \Proj^{\overline{10}}_{i_4\,i_5\, g_4\, g_5 }
  \\ \nonumber
  \Proj^{27,27d}_{g_1\,g_2\,g_3\,g_4\,g_5\,g_6}
  &=&  \frac{1}{\TR}
  \frac{\Nc (\Nc+2)}{\Nc^3+3\Nc^2-6\Nc-8}
  \Proj^{27}_{g_1\,g_2\, i_1\, i_2 } d_{i_2\,g_3\,i_3} \Proj^{27}_{i_1\,i_3\, i_4\, i_6 }
   d_{i_6\,g_6\,i_5} \Proj^{27}_{i_4\,i_5\,g_4\,g_5}
  \\ \nonumber
  \Proj^{27,27f}_{g_1\,g_2\,g_3\,g_4\,g_5\,g_6}
  &=&  \frac{1}{\TR}
  \frac{1}{\Nc+1}
  \Proj^{27}_{g_1\,g_2\, i_1\, i_2 } \ui f_{i_2\,g_3\,i_3} \Proj^{27}_{i_1\,i_3\, i_4\, i_6 }
  \ui f_{i_6\,g_6\,i_5} \Proj^{27}_{i_4\,i_5\, g_4\, g_5 }
  \\ \nonumber 
  \Proj^{8s,0}_{g_1\,g_2\,g_3\,g_4\,g_5\,g_6}
  &=& \frac{1}{\TR}
  \frac{\Nc}{2 \left(\Nc^2-4\right)} d_ {g_ 1\,g_ 2\,i_ 1}\Proj^{0}_ {i_ 1\,g_ 3\,i_ 2\,g_ 6}d_ {i_ 2\,g_ 4\,g_ 5}   
 \\ \nonumber 
  \Proj^{8a,0}_{g_1\,g_2\,g_3\,g_4\,g_5\,g_6}
  &=& -\frac{1}{\TR}
\frac{1}{2 \Nc} \ui f_{g_ 1\,g_ 2\,i_ 1}  \Proj^{0}_ {i_ 1\,g_ 3\,i_ 2\,g_ 6} \ui f_ {i_ 2\,g_ 4\,g_ 5}
  \\ \nonumber 
  \Proj^{10,0}_{g_1\,g_2\,g_3\,g_4\,g_5\,g_6}
  &=& \frac{1}{\TR}
  \frac{2 \Nc}{\Nc^2+\Nc-2} \Proj^{10}_ {g_ 1\,g_ 2\,i_ 1\,i_ 2}  \ui f_ {i_ 2\,g_ 3\,i_ 3} \Proj^{0}_ {i_ 1\,i_ 3\,i_ 4\,i_ 6} \ui f_ {i_6\,g_ 6\,i_5} \Proj^{10}_ {i_4\,i_5\,g_4\,g_5}
  \\ \nonumber 
  \Proj^{\overline{10},0}_{g_1\,g_2\,g_3\,g_4\,g_5\,g_6}
  &=& \frac{1}{\TR}
  \frac{2 \Nc}{\Nc^2+\Nc-2} \Proj^{\overline{10}}_ {g_ 1\,g_ 2\,i_ 1\,i_ 2} \ui f_ {i_ 2\,g_ 3\,i_ 3} \Proj^{0}_ {i_1\,i_ 3\,i_ 4\,i_ 6}  \ui f_ {i_6\,g_ 6\,i_5} \Proj^{\overline{10}}_ {i_ 4\,i_ 5\,g_ 4\,g_ 5}
  \\ \nonumber 
  \Proj^{0,0d}_{g_1\,g_2\,g_3\,g_4\,g_5\,g_6}
  &=& \frac{1}{\TR}
  \frac{(\Nc-2) \Nc}{\Nc^3-3 \Nc^2-6 \Nc+8}  \Proj^{0}_{g_ 1\,g_ 2\,i_ 1\,i_ 2} d_{i_ 2\,g_ 3\,i_ 3} \Proj^{0}_ {i_ 1\,i_ 3\,i_ 4\,i_ 6} d_ {i_6\,g_ 6\,i_5} \Proj^{0}_{i_ 4\,i_ 5\,g_ 4\,g_ 5}
  \\ \nonumber 
   \Proj^{0,0f}_{g_1\,g_2\,g_3\,g_4\,g_5\,g_6}
  &=& \frac{1}{\TR}
   \frac{1}{\Nc-1}  \Proj^{0}_ {g_ 1\,g_ 2\,i_ 1\,i_ 2} \ui f_ {i_ 2\,g_ 3\,i_ 3}  \Proj^{0}_ {i_ 1\,i_ 3\,i_ 4\,i_ 6}\ui f_ {i_6\,g_ 6\,i_5} \Proj^{0}_ {i_ 4\,i_ 5\,g_ 4\,g_ 5}
   \\ \nonumber 
   \Proj^{27,64=c111c111}_{g_1\,g_2\,g_3\,g_4\,g_5\,g_6}
   &=& 
   \frac{1}{\TR^3}\Tens^{27,64}_{g_1\,g_2\,g_3\,g_4\,g_5\,g_6}
   -\frac{2\Nc^2}{9(\Nc+1) (\Nc+2)}
   \Proj^{27,8}_{g_1\,g_2\,g_3\,g_4\,g_5\,g_6}
   \\ \nonumber 
   &-&\frac{4 (\Nc^2-\Nc-2)}{9 \Nc \left(\Nc+2\right)}
   \Proj^{27,27d}_{g_1\,g_2\,g_3\,g_4\,g_5\,g_6}
   \\ \nonumber
   \Proj^{10,35=c111c21}_{g_1\,g_2\,g_3\,g_4\,g_5\,g_6}   
   &=&
   \frac{1}{\TR^3}\frac{4}{3}\Tens^{10,35}_{g_1\,g_2\,g_3\,g_4\,g_5\,g_6}
   -\frac{\Nc-2}{3 \Nc} \Proj^{10,8}_{g_1\,g_2\,g_3\,g_4\,g_5\,g_6}
   -\frac{\Nc+3}{18 \Nc} \Proj^{10,10 f}_{g_1\,g_2\,g_3\,g_4\,g_5\,g_6}
   \\ \nonumber
   &+&
   \frac{\sqrt{\Nc^2-9}}{6 \Nc}
   \CG^{10f}_{g_1\,g_2\,g_3;\,i_1\,i_2}
   \CG^{10fd\,\dagger}_{i_1\,i_2;\,g_4\,g_5\,g_6}
   + \frac{\sqrt{\Nc^2-9}}{6 \Nc}
   \CG^{10fd}_{g_1\,g_2\,g_3;\,i_1\,i_2}
   \CG^{10f\,\dagger}_{i_1\,i_2;\,g_4\,g_5\,g_6}
    \\ \nonumber
    &-& 
    \frac{\Nc-3}{2 \Nc}
    \Proj^{10,10 fd}_{g_1\,g_2\,g_3\,g_4\,g_5\,g_6}
    -\frac{\Nc-2}{3 \Nc}
   \Proj^{10,27}_{g_1\,g_2\,g_3\,g_4\,g_5\,g_6}
   \\ \nonumber
   \Proj^{27,35=c111c21}_{g_1\,g_2\,g_3\,g_4\,g_5\,g_6}
   &=&
   \frac{1}{\TR^3}\frac{4}{3}\Tens^{27,35}_{g_1\,g_2\,g_3\,g_4\,g_5\,g_6}
   -\frac{ \Nc(\Nc+3)}{9 (\Nc+1) (\Nc+2)} \Proj^{27,8}_{g_1\,g_2\,g_3\,g_4\,g_5\,g_6}
   \\ \nonumber
   &-&\frac{\Nc}{3 (\Nc+2)} \Proj^{27,10}_{g_1\,g_2\,g_3\,g_4\,g_5\,g_6}
   -\frac{\Nc^2+2 \Nc-8}{18 \Nc (\Nc+2)} \Proj^{27,27d}_{g_1\,g_2\,g_3\,g_4\,g_5\,g_6}
   \\ \nonumber
   &-&  \frac{1}{\TR} \frac{1}{6 (\Nc+1)}
   \Proj^{27}_{g_1\,g_2\,i_1\,i_2}d_{i_2\,g_3\,i_3}\Proj^{27}_{i_1\,i_3\,i_4\,i_6}
    \ui f_{i_6\,g_6\,i_5}\Proj^{27}_{i_4\,i_5\,g_4\,g_5}
   \\ \nonumber
   &-&  \frac{1}{\TR} \frac{1}{6 (\Nc+1)}
   \Proj^{27}_{g_1\,g_2\,i_1\,i_2} \ui f_{i_2\,g_3\,i_3}\Proj^{27}_{i_1\,i_3\,i_4\,i_6}
   d_{i_6\,g_6\,i_5}\Proj^{27}_{i_4\,i_5\,g_4\,g_5}
   -\frac{1}{2}\Proj^{27,27f}_{g_1\,g_2\,g_3\,g_4\,g_5\,g_6}
   \\ \nonumber
   \Proj^{\tenbar,\overline{35}=c21c111}_{g_1\,g_2\,g_3\,g_4\,g_5\,g_6}   
   &=&
   \frac{1}{\TR^3}\frac{4}{3}\Tens^{\tenbar,\overline{35}}_{g_1\,g_2\,g_3\,g_4\,g_5\,g_6}
   -\frac{\Nc-2}{3 \Nc}
   \Proj^{\tenbar,8}_{g_1\,g_2\,g_3\,g_4\,g_5\,g_6}
   -\frac{\Nc+3}{18 \Nc}
   \Proj^{\tenbar,\tenbar f}_{g_1\,g_2\,g_3\,g_4\,g_5\,g_6}
   \\ \nonumber
   &-& \frac{\sqrt{\Nc^2-9}}{6 \Nc}
   \CG^{\tenbar f}_{g_1\,g_2\,g_3;\,i_1\,i_2}
   \CG^{\tenbar fd\,\dagger}_{i_1\,i_2;\,g_4\,g_5\,g_6}
   - \frac{\sqrt{\Nc^2-9}}{6 \Nc}
   \CG^{\tenbar fd}_{g_1\,g_2\,g_3;\,i_1\,i_2}
   \CG^{\tenbar f\,\dagger}_{i_1\,i_2;\,g_4\,g_5\,g_6}
    \\ \nonumber
    &-& 
    \frac{\Nc-3}{2 \Nc}
    \Proj^{\tenbar,\tenbar fd}_{g_1\,g_2\,g_3\,g_4\,g_5\,g_6}
    -\frac{\Nc-2}{3 \Nc}
   \Proj^{\tenbar,27}_{g_1\,g_2\,g_3\,g_4\,g_5\,g_6}
    \\ \nonumber
   \Proj^{27,\overline{35}=c21c111}_{g_1\,g_2\,g_3\,g_4\,g_5\,g_6}
   &=&
   \frac{1}{\TR^3}\frac{4}{3} \Tens^{27,\overline{35}}_{g_1\,g_2\,g_3\,g_4\,g_5\,g_6}
   -\frac{\Nc (\Nc+3)}{9 (\Nc+1) (\Nc+2)}
   \Proj^{27,8}_{g_1\,g_2\,g_3\,g_4\,g_5\,g_6}
   \\ \nonumber
   &-&\frac{\Nc}{3 (\Nc+2)}
   \Proj^{27,\tenbar}_{g_1\,g_2\,g_3\,g_4\,g_5\,g_6}
   -\frac{\Nc^2+2 \Nc-8}{18 \Nc (\Nc+2)}
   \Proj^{27,27d}_{g_1\,g_2\,g_3\,g_4\,g_5\,g_6}
   \\ \nonumber
   &+&  \frac{1}{\TR}\frac{1}{6 (\Nc+1)}
   \Proj^{27}_{g_1\,g_2\,i_1\,i_2}d_{i_2\,g_3\,i_3}\Proj^{27}_{i_1\,i_3\,i_4\,i_6}
    \ui f_{i_6\,g_6\,i_5}\Proj^{27}_{i_4\,i_5\,g_4\,g_5}
   \\ \nonumber
   &+&  \frac{1}{\TR}\frac{1}{6 (\Nc+1)}
   \Proj^{27}_{g_1\,g_2\,i_1\,i_2} \ui f_{i_2\,g_3\,i_3}\Proj^{27}_{i_1\,i_3\,i_4\,i_6}
   d_{i_6\,g_6\,i_5}\Proj^{27}_{i_4\,i_5\,g_4\,g_5}
   -\frac{1}{2}\Proj^{27,27f}_{g_1\,g_2\,g_3\,g_4\,g_5\,g_6}  
   \\ \nonumber
   \Proj^{10,c21c21}_{g_1\,g_2\,g_3\,g_4\,g_5\,g_6}
   &=& 
   \frac{1}{\TR^3}\frac{16}{9}\Tens^{10,c21c21}_{g_1\,g_2\,g_3\,g_4\,g_5\,g_6}
   -\frac{1}{3}\Proj^{10,8}_{g_1\,g_2\,g_3\,g_4\,g_5\,g_6}
   -\frac{4}{9}\Proj^{10,10f}_{g_1\,g_2\,g_3\,g_4\,g_5\,g_6}
   \\ \nonumber
   &-& \frac{2 \left(\Nc+1\right)}{3 \Nc}
   \Proj^{10,27}_{g_1\,g_2\,g_3\,g_4\,g_5\,g_6}
   -\frac{2 \left(\Nc-1\right)}{3 \Nc}
   \Proj^{10,0}_{g_1\,g_2\,g_3\,g_4\,g_5\,g_6}
   \\ \nonumber 
   \Proj^{\tenbar,c21c21}_{g_1\,g_2\,g_3\,g_4\,g_5\,g_6}
   &=& 
      \frac{1}{\TR^3} \frac{16}{9} \Tens^{\tenbar,c21c21}_{g_1\,g_2\,g_3\,g_4\,g_5\,g_6}
      -\frac{1}{3}\Proj^{\tenbar,8}_{g_1\,g_2\,g_3\,g_4\,g_5\,g_6}
      -\frac{4}{9}\Proj^{\tenbar,\tenbar f}_{g_1\,g_2\,g_3\,g_4\,g_5\,g_6}
      \\ \nonumber
      &-&\frac{2 \left(\Nc+1\right)}{3 \Nc}
      \Proj^{\tenbar,27}_{g_1\,g_2\,g_3\,g_4\,g_5\,g_6}
      -\frac{2 \left(\Nc-1\right)}{3 \Nc}
   \Proj^{\tenbar,0}_{g_1\,g_2\,g_3\,g_4\,g_5\,g_6}
   \\ \nonumber
   \Proj^{27,c21c21}_{g_1\,g_2\,g_3\,g_4\,g_5\,g_6}
   &=& 
   \frac{1}{\TR^3}
   \frac{16}{9}\Tens^{27,c21c21}_{g_1\,g_2\,g_3\,g_4\,g_5\,g_6}
   -\frac{(\Nc+3) (5 \Nc+6)}{9 (\Nc+1) (\Nc+2)}
   \Proj^{27,8}_{g_1\,g_2\,g_3\,g_4\,g_5\,g_6}  
   \\\nonumber
   &-&\frac{2 (\Nc+3)}{3 (\Nc+2)}
   \Proj^{27,10}_{g_1\,g_2\,g_3\,g_4\,g_5\,g_6}
   -\frac{2 (\Nc+3)}{3 (\Nc+2)}
   \Proj^{27,\tenbar}_{g_1\,g_2\,g_3\,g_4\,g_5\,g_6}
   \\\nonumber
   &-&\frac{4(\Nc+1) (\Nc+4)}{9 \Nc (\Nc+2)}
   \Proj^{27,27d}_{g_1\,g_2\,g_3\,g_4\,g_5\,g_6}
   \\\nonumber
   \Proj^{0,c21c21}_{g_1\,g_2\,g_3\,g_4\,g_5\,g_6}
   &=& 
   \frac{1}{\TR^3}\frac{16}{9}
   \Tens^{0,c21c21}_{g_1\,g_2\,g_3\,g_4\,g_5\,g_6}
   +\frac{(6-5 \Nc) (\Nc-3)}{9 (\Nc-2) (\Nc-1)}
   \Proj^{0,8}_{g_1\,g_2\,g_3\,g_4\,g_5\,g_6}
   \\\nonumber
   &-&\frac{2 (\Nc-3)}{3 (\Nc-2)}
   \Proj^{0,10}_{g_1\,g_2\,g_3\,g_4\,g_5\,g_6}
   -\frac{2 (\Nc-3)}{3 (\Nc-2)}
   \Proj^{0,\tenbar}_{g_1\,g_2\,g_3\,g_4\,g_5\,g_6}
   \\\nonumber
   &-&\frac{4(\Nc-4) (\Nc-1)}{9 (\Nc-2) \Nc}
   \Proj^{0,0d}_{g_1\,g_2\,g_3\,g_4\,g_5\,g_6}
   \\\nonumber
   \Proj^{10,c111c3}_{g_1\,g_2\,g_3\,g_4\,g_5\,g_6}
   &=& 
   \frac{1}{\TR^3}\Tens^{10,c111c3}_{g_1\,g_2\,g_3\,g_4\,g_5\,g_6}-
   \frac{4}{9}\Proj^{10,10f}_{g_1\,g_2\,g_3\,g_4\,g_5\,g_6}
   \\ \nonumber 
   \Proj^{\overline{10},c3c111}_{g_1\,g_2\,g_3\,g_4\,g_5\,g_6}
   &=& 
   \frac{1}{\TR^3}\Tens^{\overline{10},c3c111}_{g_1\,g_2\,g_3\,g_4\,g_5\,g_6}-
   \frac{4}{9}\Proj^{\overline{10},\overline{10}f}_{g_1\,g_2\,g_3\,g_4\,g_5\,g_6}
   \\ \nonumber 
   \Proj^{0,c21c3}_{g_1\,g_2\,g_3\,g_4\,g_5\,g_6}  
   &=&
     \frac{1}{\TR^3}\frac{4}{3}\Tens^{0,c21c3}_{g_1\,g_2\,g_3\,g_4\,g_5\,g_6}
     -\frac{(\Nc-3) \Nc}{9 (\Nc-2) (\Nc-1)}
     \Proj^{0,8}_{g_1\,g_2\,g_3\,g_4\,g_5\,g_6}
     \\ \nonumber
     &-&\frac{\Nc}{3 (\Nc-2)}
     \Proj^{0,10}_{g_1\,g_2\,g_3\,g_4\,g_5\,g_6}
     -\frac{\Nc^2-2 \Nc-8}{18 (\Nc-2) \Nc}
     \Proj^{0,0d}_{g_1\,g_2\,g_3\,g_4\,g_5\,g_6}
     \\ \nonumber
     &+& \frac{1}{\TR} \frac{1}{6 (\Nc-1)}
     \Proj^{0}_{g_ 1\,g_ 2\,i_ 1\,i_ 2} d_{i_ 2\,g_ 3\,i_ 3} \Proj^{0}_ {i_ 1\,i_ 3\,i_ 4\,i_ 6} 
     \ui f_ {i_ 6\,g_ 6\,i_ 5} \Proj^{0}_ {i_ 4\,i_ 5\,g_ 4\,g_ 5}
     \\ \nonumber
     &+& \frac{1}{\TR} \frac{1}{6 (\Nc-1)}
     \Proj^{0}_{g_ 1\,g_ 2\,i_ 1\,i_ 2} \ui f_{i_ 2\,g_ 3\,i_ 3} \Proj^{0}_ {i_ 1\,i_ 3\,i_ 4\,i_ 6} 
     d_ {i_ 6\,g_ 6\,i_ 5} \Proj^{0}_ {i_ 4\,i_ 5\,g_ 4\,g_ 5}
     -
     \frac{1}{2}
     \Proj^{0,0f}_{g_1\,g_2\,g_3\,g_4\,g_5\,g_6}
   \\ \nonumber 
   \Proj^{10,c21c3}_{g_1\,g_2\,g_3\,g_4\,g_5\,g_6}  
   &=&
   \frac{1}{\TR^3}\frac{4}{3}\Tens^{10,c21c3}_{g_1\,g_2\,g_3\,g_4\,g_5\,g_6}
   - \frac{\Nc+2}{3 \Nc}
   \Proj^{10,8}_{g_1\,g_2\,g_3\,g_4\,g_5\,g_6}
   \\ \nonumber
   &-& \frac{\Nc-3}{18 \Nc}
   \Proj^{10,10 f}_{g_1\,g_2\,g_3\,g_4\,g_5\,g_6}
   - \frac{\sqrt{\Nc^2-9}}{6 \Nc}
   \CG^{10f}_{g_1\,g_2\,g_3;\,i_1\,i_2}
   \CG^{10fd\,\dagger}_{i_1\,i_2;\,g_4\,g_5\,g_6}
   \\ \nonumber
   &-&\frac{\sqrt{\Nc^2-9}}{6 \Nc}
   \CG^{10fd}_{g_1\,g_2\,g_3;\,i_1\,i_2}
   \CG^{10f\,\dagger}_{i_1\,i_2;\,g_4\,g_5\,g_6}
   -\frac{\Nc+3}{2 \Nc}
   \Proj^{10,10 fd}_{g_1\,g_2\,g_3\,g_4\,g_5\,g_6}
   -\frac{\Nc+2}{3 \Nc}
   \Proj^{10,0}_{g_1\,g_2\,g_3\,g_4\,g_5\,g_6}
   \\ \nonumber 
   \Proj^{0,c3c21}_{g_1\,g_2\,g_3\,g_4\,g_5\,g_6}  
   &=&
     \frac{1}{\TR^3}\frac{4}{3}\Tens^{0,c3c21}_{g_1\,g_2\,g_3\,g_4\,g_5\,g_6}
     -\frac{(\Nc-3) \Nc}{9 (\Nc-2) (\Nc-1)}
     \Proj^{0,8}_{g_1\,g_2\,g_3\,g_4\,g_5\,g_6}
     \\ \nonumber
     &-&\frac{\Nc}{3 (\Nc-2)}
     \Proj^{0,\tenbar}_{g_1\,g_2\,g_3\,g_4\,g_5\,g_6}
     -\frac{\left(\Nc^2-2\Nc -8\right)}{18 (\Nc-2) \Nc}
     \Proj^{0,0d}_{g_1\,g_2\,g_3\,g_4\,g_5\,g_6}
     \\ \nonumber
     &-& \frac{1}{\TR}\frac{1}{6 (\Nc-1)}
     \Proj^{0}_{g_ 1\,g_ 2\,i_ 1\,i_ 2} d_{i_ 2\,g_ 3\,i_ 3} \Proj^{0}_ {i_ 1\,i_ 3\,i_ 4\,i_ 6} 
     \ui f_ {i_ 6\,g_ 6\,i_ 5} \Proj^{0}_ {i_ 4\,i_ 5\,g_ 4\,g_ 5}
     \\ \nonumber
     &-& \frac{1}{\TR}\frac{1}{6 (\Nc-1)}
     \Proj^{0}_{g_ 1\,g_ 2\,i_ 1\,i_ 2} \ui f_{i_ 2\,g_ 3\,i_ 3} \Proj^{0}_ {i_ 1\,i_ 3\,i_ 4\,i_ 6} 
     d_ {i_ 6\,g_ 6\,i_ 5} \Proj^{0}_ {i_ 4\,i_ 5\,g_ 4\,g_ 5}
     -\frac{1}{2}
     \Proj^{0,0f}_{g_1\,g_2\,g_3\,g_4\,g_5\,g_6}
   \\ \nonumber 
   \Proj^{\tenbar,c3c21}_{g_1\,g_2\,g_3\,g_4\,g_5\,g_6}  
   &=&
   \frac{1}{\TR^3}\frac{4}{3}\Tens^{\tenbar,c3c21}_{g_1\,g_2\,g_3\,g_4\,g_5\,g_6}
   -\frac{\Nc+2}{3 \Nc}
   \Proj^{\tenbar,8}_{g_1\,g_2\,g_3\,g_4\,g_5\,g_6}
   \\ \nonumber
   &-& \frac{\Nc-3}{18 \Nc}
   \Proj^{\tenbar,\tenbar f}_{g_1\,g_2\,g_3\,g_4\,g_5\,g_6}
   +\frac{\sqrt{\Nc^2-9}}{6 \Nc}
   \CG^{\tenbar f}_{g_1\,g_2\,g_3;\,i_1\,i_2}
   \CG^{\tenbar fd\,\dagger}_{i_1\,i_2;\,g_4\,g_5\,g_6}
   \\ \nonumber
   &+&\frac{\sqrt{\Nc^2-9}}{6 \Nc}
   \CG^{\tenbar fd}_{g_1\,g_2\,g_3;\,i_1\,i_2}
   \CG^{\tenbar f\,\dagger}_{i_1\,i_2;\,g_4\,g_5\,g_6}
  -\frac{\Nc+3}{2 \Nc}
  \Proj^{\tenbar,\tenbar fd}_{g_1\,g_2\,g_3\,g_4\,g_5\,g_6}
  -\frac{\Nc+2}{3 \Nc}
   \Proj^{\tenbar,0}_{g_1\,g_2\,g_3\,g_4\,g_5\,g_6}
   \\ \nonumber 
   \Proj^{0,c3c3}_{g_1\,g_2\,g_3\,g_4\,g_5\,g_6}
   &=& 
     \frac{1}{\TR^3}\Tens^{0,c3c3}_{g_1\,g_2\,g_3\,g_4\,g_5\,g_6}
     -\frac{2\Nc^2}{9 (\Nc-2) (\Nc-1)}\Proj^{0,8}_{g_1\,g_2\,g_3\,g_4\,g_5\,g_6}
     \\ \nonumber 
     &-&\frac{4(\Nc^2+\Nc-2)}{9 (\Nc-2) \Nc}\Proj^{0,0d}_{g_1\,g_2\,g_3\,g_4\,g_5\,g_6}
  \\ 
\end{eqnarray}
where the tensors $\Tens^{M_{12},M_{123}}$ are constructed as indicated
in \eqref{eq:new_projector}, and 
\begin{eqnarray}
  \CG^{10f\,\dagger }_{i_1\,i_2;\,g_4\,g_5\,g_6}&=&
  \private{\sqrt{\frac{dim10}{B1010f}}}
  \frac{1}{\sqrt{\Nc \TR}}
  \Proj^{10}_{i_1\,i_2\, i_3\, i_4} \ui f_{i_4\,g_6\,i_5}
  \Proj^{10}_{i_3\,i_5\, g_4\, g_5}
  \nonumber\\
  \CG^{\overline{10}f\,\dagger}_{i_1\,i_2;\,g_4\,g_5\,g_6}&=&
   \private{\sqrt{\frac{dim10}{B1010f}}}
  \frac{1}{\sqrt{\Nc \TR}}
  \Proj^{\overline{10}}_{i_1\,i_2\, i_3\, i_4} \ui f_{i_4\,g_6\,i_5}
  \Proj^{\overline{10}}_{i_3\,i_5\, g_4\, g_5}
  \nonumber\\
  \nonumber\\
  \CG^{10fd\,\dagger}_{i_1\,i_2;\,g_4\,g_5\,g_6}&=&
  \private{\sqrt{\frac{dim10bar}{B10bar10barf}}}
  \frac{\sqrt{\Nc}}{\sqrt{\TR (\Nc^2-9)}}
  \Proj^{10}_{i_1\,i_2\, i_3\, i_4} 
  (d_{i_4\,g_6\,i_5}-\frac{1}{\Nc}\ui f_{i_4\,g_6\,i_5})
  \Proj^{10}_{i_3\,i_5\, g_4\, g_5}
  \nonumber\\
  \CG^{\overline{10}fd\,\dagger}_{i_1\,i_2;\,g_4\,g_5\,g_6}&=&
  \private{\sqrt{\frac{dim10bar}{B10bar10barfd}}}
  \frac{\sqrt{\Nc}}{\sqrt{\TR (\Nc^2-9)}}
  \Proj^{\overline{10}}_{i_1\,i_2\, i_3\, i_4} 
  (d_{i_4\,g_6\,i_5}+\frac{1}{\Nc}\ui f_{i_4\,g_6\,i_5})
  \Proj^{\overline{10}}_{i_3\,i_5\, g_4\, g_5}\;.
  \label{eq:DecupletTensors}
\end{eqnarray}
To understand the presence of $\TR$ in \eqref{eq:3gProjectors} we note that 
expressed in terms of pure traces,
such as for example $\tr[t^{g_2}t^{g_5}t^{g_6}t^{g_4}t^{g_3}t^{g_1}]$,
the $\TR$ factor would enter as $1/\TR^3$,
to compensate for the contraction of three gluons (giving $\TR^3$)
when squaring the projection operator. However, every internal 
index gives rise to an extra factor $\TR$ when contracted, and  
the $\ui f_{abc}$ and $d_{abc}$
come with a factor $1/\TR$, as can be seen from eqs. (\ref{eq:f}) and (\ref{eq:d_def}). 
Similarly, the
two gluon projection operators contain a compensating factor $1/\TR^2$. 
When all this is 
accounted for the $\TR$ appears as in \eqref{eq:3gProjectors} 
and \eqref{eq:DecupletTensors}.
 
\section{Properties of some projection operators}
\label{sec:proj_prop}

In this appendix we collect and prove some properties of the gluon
projection operators in \secref{sec:Gluons}.

First we prove \eqref{eq:back_in_hist}.
We explicitly verify that $\Proj^{M'}$ is a projector by squaring,
\begin{eqnarray}
\label{eq:back_in_hist_squared}
 \Big(\Proj^{\hdots,M',M,M'}\Big)^2
 = \left( \frac{\dim M'}{\dim M} \right)^2\
 \parbox{8cm}{\epsfig{file=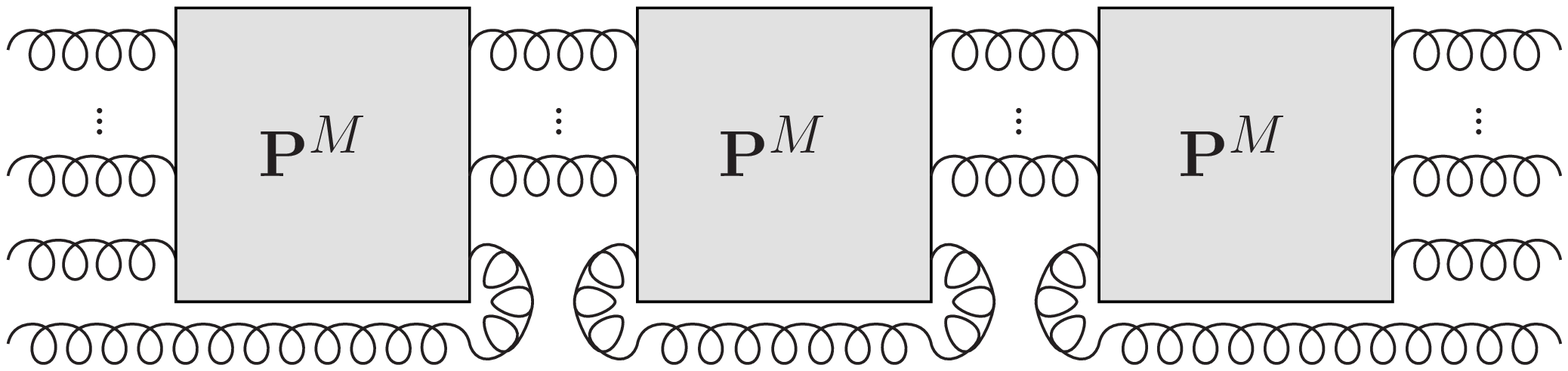,width=8cm}} \, .
 \nonumber \\
\end{eqnarray}
Due to property (ii) the term in the middle has to be proportional to
$\Proj^{M'}: A^{\otimes(\ng-2)} \to A^{\otimes(\ng-2)}$, i.e.
\begin{equation}
  \parbox{3.5cm}{\epsfig{file=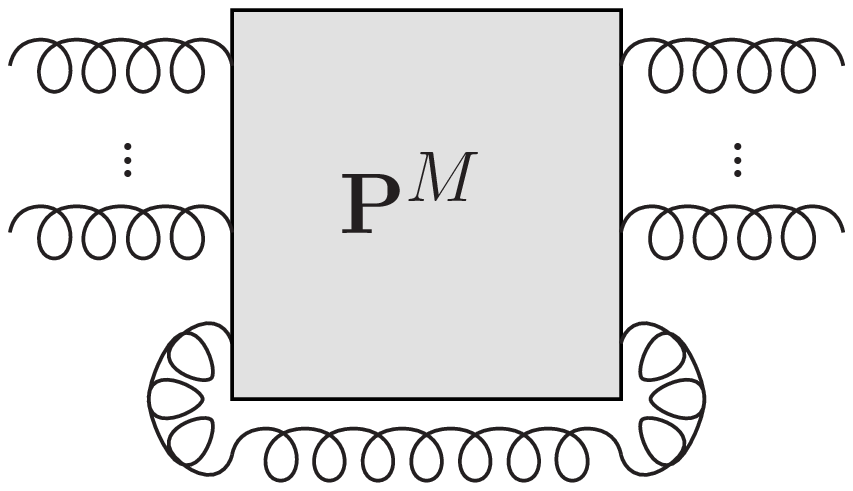,width=3.5cm}} 
  = \ \alpha \ 
    \parbox{3cm}{\epsfig{file=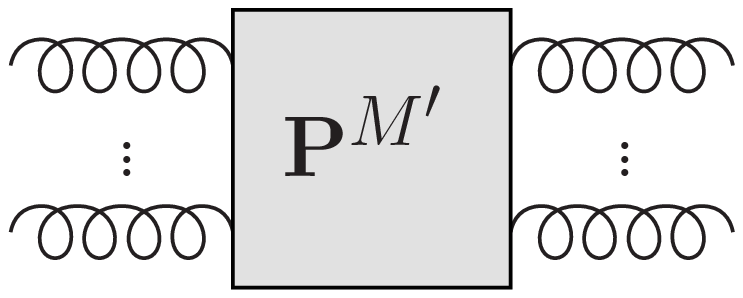,width=3cm}} \, .
\end{equation}
The constant can be found by taking the trace, $\dim M = \alpha \dim
M'$, and thus $\alpha = \dim M / \dim M'$. Substituting into
\eqref{eq:back_in_hist_squared} we obtain
$\big(\Proj^{\hdots,M,M'}\big)^2 = \Proj^{\hdots,M,M'}$ as desired.

We now turn to \eqref{eq:new_to_same_nf}. Letting 
\begin{equation}
 \Proj^{\hdots,M,M'} 
 = \mathcal{N}(M,M')
 \parbox{7cm}{\epsfig{file=Figures/new_to_same_nf.eps,width=7cm}} \, ,
\label{eq:new_to_same_nfN1}
\end{equation}
where $\mathcal{N}(M,M')$ denotes a normalization factor, we have 
\begin{eqnarray}
 &&(\Proj^{\hdots,M,M'})^2=\mathcal{N}^2(M,M') \times\\
 && \quad \times 
 \parbox{14cm}{\epsfig{file=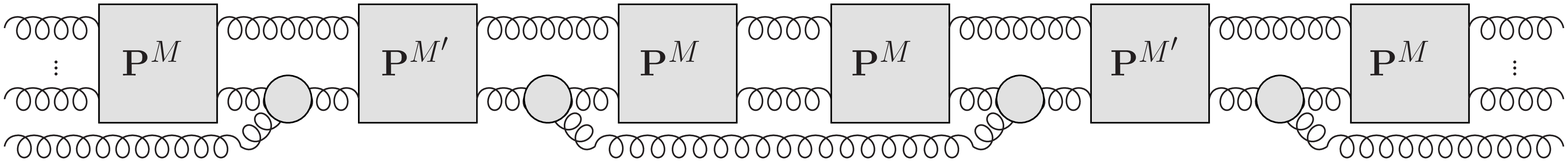,width=14cm}}  
 \nonumber\, .
\label{eq:new_to_same_nfN2}
\end{eqnarray}
After employing $(\Proj^{M})^2=\Proj^{M}$ we observe that the middle
part can be written as
\begin{eqnarray}
  \raisebox{-1mm}{\parbox{7cm}{\epsfig{file=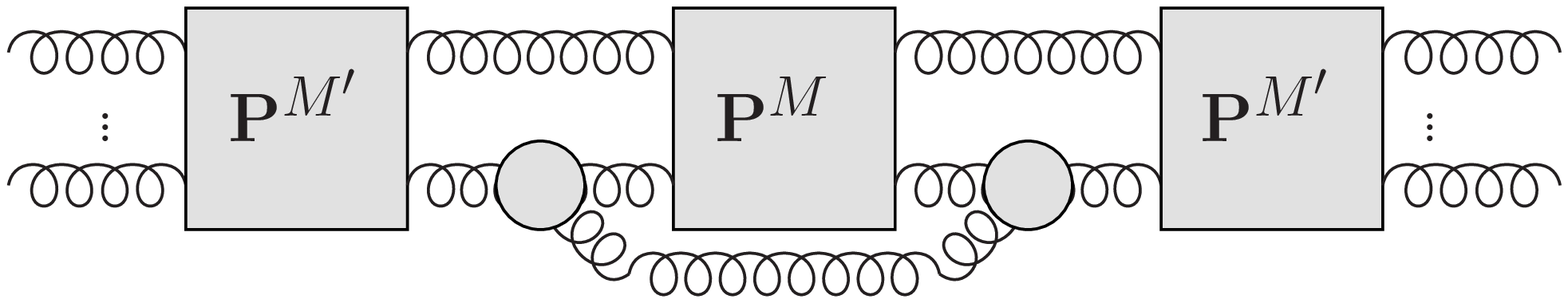,width=7cm}}}
  =\alpha
  \parbox{2.7cm}{\epsfig{file=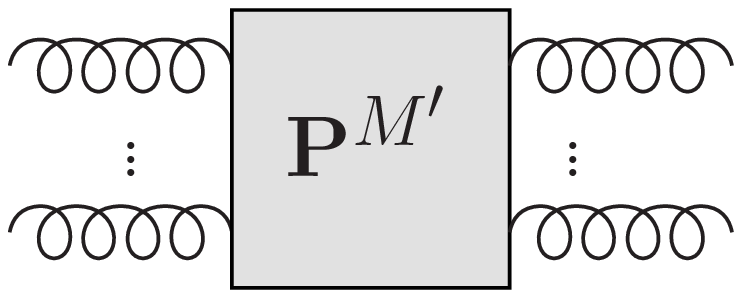,width=2.7cm}} \,
\label{eq:new_to_same_nfN3}
\end{eqnarray}
for some $\alpha$. We find $\alpha$ and thus $\mathcal{N}(M,M')$ by
taking the trace
\begin{eqnarray}
  B(M,M')=
    \parbox{6cm}{\epsfig{file=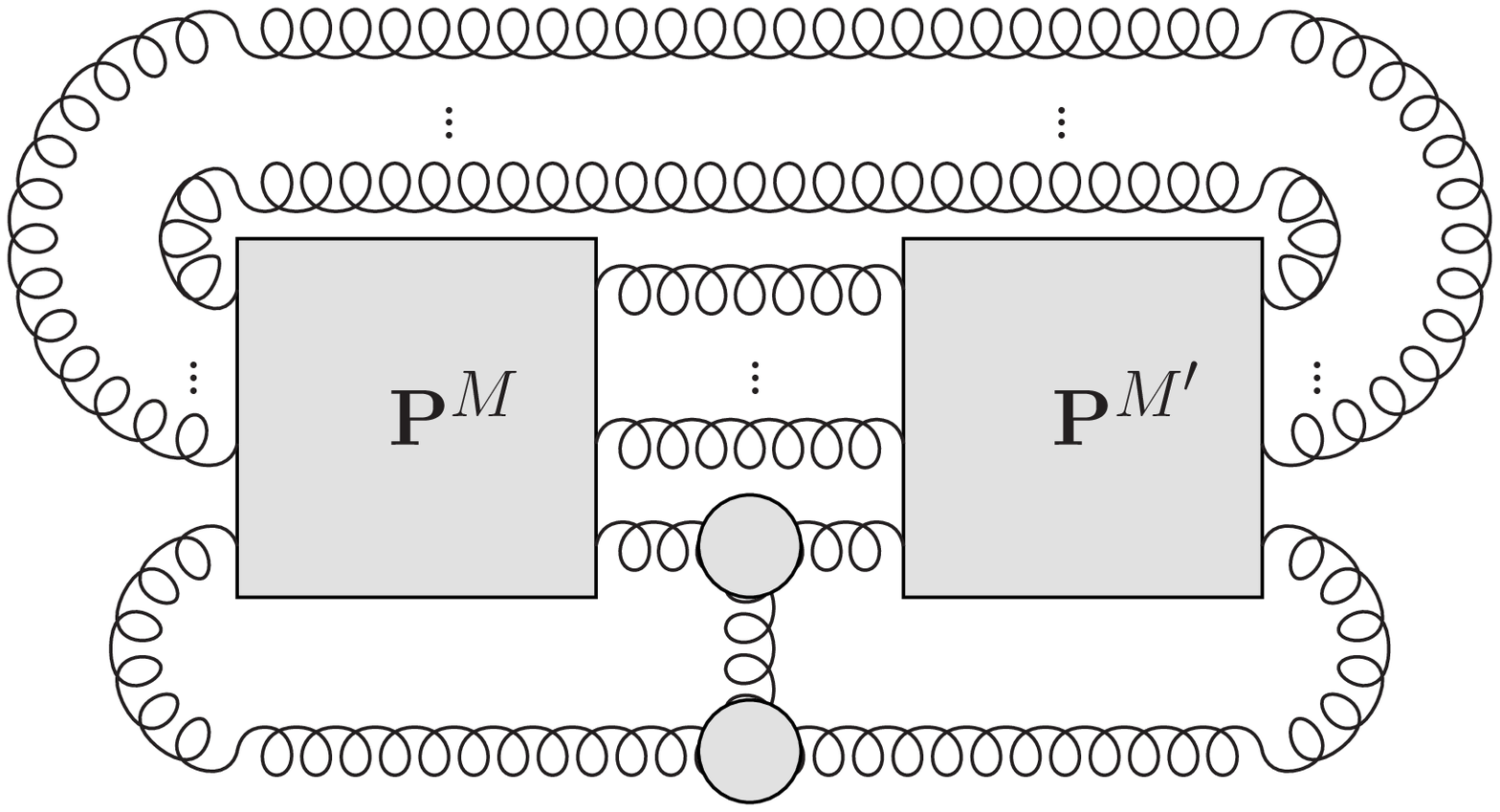,width=6cm}}
     =\alpha \;\mbox{dim}(M')\;,
  \label{eq:new_to_same_nfN4}
\end{eqnarray}
i.e. $\mathcal{N}(M,M')=\mbox{dim}(M')/B(M,M').$

Next we show that the projector (\ref{eq:PMMfd}) can also be written
in the form (\ref{eq:new_to_same_nf}). Expanding the products in
\eqref{eq:TMMfd} we have
\begin{equation}
\begin{split}
  \Tens^{\hdots,M,Mfd} 
  &= \Proj^{\hdots,M,Md} 
    - \Proj^{\hdots,M,Mf}\Proj^{\hdots,M,Md} 
    - \Proj^{\hdots,M,Md}\Proj^{\hdots,M,Mf} \\ & \quad 
    + \Proj^{\hdots,M,Mf}\Proj^{\hdots,M,Md}\Proj^{\hdots,M,Mf} 
    \, .
\end{split}
\end{equation}
The second term on the r.h.s. reads
\begin{equation}
\begin{split}
  \Proj^{\hdots,M,Mf}\Proj^{\hdots,M,Md}
  &= \frac{(\dim M)^2}{B_f(M) \, B_d(M)} 
  \times \\ & \quad \times 
  \parbox{11cm}{\epsfig{file=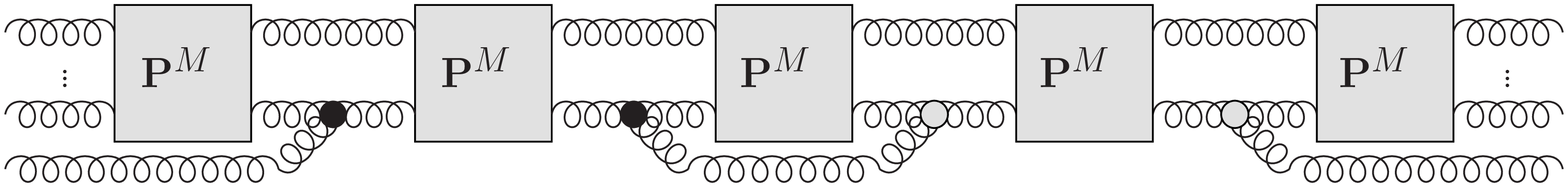,width=11cm}} \, . 
\end{split}
\label{eq:PMfPMd}
\end{equation}
Here and in the following we denote by $B_f(M)$ the bubble diagram
(\ref{eq:new_to_same_nfN4}) with $M'=M$ and where both gray circles
are chosen as $\ui f$, and by $B_d(M)$ the same diagram with both gray
circles being $d$.
The middle part in \eqref{eq:PMfPMd} is proportional to $\Proj^{\hdots,M}$,
\begin{equation}
  \raisebox{-1mm}{\parbox{7cm}{\epsfig{file=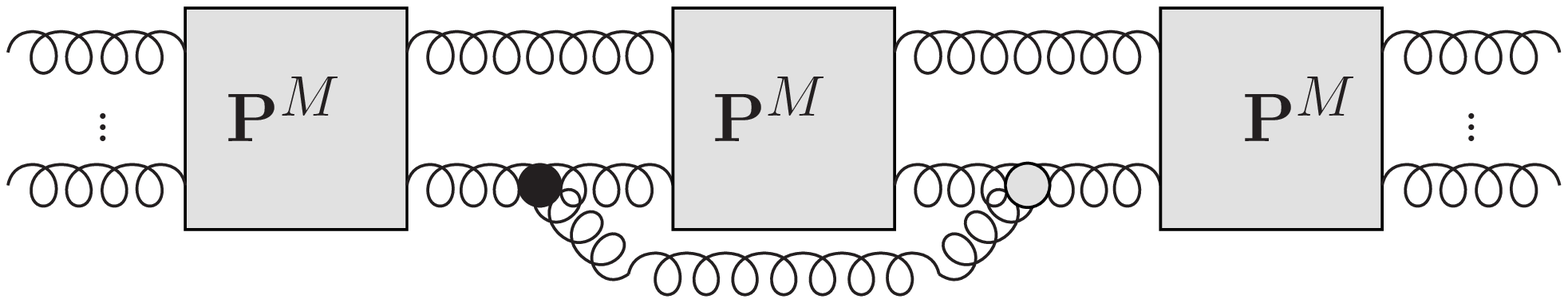,width=7cm}}} 
  = \alpha \parbox{2.7cm}{\epsfig{file=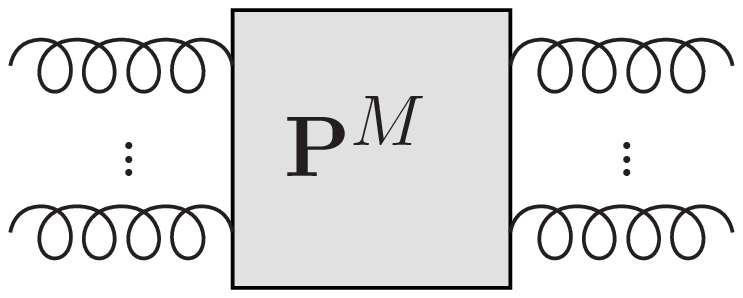,width=2.7cm}}
  \, .
\end{equation}
 We find $\alpha$ by taking the trace, 
\begin{equation}
  \alpha = \frac{C(M)}{\dim M} \, , 
\end{equation}
where $C(M)$ denotes the bubble diagram (\ref{eq:new_to_same_nfN4})
with $M'=M$ and where one gray circle represents $\ui f$ and the other
one equals $d$. Note that $\alpha$ is real. By similar calculations we
also find the other terms,
\begin{equation}
\begin{split}
  \Tens^{\hdots,M,Mfd} 
  &= \frac{\dim M}{B_d(M)} \Bigg( 
     \parbox{6cm}{\epsfig{file=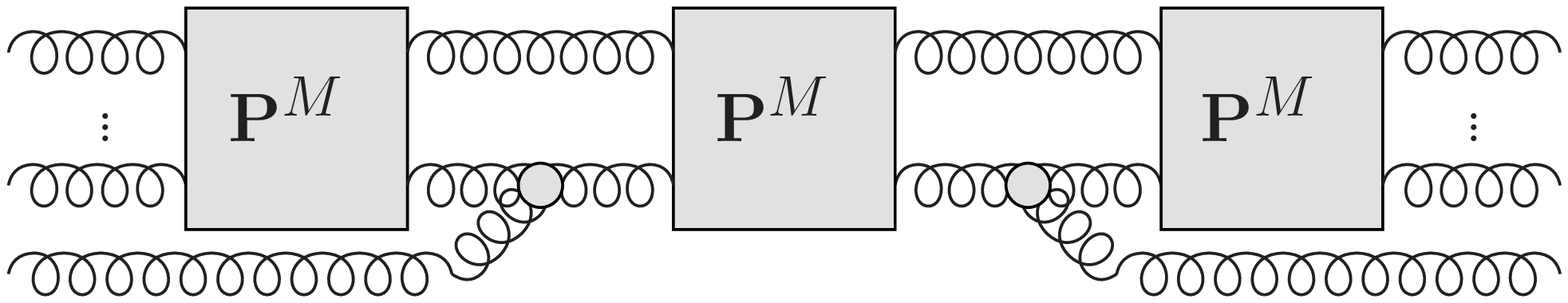,width=6cm}} 
     \\[1ex] & \qquad\qquad\qquad
     - \frac{C(M)}{B_f(M)} 
     \parbox{6cm}{\epsfig{file=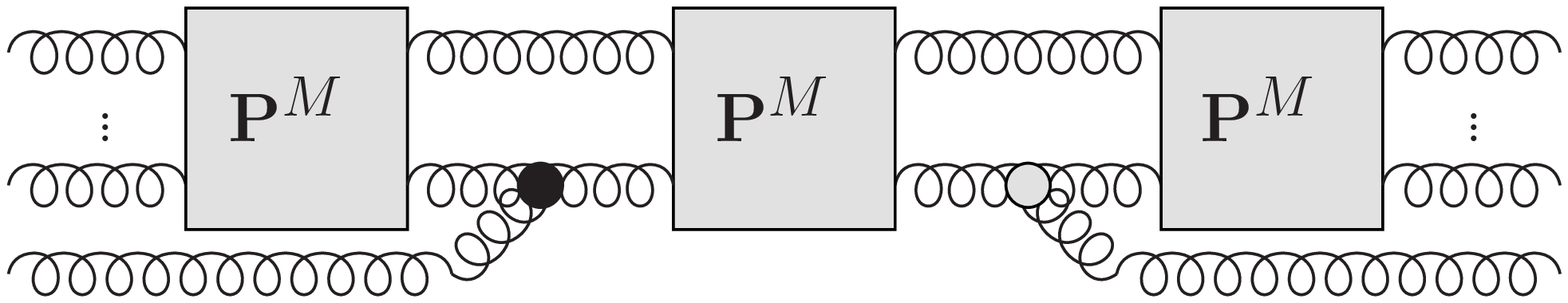,width=6cm}} 
     \\[1ex] & \qquad\qquad\qquad
      - \frac{C(M)}{B_f(M)} 
     \parbox{6cm}{\epsfig{file=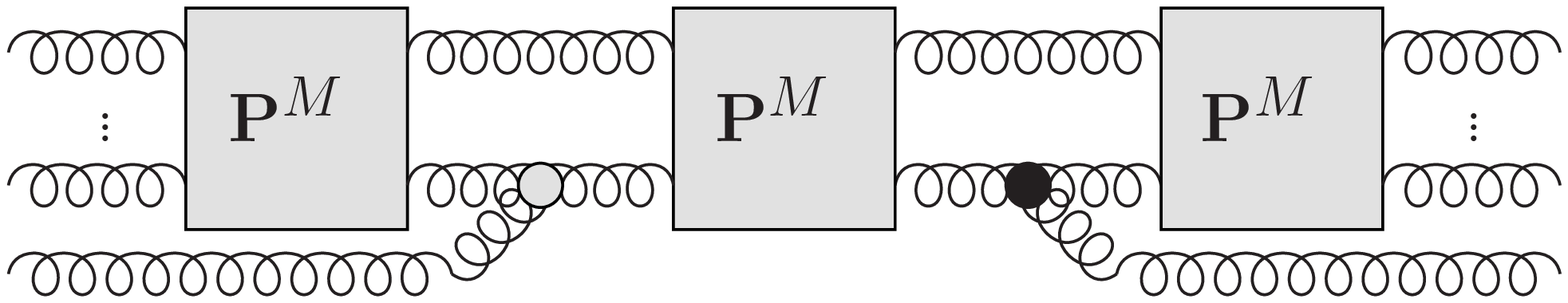,width=6cm}} 
     \\[1ex] & \qquad\qquad\qquad
      + \left(\frac{C(M)}{B_f(M)}\right)^2
     \parbox{6cm}{\epsfig{file=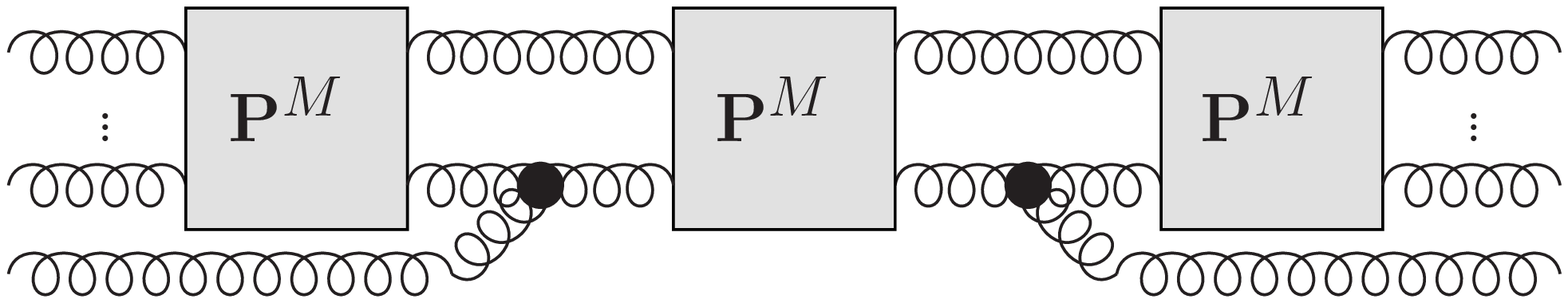,width=6cm}} 
     \Bigg) \\[1ex]
  &= \frac{\dim M}{B_d(M)} 
     \parbox{6cm}{\epsfig{file=Figures/new_to_same_nf.eps,width=6cm}} \, ,
\end{split}
\end{equation}
where on the last line the big gray circles are given by 
\begin{equation}
\label{eq:linear_comb_fd}
  \raisebox{.5mm}{\parbox{2cm}{\epsfig{file=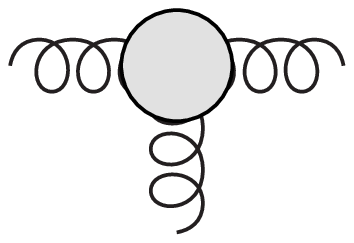,width=2cm}}}
  = \parbox{2cm}{\epsfig{file=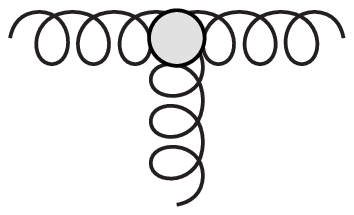,width=2cm}} 
    - \frac{C(M)}{B_f(M)} 
      \parbox{2cm}{\epsfig{file=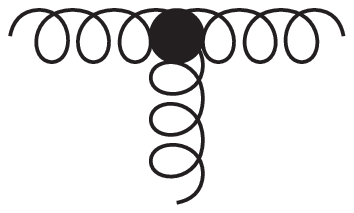,width=2cm}} \, .
\end{equation}
Upon normalization we have cast $\Proj^{\hdots,M,Mfd}$ in the form
(\ref{eq:new_to_same_nf}) with $M'=M$ and where all big gray circles,
also those inside $B(M,M)$, are given by \eqref{eq:linear_comb_fd}.
 
\section{Invariant tensors, Schur's lemma and color conservation}
\label{sec:invariant_tensors}

We briefly summarize properties of invariant tensors and
their relation to the multiplet version of color conservation 
which follows from Schur's Lemma.

Let $v_j \in V$, $j=1,\hdots,\Nq$, $w_k \in \overline{V}$,
$k=1,\hdots,\Nqbar$, and $u_l \in A$, $l=1,\hdots,\Ng$. Under $\gre
\in \SU(\Nc)$ the $v_j$ transform in the defining representation, $v_j
\mapsto \gre v_j$, the $w_k$ transform in the complex conjugate of the
defining representation, $w_k \mapsto \gre^* w_k$ and the $u_l$
transform in the adjoint representation, $u_l \mapsto \Ad(\gre) u_l$.
A tensor $\Tens \in V^{\otimes\Nq} \otimes
\overline{V}^{\otimes\Nqbar} \otimes A^{\otimes\Ng}$ is called {\it
  invariant} if
\begin{equation}
\label{eq:invariant_tensor}
\begin{split}
  &\left\langle \Tens \,\Big|\, 
  v_1 \otimes \cdots \otimes v_{\Nq} \otimes 
  w_1 \otimes \cdots \otimes w_{\Nqbar} \otimes 
  u_1 \otimes \cdots \otimes u_{\Ng} \right\rangle 
  \\ &=
    \left\langle \Tens \,\Big|\, 
  \gre v_1 \otimes \cdots \otimes \gre v_{\Nq} \otimes 
  \gre^* w_1 \otimes \cdots \otimes \gre^* w_{\Nqbar} \otimes 
  \Ad(\gre) u_1 \otimes \cdots \otimes \Ad(\gre) u_{\Ng} \right\rangle 
\end{split}
\end{equation}
$\forall\ \gre \in \SU(\Nc)$ and $\forall\ v_j,w_k,u_l$, where we use
the scalar product of \secref{sec:ColorSpace}. Another way to
  express invariance is to say that $\Tens$ is a (color) singlet,
  i.e.\ that it transforms under the trivial representation of
  $\SU(\Nc)$. Viewing $\Tens$ as a linear map, e.g., as $\Tens: V^{\otimes\Nqbar}
\otimes A^{\otimes\ng} \to V^{\otimes\Nq} \otimes
A^{\otimes(\Ng-\ng)}$, the invariance condition can be rewritten as
\begin{equation}
\label{eq:invariant_tensor_2}
\begin{split}
  &\left\langle \Tens( w_1^* \otimes \cdots \otimes w_{\Nqbar}^* \otimes 
                       u_1 \otimes \cdots \otimes u_{\ng}) \,\Big|\,
  v_1 \otimes \cdots \otimes v_{\Nq} \otimes 
  u_{\ng+1} \otimes \cdots \otimes u_{\Ng} \right\rangle
  \\ &= 
  \left\langle \Tens \, \Gamma^a(\gre) ( 
    w_1^* \otimes \cdots \otimes w_{\Nqbar}^* \otimes 
    u_1 \otimes \cdots \otimes u_{\ng}) \,\Big|\,
  \Gamma^b(\gre) (v_1 \otimes \cdots \otimes v_{\Nq} \otimes 
  u_{\ng+1} \otimes \cdots \otimes u_{\Ng}) \right\rangle
\end{split}
\end{equation}
where $\Gamma^a$ and $\Gamma^b$ are product representations of $\SU(\Nc)$,
\begin{equation}
  \Gamma^a(\gre) = \gre^{\otimes\Nqbar} \otimes \Ad(\gre)^{\otimes\ng} \, , \qquad 
  \Gamma^b(\gre) = \gre^{\otimes\Nq} \otimes \Ad(\gre)^{\otimes(\Ng-\ng)} \, .
\end{equation}
Using unitarity of $\Gamma^b(\gre)$ we can further rewrite the invariance
condition as,
\begin{equation}
\label{eq:Gamma^a&Gamma^b}
  \langle \Tens (\cdots) \mid \cdots \rangle
  = \langle \Gamma^b(\gre)^\dag \, \Tens \, \Gamma^a(\gre) (\cdots) 
    \mid \cdots \rangle
  = \langle \Gamma^b(\gre)^{-1} \, \Tens \, \Gamma^a(\gre) (\cdots) 
    \mid \cdots \rangle \, ,
\end{equation}
where the dots stand for the tensor products of
\eqref{eq:invariant_tensor_2}. Since the condition has to hold for any
$v_j,w_k,u_l$ an invariant tensor $\Tens$ commutes with the representations
$\Gamma^a$ and $\Gamma^b$ in the sense that
\begin{equation}
  \Tens \, \Gamma^a(\gre) = \Gamma^b(\gre) \, \Tens 
  \quad \forall\ \gre \in \SU(\Nc) \, .
\end{equation}

In the case of irreducible representations $\Gamma^a$ and $\Gamma^b$,
Schur's lemma, see e.g. \cite[sec.~3-14]{Ham62}, states that $\Tens$
has to be a scalar multiple of the identity if the two representations
are equivalent and that $\Tens$ vanishes if they are not
equivalent. In general, the representations (\ref{eq:Gamma^a&Gamma^b})
are not irreducible, but their carrier spaces $V^{\otimes\Nqbar}
\otimes A^{\otimes\ng}$ and $V^{\otimes\Nq} \otimes
A^{\otimes(\Ng-\ng)}$ can be decomposed into a direct sum of
multiplets, $M_1 \oplus M_2 \oplus \hdots$. Introducing bases, such
that the first $\dim(M_1)$ vectors span $M_1$, the next $\dim(M_2)$
span $M_2$ etc., the invariant tensor $\Tens$ gets a block
structure. Due to Schur's lemma all blocks which would map elements of
a multiplet $M$ to a multiplet $M'$ carrying a non-equivalent
irreducible representation vanish identically. For instance, $\Tens$
can map an octet state to any other octet state -- and there can be
several octets in the decompositions of $V^{\otimes\Nqbar} \otimes
A^{\otimes\ng}$ and $V^{\otimes\Nq} \otimes A^{\otimes(\Ng-\ng)}$ --
but it can never map an octet state to a decuplet state. In the
context of QCD we refer to this property as color conservation.

Examples for $\SU(\Nc)$-invariant tensors are $\delta_{q_1}^{q_2}$
(quark or anti-quark lines), $\delta_{g_1g_2}$ (gluon lines), the
three-gluon vertices $\ui f_{g_1g_2g_3}$ and $d_{g_1g_2g_3}$ or the
generators $(t^{g})_{q_1}^{q_2}$ of the defining representation,
see e.g.\ \cite{Dittner:1971fy,Dittner:1972hm,Cvi76}. Tensor
products and contractions of invariant tensors are again invariant
tensors, i.e.\ {\it all} tensors appearing in our constructions of
projectors and basis vectors are invariant tensors.
 
\section{Exponential scaling of number of basis vectors}
\label{sec:NumberOfProjectors}

We here show that for finite $\Nc$ the number of projection
operators required for $\ng \to \ng$ gluons grows at most
exponentially in $\ng$.  To see this, we note that starting in any
multiplet $M$ we have, for finite $\Nc$, according to
\eqref{ResultingMultiplets}, at most a finite number of new multiplets
$N_\mathrm{max}$. For $\Nc=3$, the possibilities may, in
accordance with \eqref{ResultingMultiplets}, be enumerated as
\begin{eqnarray}
  \begin{tiny}
  \young(11,\bullet,2)\,,\;
  \young(11,2,\bullet)\,,\;
  \young(\bullet,11,2)\,,\;
  \young(1,12,\bullet)\,,\;
  \young(\bullet,1,12)\,,\;
  \young(1,\bullet,12)\,,\;
  \young(1,1,2)\,,\;
  \young(1,2,1)\,.
  \nonumber
  \end{tiny}
\label{SU3ResultingMultiplets}
\end{eqnarray}
For $\Nc=3$ we thus have $N_\mathrm{max}=8$, and the number of
projection operators increases at most by a factor of 8 for each new
gluon. In fact, it grows much slower for a small number of gluons, as
then most placements above are forbidden. However, we observe that it
approaches this increase for many gluons. In general we have
$N_\mathrm{max}=(\Nc+1)(\Nc-1)$ as there are $\Nc$ ways of placing the
only 2-row, and, for each option, $(\Nc-1)$ ways of placing the
$\tyoung{\bullet}$. Finally, when there is no $\tyoung{\bullet}$
(rightmost cases above), there are $\Nc-1$ ways of placing the second $1$.
As we always, irrespective of the
shape of the starting multiplet, have a finite $N_\mathrm{max}$ the
number of projection operators grows at most as 
$N_{\mathrm{max}}^{\ng}$.

Having this upper bound for the growth of the number of projection
operators we note that the number of basis vectors in the $2\ng$ space
grows at most as 
$N_\mathrm{max}^{2\ng}$. 
(This would correspond to allowing transitions between all multiplets.)
Again, in reality, it grows much slower for a small number of gluons.

Finally, we note that for processes involving quarks, each
$\qqbar$-pair is either in an octet, in which case the above counting
for the gluon case carries over, or in a singlet, corresponding to the
case of one less gluon above.  We thus find that also in the general
case of both, quarks and gluons, the number of projectors and basis
vectors grows (at most) exponentially.
 
\section{Orthogonal basis for six gluons in $\SU(\Nc)$}
\label{sec:6gBasis}

In this appendix we discuss in detail how to construct a basis for the 
six gluon color space. As always we utilize the method of subgrouping. 
In order to construct all basis vectors we have to combine every instance of a 
multiplet $M$ on the incoming side with every instance of $M$ on the 
outgoing side. 
One way of splitting the gluons is to consider 
$g_1 \, g_2 \, g_3 \to g_4 \, g_5 \, g_6$.
Then, clearly, the possible multiplets are the same on the incoming and outgoing side.
The multiplets with first occurrence up to two are in fact already listed
in \tabref{table:1qqbar4g}. In addition to 
these multiplets there are also new multiplets, as listed in 
\tabref{table:3gNew}. In total this gives rise to 265 basis vectors,
in agreement with subfactorial$(6)$, \eqref{eq:Nv}.
For $\Nc=3$ this number is reduced to 145.

\TABLE[t]{\small
\centering 
\begin{tabular}{c c c c c c c c c c} 
\hline\hline\\[-2.5ex] 
$\SU(3)$ dim & 64 & 35 & $\overline{35}$ & 0\\
Multiplet & c111c111 & c111c21 & c21c111 & c21c21 \\
\hline
\hline\\[-2.5ex]
Out $g_4 g_5 g_6 $ 
& $((45)^{27}6)^{64}$  
& $((45)^{10}6)^{35}$
& $((45)^{\overline{10}}6)^{\overline{35}}$  
& $((45)^{10}6)^{c21c21}$ 
&\\ 
& 
& $((45)^{27}6)^{35}$
& $((45)^{27}6)^{\overline{35}}$  
& $((45)^{\overline{10}}6)^{c21c21}$ 
&\\ 
& 
& 
& 
& $((45)^{27}6)^{c21c21}$ 
&\\ 
& 
& 
& 
& $((45)^{0}6)^{c21c21}$ 
&\\ 
\hline \hline\\[-2.5ex]
$\SU(3)$ dim & 0 & 0 & 0 & 0 &0\\
Multiplet & c111c3 & c3c111 & c21c3 & c3c21 & c3c3\\ [0.5ex] 
\hline\\[-2.5ex]
Out $g_4 g_5 g_6 $ 
& $((45)^{10}6)^{c111c3}$  
& $((45)^{\overline{10}}6)^{c3c111}$
& $((45)^{10}6)^{c21c3}$  
& $((45)^{\overline{10}}6)^{c3c21}$ 
& $((45)^{0}6)^{c3c3}$ 
\\ 
& 
& 
& $((45)^{0}6)^{c21c3}$  
& $((45)^{0}6)^{c3c21}$ 
& 
\\ 
\hline 
\end{tabular}
\caption{\label{table:3gNew} 
The new multiplets for $3g \to 3g$, appearing in addition to those
listed in \tabref{table:1qqbar4g}.
Putting together the information here and in \tabref{table:1qqbar4g}
we note that there are 
$2^2+9^2+6^2+6^2+6^2+6^2+1+2^2+2^2+ 4^2+1+1+2^2+2^2+1=265$ bases vectors,
reducing to $2^2+8^2+4^2+4^2+6^2+1+2^2+2^2=145$ for SU(3).
}
}
To construct the corresponding basis we note that we may divide the basis
vectors into four different categories, according to their first 
occurrence zero, one, two and three. 
For the singlets we effectively have zero gluons passing from the 
incoming to the outgoing side. They are thus of the form 

\begin{equation}
  \CG^{8a/s,1}_{g_1\,g_2\,g_3}\CG^{8a/s,1\,\dagger }_{g_4\,g_5\,g_6}
  \propto  \parbox{5cm}{\epsfig{file=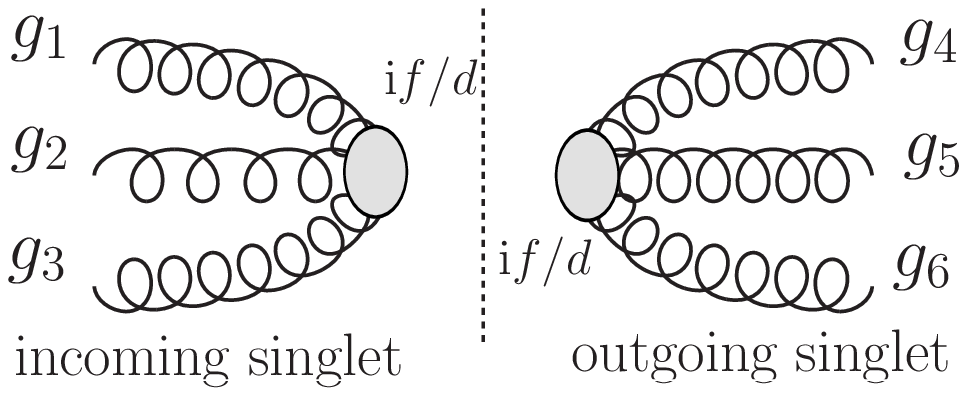,width=5cm}} \, 
\end{equation}
where -- as opposed to in the projector case, c.f.\;\eqref{eq:P=CCdag} --
the color structure on the incoming side in general differs from that on the 
outgoing side, giving rise to a total of four singlets.

Similarly, in order to construct the basis vectors corresponding to multiplets
with first occurrence one (octets), we note that they may be drawn 
having one gluon passing from the incoming to the outgoing side, for example 
\begin{equation}
  \CG^{M_{2},8}_{g_1\,g_2\,g_3;\,i_1}\CG^{ 8a/s,8a/s\,\dagger}_{i_1;g_4\,g_5\,g_6}
  \propto  \parbox{6.5cm}{\epsfig{file=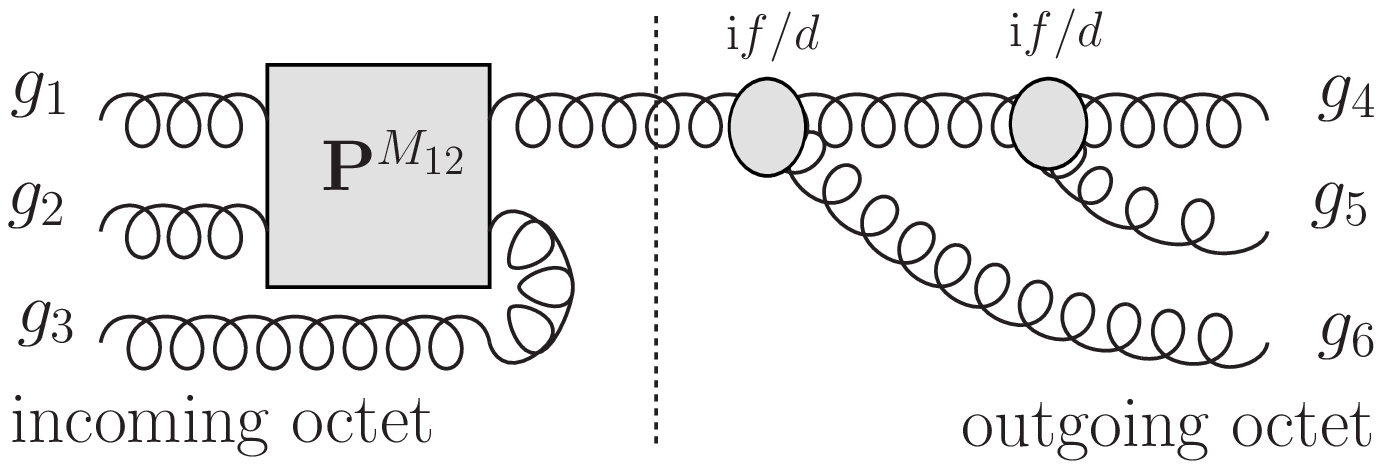,width=6.5cm}} . 
\end{equation}
Moreover, the first occurrence two projectors can be treated in the 
same way, e.g
\begin{eqnarray}
  & &\CG^{M_{2},M_{3}}_{g_1\,g_2\,g_3;\,i_1\,i_2}\CG^{8a/s,M_{3}\,\dagger}_{i_1\,i_2;\,g_4\,g_5\,g_6}
  \propto
  \parbox{10cm}{\epsfig{file=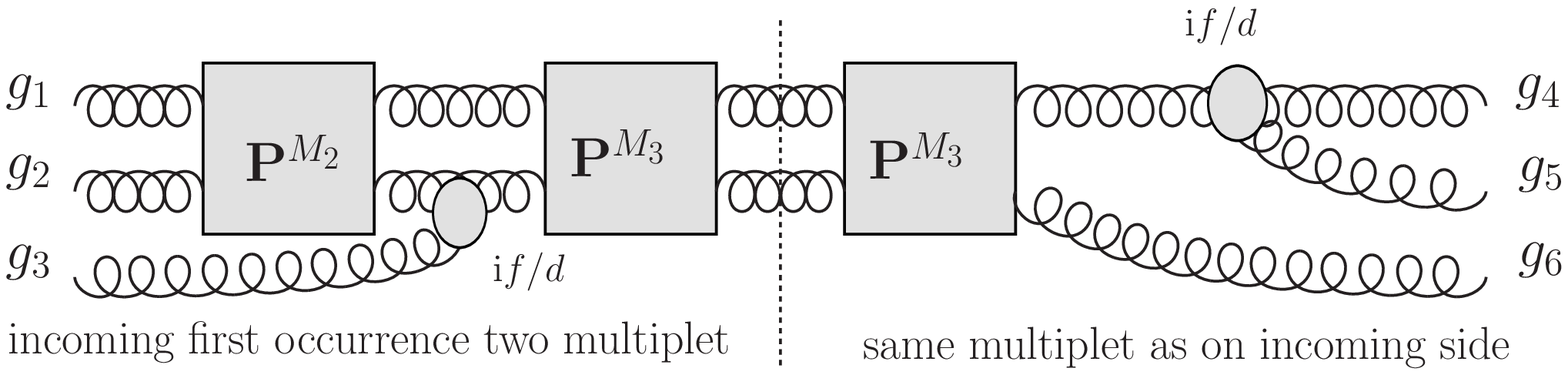,width=10cm}}.\nonumber\\
\end{eqnarray}
For the first occurrence zero, one, and two case, 
the normalization of the basis vectors is always given by \eqref{eq:VecNorm}.

For multiplets with first occurrence three and higher we 
encounter a new situation. Even for $\Nf$ gluons 
there are several instances of the same multiplet corresponding to different
construction histories. Therefore the basis vectors are in general
{\it not } proportional to

\begin{eqnarray}
  \CG^{M_{2},M_{3}}_{g_1\,g_2\,g_3;\,i_1\,i_2\,i_3}\CG^{ M'_{2},M_{3}\,\dagger}_{i_1\,i_2\,i_3;\,g_4\,g_5\,g_6}
  \label{eq:6gf3}
\end{eqnarray}
for all first occurrence 3 projectors. 
They can only be constructed in this way if $M_{2}=M'_{2}$, 
in which case the six gluon vector is
proportional to the corresponding $ggg\to ggg$ projector,
and the normalization is given by \eqref{eq:HermProjSquare}. 
If $M_{2}\ne M'_{2}$ the above color structure vanishes, due to the 
construction history property (iii) in \secref{sec:Gluons}.
In order to find the corresponding vector we instead permute the gluon 
lines before contracting the projectors

\begin{eqnarray}
  \Proj^{M_{2},M_{3}}_{g_1\,g_2\,g_3\,i_1\,g_6\,i_2}\Proj^{M'_{2}}_{i_1\,i_2\,g_4\,g_5}
  \propto
  \parbox{8cm}{\epsfig{file=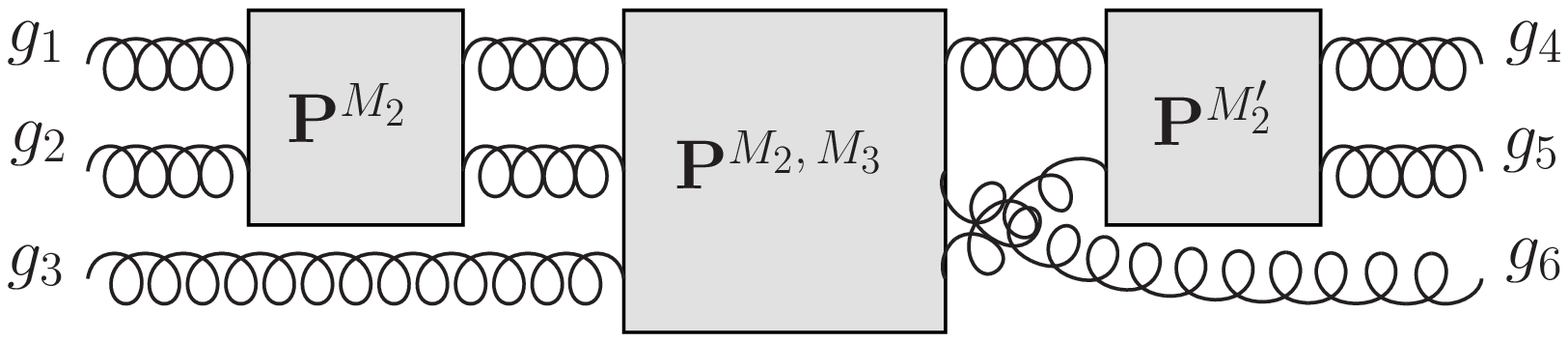,width=8cm}}.
  \label{eq:6gf3v2}
\end{eqnarray}
The electronically attached six gluon vectors corresponding to first occurrence 
three multiplets are constructed in this way.
Having more than three gluons, there are yet more possibilities of having 
transitions between different instances of the same multiplets on the 
incoming and outgoing side.
In this case we remark that having all projectors 
at hand, we can always construct the basis
by pursuing the same strategy as in \eqref{eq:any_old}.

\section{Multiplet basis for two $\qqbar$ pairs and two gluons}
\label{sec:2qqbar2gVectors}

Here the  multiplet basis for $\q_1 \qbar_2 g_3 \to q_4 \qbar_5 g_6$,
discussed in \secref{sec:2qqbar2g} is given:

\begin{eqnarray}
\Vec^{8,1;8,1}_{q_1\,q_2\,g_3\,q_4\,q_5\,g_6}&=& 
\frac{1}{\TR}
\frac{1}{\Nc^2-1}
 (t^{g_3})^{q_1}_{q_2}\; (t^{g_6})^{q_5}_{q_4}\nonumber \\
\Vec^{1,8;1,8}_{q_1\,q_2\,g_3\,q_4\,q_5\,g_6}&=& 
\frac{1}{\Nc \sqrt{\Nc^2-1}}
\delta^{q_1}_{q_2}\; \delta_{g_3\,g_6}\; \delta^{q_5}_{q_4} \nonumber \\
\Vec^{1,8;8,8s}_{q_1\,q_2\,g_3\,q_4\,q_5\,g_6}&=& 
\frac{1}{\TR}
\frac{1}{\sqrt{2(\Nc^4-5\Nc^2+4)}}
\delta^{q_1}_{q_2}\;d_{g_3\,g_6\,i_1}\; (t^{i_1})^{q_5}_{q_4}\nonumber \\
\Vec^{1,8;8,8a}_{q_1\,q_2\,g_3\,q_4\,q_5\,g_6}&=& 
\frac{1}{\TR}
\frac{1}{\Nc\sqrt{2(\Nc^2-1)}}
\delta^{q_1}_{q_2}\;\ui f_{g_3\,g_6\,i_1}\; (t^{i_1})^{q_5}_{q_4}\nonumber \\
\Vec^{8,8s;1,8}_{q_1\,q_2\,g_3\,q_4\,q_5\,g_6}&=&
\frac{1}{\TR}
\frac{1}{\sqrt{2(\Nc^4-5\Nc^2+4)}}
(t^{i_1})^{q_1}_{q_2}\; d_{i_1\,g_3\,g_6}\; \delta^{q_5}_{q_4}\nonumber \\
\Vec^{8,8s;8,8s}_{q_1\,q_2\,g_3\,q_4\,q_5\,g_6}&=&
\frac{1}{\TR^2}
\frac{N}{2(\Nc^2-4)\sqrt{\Nc^2-1}}
(t^{i_1})^{q_1}_{q_2} \; d_{i_1\,g_3\,i_2}\; d_{i_2\,g_6\,i_3}\; (t^{i_3})^{q_5}_{q_4}\nonumber \\
\Vec^{8,8s;8,8a}_{q_1\,q_2\,g_3\,q_4\,q_5\,g_6}&=&
\frac{1}{\TR^2}
\frac{1}{2\sqrt{\Nc^4-5\Nc^2+4}}
(t^{i_1})^{q_1}_{q_2} \; d_{i_1\,g_3\,i_2}\; \ui f_{i_2\,g_6\,i_3}\; (t^{i_3})^{q_5}_{q_4}\nonumber \\
\Vec^{8,8a;1,8}_{q_1\,q_2\,g_3\,q_4\,q_5\,g_6}&=&
\frac{1}{\TR}
\frac{1}{\Nc \sqrt{2(\Nc^2-1)}}
(t^{i_1})^{q_1}_{q_2}\; \ui f_{i_1\,g_3\,g_6}\; \delta^{q_5}_{q_4}\nonumber \\
\Vec^{8,8a;8,8s}_{q_1\,q_2\,g_3\,q_4\,q_5\,g_6}&=&
\frac{1}{\TR^2}
\frac{1}{2\sqrt{\Nc^4-5\Nc^2+4}}
(t^{i_1})^{q_1}_{q_2} \; \ui f_{i_1\,g_3\,i_2}\; d_{i_2\,g_6\,i_3}\; (t^{i_3})^{q_5}_{q_4}\nonumber \\
\Vec^{8,8a;8,8a}_{q_1\,q_2\,g_3\,q_4\,q_5\,g_6}&=&
\frac{1}{\TR^2}
\frac{1}{2 \Nc \sqrt{\Nc^2-1}}
(t^{i_1})^{q_1}_{q_2} \; \ui f_{i_1\,g_3\,i_2}\; \ui f_{i_2\,g_6\,i_3}\; (t^{i_3})^{q_5}_{q_4}\nonumber \\
\Vec^{8,10;8,10}_{q_1\,q_2\,g_3\,q_4\,q_5\,g_6}&=&
\frac{1}{\TR}
\frac{2}{\sqrt{\Nc^4-5\Nc^2+4}}
(t^{i_1})^{q_1}_{q_2} \; \Proj^{10}_{g_3\;i_1 g_6\,i_2} (t^{i_2})^{q_5}_{q_4}\nonumber \\
\Vec^{8,\overline{10};8,\overline{10}}_{q_1\,q_2\,g_3\,q_4\,q_5\,g_6}&=&
\frac{1}{\TR}
\frac{2}{\sqrt{\Nc^4-5\Nc^2+4}}
(t^{i_1})^{q_1}_{q_2} \; \Proj^{\overline{10}}_{g_3\;i_1\,g_6\,i_2} (t^{i_2})^{q_5}_{q_4}\nonumber \\
\Vec^{8,27;8,27}_{q_1\,q_2\,g_3\,q_4\,q_5\,g_6}&=&
\frac{1}{\TR}
\frac{2}{\Nc\sqrt{\Nc^2+2\Nc-3}}
(t^{i_1})^{q_1}_{q_2} \; \Proj^{27}_{g_3\;i_1\,g_6\,i_2} (t^{i_2})^{q_5}_{q_4}\nonumber \\
\Vec^{8,0;8,0}_{q_1\,q_2\,g_3\,q_4\,q_5\,g_6}&=&
\frac{1}{\TR}
\frac{2}{\Nc\sqrt{\Nc^2-2\Nc-3}}
(t^{i_1})^{q_1}_{q_2} \; \Proj^{0}_{g_3\;i_1\,g_6\,i_2} (t^{i_2})^{q_5}_{q_4}.
\end{eqnarray}
The normalization is fixed such that the norm square is 1.
Note that for the basis vectors with the same construction history
on the incoming and outgoing side, the vectors are proportional to
the corresponding projector, and the normalization is given by
\eqref{eq:HermProjSquare}.

\bibliographystyle{JHEP} 
\bibliography{Refs} 

\end{document}